\documentclass{pinchcrn}
\pdfoutput=1

\usepackage{graphicx,epsfig}
\usepackage{braket,slashed,mathrsfs,bbm,overpic}
\usepackage{bm,amsmath,amssymb}

\definecolor{refkey}{rgb}{1.0,0.0,0.0}
\definecolor{labelkey}{rgb}{0,0.5,0.0}

\usepackage{tikz}
\usetikzlibrary{arrows,decorations.pathmorphing,backgrounds,positioning,fit,petri,automata,shadows,calendar,mindmap,
decorations.markings,calc}

\usepackage{cancel}
\usepackage[normalem]{ulem}

\newcommand\vev[1]{\left \langle #1 \right \rangle}
\newcommand\abs[1]{\left | #1 \right |}
\newcommand{\MSbar}{\ensuremath{\overline{\text{MS}}}}

\newcommand\nn{\nonumber \\ }
\newcommand\rd{\text{d}}
\newcommand\eUV{\epsilon_\text{UV}}
\newcommand\eIR{\epsilon_\text{IR}}

\def\d{\mathsf{d}}
\def\opdim{\mathscr{D}}

\def\dd{{\rm d}}

\newcommand{\beq}{\begin{equation} }
\newcommand{\eeq}{\end{equation}}
\newcommand{\brackets}[1]{\left ( #1 \right )}

\newcounter{originalchapter}
\newenvironment{Lexercise}[2]{
\setcounter{originalchapter}{\value{chapter}}
\setcounter{chapter}{#1}
\setcounter{exercise}{#2 - 1}
\begin{exercise}
\setcounter{chapter}{\theoriginalchapter}
}
{
\end{exercise}
}
\newenvironment{Lexercisenb}[2]{
\setcounter{originalchapter}{\value{chapter}}
\setcounter{chapter}{#1}
\setcounter{exercise}{#2 - 1}
\begin{exercisenb}
\setcounter{chapter}{\theoriginalchapter}
}
{
\end{exercisenb}
}

\numberwithin{equation}{section}

\usepackage{float}

\usepackage{placeins}
\makeatletter
\DeclareRobustCommand*\cal{\@fontswitch\relax\mathcal}
\makeatother
\renewcommand{\k}{\mathbf{k}}
\def\wv#1{\widetilde{#1}}
\def\bea{\begin{eqnarray}} \def\eea{\end{eqnarray}}
\def\bean{\begin{eqnarray*}} \def\eean{\end{eqnarray*}}
\def\eq#1{(\ref{#1})} \def\nonu{\nonumber}
\def\bm#1{\text{\boldmath$#1$}}
\def\l{\left}			\def\r{\right}

\usepackage{amsfonts}
\usepackage{amssymb}
\usepackage{physics}
\usepackage{empheq}
\usepackage[force]{feynmp-auto} 

\newcommand{\tb}[1]{\textbf{#1}}
\newcommand{\tx}[1]{\text{#1}}
\def\mean#1{\left< #1 \right>}

\newcommand{\xx}{\frac{\xi}{\sigma^2}}

\definecolor{goodred}{rgb}{.8,0,0}

\title{
\centerline{\vspace{2cm}\hspace{0.7cm}\hbox{\includegraphics[width=18cm]{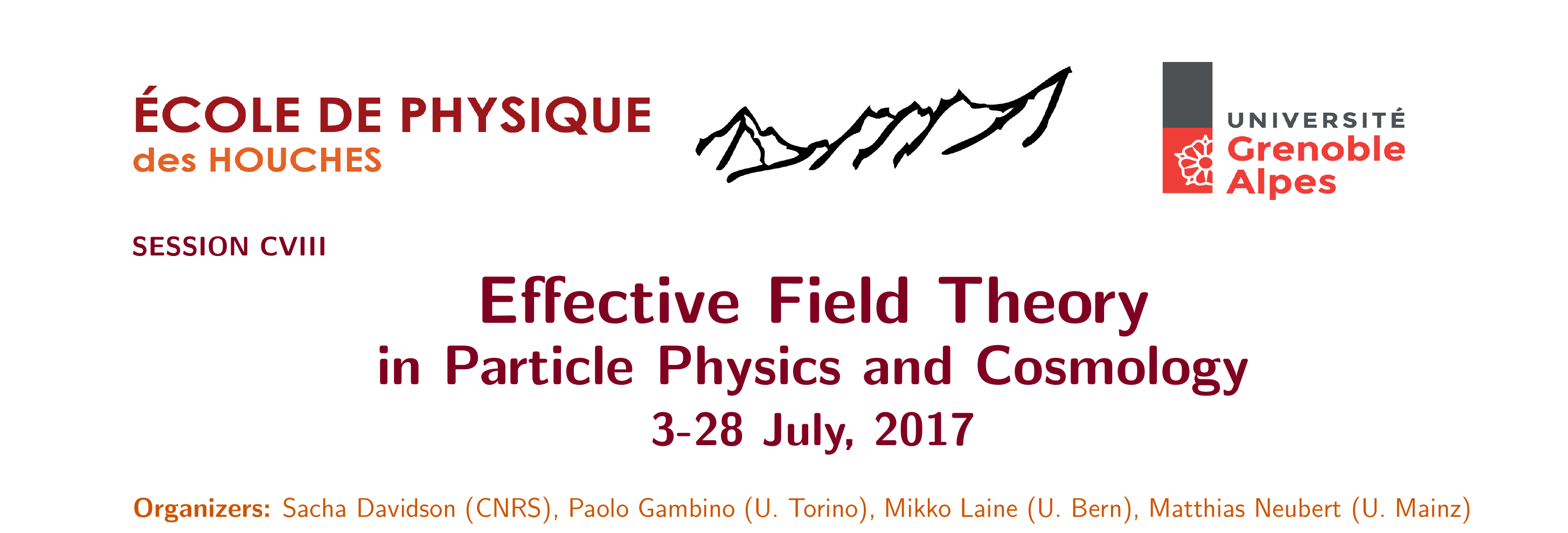}}}
Solutions to Problems at Les Houches Summer School on EFT}

\author{Marcel~Balsiger$^a$, Marios~Bounakis$^b$, Mehdi~Drissi$^{c,d}$, John~Gargalionis$^e$, Erik~Gustafson$^f$, Greg~Jackson$^a$, Matthew~Leak$^g$, Christopher~Lepenik$^h$, Scott~Melville$^{i,j}$, Daniel~Moreno$^k$, Michele~Tammaro$^l$, Selim~Touati$^m$, Timothy~Trott$^n$}
\affiliation{$a$\hspace{0.3cm}Albert Einstein Center for Fundamental Physics, Institut f\"ur Theoretische Physik, \\Universit\"at Bern, Sidlerstrasse 5, CH-3012 Bern, Switzerland\\ \vspace{0.1cm}
$b$\hspace{0.3cm}School of Mathematics, Statistics and Physics, Newcastle University, \\Newcastle Upon Tyne, NE1 7RU, United Kingdom\\ \vspace{0.1cm}
$c$\hspace{0.3cm}IRFU, CEA, Universit\'e Paris-Saclay, F-91191 Gif-sur-Yvette, France\\ \vspace{0.1cm}
$d$\hspace{0.3cm}Department of Physics, University of Surrey, Guildford GU2 7XH, United Kingdom \\ \vspace{0.1cm}
$e$\hspace{0.3cm}ARC Centre of Excellence for Particle Physics at the Terascale, School of Physics, The~University of Melbourne, Victoria 3010, Australia \\ \vspace{0.1cm}
$f$\hspace{0.3cm}Department of Physics and Astronomy, The University of Iowa, Iowa City, IA 52242, USA\\ \vspace{0.1cm}
$g$\hspace{0.3cm}Theoretical Physics Division, Department of Mathematical Sciences, \\University of Liverpool, Liverpool L69 3BX, United Kingdom\\ \vspace{0.1cm}
$h$\hspace{0.3cm}University of Vienna, Faculty of Physics, Boltzmanngasse 5, A-1090 Wien, Austria\\ \vspace{0.1cm}
$i$\hspace{0.3cm}DAMTP, University of Cambridge, Wilberforce Road, Cambridge, \\CB3 0WA, United Kingdom\\ \vspace{0.1cm}
$j$\hspace{0.3cm}Emmanuel College, University of Cambridge, St Andrew’s Street, Cambridge, \\CB2 3AP, United Kingdom\\ \vspace{0.1cm}
$k$\hspace{0.3cm}Grup de F\'\i sica Te\`orica, Dept. F\'\i sica and IFAE-BIST, Universitat Aut\`onoma de Barcelona, E-08193 Bellaterra (Barcelona), Spain\\ \vspace{0.1cm}
$l$\hspace{0.3cm}Department of Physics, University of Cincinnati, Cincinnati, Ohio 45221, USA\\ \vspace{0.1cm}
$m$\hspace{0.3cm}Laboratoire de Physique Subatomique et de Cosmologie, Universit\'e Grenoble-Alpes, CNRS/IN2P3, Grenoble INP, 38000 Grenoble, France\\ \vspace{0.1cm}
$n$\hspace{0.3cm}Department of Physics, University of California, Santa Barbara, CA 93106, USA
}

\begin{document}

\maketitle

\tableofcontents

\maintext

\chapter{~Solutions}
This work details worked solutions to the various problems set by the lecturers
during the course of the Les Houches summer school 2017 on effective field theories in particle physics and cosmology and is based on the final chapter of \cite{LesHouchesBook}.
Further exercises that were added after the school are not solved here, and are left as a challenge for the enterprising reader.

\section{Preface}

The topic of the CVIII session of
the \'Ecole de physique des Houches, held in July 2017, was
Effective Field Theory (EFT) in Particle Physics and Cosmology.

In both particle physics and cosmology, our current best understanding
is captured by a ``Standard Model'' which describes all known interactions,
but which is nevertheless unsatisfactory from a number of points of view.
The hope is that eventually experimental deviations from the Standard
Model are found, which then indicate a road to physics Beyond
the Standard Model (BSM). Ideally this leads, amongst others,
to a verifiable explanation for the origin of Dark Matter.

In order to be sure that a given deviation is really an indication
of BSM physics, Standard Model predictions must be robust and accurate.
This represents a non-trivial challenge, as many
particles, with varying masses and momenta, can participate in
the interactions. Their effects, independently of whether the particles
appear as real or virtual states, need to be systematically accounted for.
As EFTs help disentangling the effects of physics at different scales, they simplify considerably this task.
These scales could be masses (for instance, the light masses of the particles of the Standard Model versus the heavy masses of yet undiscovered particles), momenta (for instance, the hard, collinear and soft momenta playing a role in jets produced in high-energy collisions), or length scales (for instance, the lattice spacing appearing in numerical simulations versus the pion Compton wavelength of interest to low-energy hadronic interactions).

The underlying idea of EFT is that at each scale the relevant physics can be parametrised with appropriate variables, which may change with the scale.
In order to achieve precise predictions there is often no need to know the underlying exact theory: one can work with simpler field theories, describing only the degrees of freedom
relevant at a certain scale, while performing a systematic expansion in
one or more small parameters, generally ratios of well-separated scales.
Hence EFTs are essential tools both for precision analyses within known multi scale theories, such as the Standard Model, and also for a concise parameterisation of hypothetical BSM models.

The goal of this school was
to offer a broad introduction to the foundations and modern applications
of Effective Field Theory in many of its incarnations.
The basic foundations were laid out in two lecture series,
by Matthias Neubert\cite{Neubert:2019mrz}
and Aneesh Manohar\cite{Manohar:2018aog}, which review
the field-theoretic background of renormalization, power counting,
and operator classification.
Effective Field Theories for treating systems with
spontaneously broken global symmetries
are introduced by Antonio Pich \cite{Pich:2018ltt},
and applications of EFT to the analysis of inflation
and Large Scale Structure formation were
respectively presented by Cliff Burgess \cite{Burgess:2017ytm},
and Tobias Baldauf \cite{Baldauf:lectures}.
Ubirajara van Kolck introduced some
uses of Effective Field Theory in nuclear physics\cite{vanKolck:2019vge}.
In all these lecture series, problems were proposed
to the students, who wrote up the solutions as
this document. There were additional lectures,
without formal problems, by Thomas Mannel about Heavy Quark EFT,
about Soft Collinear Effective Theory by Thomas Becher,
about flavour physics by Luca Silvestrini,
Dark Matter by Junji Hisano and lattice by
Rainer Sommer.

\newpage

\section{Introduction to Renormalisation and the Renormalisation Group (Neubert)}

The lectures \cite{Neubert:2019mrz} provide an overview of renormalization in quantum field theories, with particular attention paid to effective quantum field theories---in which the renormalization of composite operators, operator mixing under scale evolution, and the resummation of large logarithms are important.

This section contains three introductory problems, which cover propagators and 1PI diagrams, superficial degrees of divergence and Renormalization Group Equations.

\begin{Lexercise}{1}{4}

Express the propagator of a spin-1 field in terms of 1PI self-energy diagrams.

\medskip

\noindent {\bf SOLUTION:}

From the Lagrangian, we can derive the classical equations of motion,
\begin{equation}
\mathcal{L} = - \frac{1}{4} F_{\mu \nu}^2 + \frac{1}{\xi} ( \partial_\mu A^\mu )^2 + J_\mu A^\mu \;\; \implies \;\; \left[ p^2 \eta_{\mu \nu} - \left( 1- \frac{1}{\xi} \right) p_\mu p_\nu \right] A_\nu = J_\nu
\end{equation}
Inverting this provides the Feynman propagator (classical Green's function),
\begin{equation}
P_{\mu \nu} = \frac{-i}{p^2} \left[ \eta_{\mu \nu} - ( 1- \xi) \frac{p_\mu p_\nu}{p^2} \right]
\end{equation}
where the $+ i \epsilon$ Feynman prescription for the poles is implicit.

The 1PI diagrams must satisfy the Ward identity by gauge invariance, and so we can write them as,
\begin{equation}
p^2 \Pi ( p ) X^{\mu \nu} \;\;\;\; \text{where} \;\;\;\; X^{\mu \nu} = \eta^{\mu \nu} - p^\mu p^\nu / p^2
\end{equation}
where $X^{\mu \nu}$ is a projection onto the physical (transverse) polarization states. As a projection, it is idempotent,
$$ X^{\mu \nu} \eta_{\nu \alpha} X^{\alpha \beta} = X^{\mu \beta} $$
and it also annihilates the gauge-fixing term as $X^{\mu \nu} p_\mu = 0$. It is then straightforward to resum a geometric series of 1PI blobs,
\begin{align}
\Pi^{\mu \nu} &= P^{\mu \nu} + P^{\mu \alpha} X_{\alpha \beta} P^{\beta \nu} + P X P X P + .... \\
&= \frac{-i \xi p^\mu p^\nu}{ p^4} - \frac{i}{p^2} \sum_{n=0}^{\infty} \left( \Pi ( p^2) \right)^n X^{\mu \nu} \\
&= \frac{-i \xi p^\mu p^\nu}{ p^4 } - \frac{i}{p^2 (1 - \Pi (p^2) )} \left[ \eta^{\mu \nu} - \frac{p^\mu p^\nu}{p^2} \right]
\end{align}

Having extracted the factor of $k^2 X^{\mu \nu}$ from the 1PI self-energy diagrams, we find that the remaining $\Pi (k^2)$ function shifts the residue of the pole at $k^2=0$ in the resummed propagator to $1/(1-\Pi(0))$. This is exactly analogous to the residue renormalization for scalars.

\end{Lexercise}

\begin{Lexercisenb}{2}{1}
Find the superficial degree of divergence for 1PI QCD Feynman graphs.

\medskip

\noindent {\bf SOLUTION:}

Counting powers of loop momenta, we define,
\begin{equation}
D = 4 L - P_q - 2 P_g - P_c + V_{3g}
\end{equation}
as the Grassmann-valued quarks and ghosts have propagators with only one power of momenta, while the bosonic gluons has two powers of momenta. The three gluon vertex must also contain one power of momenta (by Lorentz invariance).

From the Euler characteristic,
$$ L = P_q + P_g + P_c - ( V_{qg} + V_{3g} + V_{4g} + V_{cg} ) + 1 $$
and from counting the total numbers of quark, ghost and gluon legs,
\begin{align}
2 P_q + N_q &= 2 V_{qg} \\
2 P_c + N_c &= V_{cg} \\
2 P_g + N_g &= V_{qg} + V_{cg} + 3 V_{3g} + 4 V_{4g}
\end{align}
Putting these together, one finds that,
\begin{equation}
D = 4 - N_g - \frac{3}{2} ( N_q + N_c )
\end{equation}
where $N_X$ are the numbers of external particles. Although all physical diagrams have $N_c = 0$, it is useful to keep it explicitly for analyzing subdiagrams.

\end{Lexercisenb}

\begin{Lexercisenb}{3}{1}

From the RGEs,
\begin{align}
\frac{1}{\alpha_s} \beta &= - 2 \epsilon - Z^{-1}_{\alpha} \frac{d}{d \text{ln} \mu} Z_{\alpha} \\
\gamma_m &= - \frac{1}{Z_m'} \frac{d}{d \text{ln} \mu} Z_{m}'
\end{align}
in the MS scheme,
\begin{align}
\frac{1}{\alpha_s} \beta &= - 2 \epsilon - \beta Z^{-1}_{\alpha} \frac{d}{d \alpha_s} Z_{\alpha} \\
\gamma_m &= - \frac{\beta}{Z_m'} \frac{d}{d \alpha_s } Z_{m}'
\end{align}
derive expressions for $\beta ( \alpha_s, \epsilon ), Z_\alpha ( \alpha_s, \epsilon ), \gamma ( \alpha_s, \epsilon ) , Z_\alpha ( \alpha_s, \epsilon )$ as perturbative series in $\epsilon$.

\medskip

\noindent {\bf SOLUTION:}

Assume that $\beta$ is a smooth function of $\epsilon$,
\begin{equation}
\beta = \sum_{k=0}^\infty \epsilon^k \beta^{(k)}
\end{equation}
The renormalization $Z_{\alpha}$ receives pole contributions $1/\epsilon^k$ from $k$-loop diagrams, i.e.
\begin{equation}
Z = 1 + \sum_{k=1}^{\infty } \epsilon^{-k} Z_\alpha^{(k)}
\end{equation}
where $Z_\alpha^{(k)} \sim \alpha_s^k$.
The RGEs can be written,
\begin{equation}
\beta ( \alpha, \epsilon ) = \frac{ - 2 \epsilon \alpha_s Z_{\alpha} }{ \left( 1 + \alpha_s \frac{d}{d \alpha_s} \right) Z_\alpha }
\end{equation}
and hold for all $\epsilon$. We can therefore equate the coefficients of each power of $\epsilon$,
\begin{equation}
\sum_{k=0}^\infty \epsilon^k \beta^{(k)} = - 2 \epsilon \alpha_s + 2 \alpha_s^2 \frac{d}{d \alpha_s} Z_\alpha^{(1)} + \mathcal{O} \left( \frac{1}{ \epsilon } \right)
\end{equation}
and conclude that,
\begin{equation}
\beta (\alpha_s , \epsilon ) = 2 \alpha_s^2 \frac{d}{d \alpha_s} Z_\alpha^{(1)} (\alpha_s) - 2 \alpha_s \epsilon .
\end{equation}
Substituting this into the RGE, we find an infinite number of consistency conditions which fix all $Z_{\alpha}^{(k>1)}$ in terms of $Z_{\alpha}^{(1)} ( \alpha_s )$,
\begin{align}
\frac{ d Z_{\alpha}^{(k+1)} }{ d\alpha_s} = \frac{d Z_\alpha^{(1)}}{d \alpha_s} \left[ 1 + \alpha_s \frac{ d }{ d\alpha_s} \right] Z_\alpha^{(k)} \;\;\;\; \forall \;\;\;\; k \geq 1
\end{align}
Note that by taking repeated $\alpha_s$ derivatives, this gives every $\alpha_s^n$ coefficient of $Z_{\alpha}^{(k+1)}$ (we don't have $Z_{\alpha_s}^{(k+1)} |_{\alpha_s =0}$, but we know that this is zero).

Similarly, assume that $\gamma_m$ is a smooth function of $\epsilon$,
\begin{equation}
\gamma_m = \sum_{k=0}^\infty \epsilon^k \gamma_m^{(k)}
\end{equation}
and then the positive powers of $\epsilon$ in the RGE give,
\begin{equation}
\gamma_m = 2 \alpha_s \frac{d}{d \alpha_s} Z_m'^{(1)} ( \alpha_s )
\end{equation}
and similarly we have consistency relations,
\begin{equation}
\frac{d Z_m'^{(k+1)} }{d \alpha_s} =
\left[ \frac{d Z_m'^{(1)} }{d \alpha_s} + \alpha_s \frac{d Z_\alpha^{(1)} }{d \alpha_s} \frac{d}{d \alpha_s} \right] Z_m'^{(k)}
\end{equation}

Physically, this is because the higher order poles in $Z$ come from multiple subdivergences, rather than than a genuine $k$ loop divergence.

\end{Lexercisenb}

\section{Introduction to Effective Field Theories (Manohar)}

The lectures \cite{Manohar:2018aog} covered introductory material on EFTs as used in
high-energy physics to compute experimentally observable quantities.
The following exercise solutions treat power counting, loop
corrections, field redefinitions and their relation to the equations
of motion, decoupling of heavy particles, naive dimensional analysis,
and the Standard Model Effective Field Theory (SMEFT) and many more
basic concepts.

Useful references for solving the problems are

\begin{itemize}

\item EFT: \cite{Manohar:1996cq,Manohar:2003vb}

\item Power Counting: \cite{Manohar:1983md,Gavela:2016bzc}

\item Matching in HQET and field redefinitions: \cite{Manohar:1997qy}

\item Invariants: \cite{Jenkins:2009dy,Hanany:2010vu,Lehman:2015via,Henning:2015alf,Henning:2017fpj}

\item SMEFT: \cite{Jenkins:2013zja,Jenkins:2013wua,Alonso:2013hga,Alonso:2014rga}

\end{itemize}

Equations from the lecture notes are referred to with a prefix L, e.g. eqn~(L1.1).

\begin{Lexercise}{1}{1}

Show that for a \emph{connected} graph, $V-I+L=1$, where $V$ is the number of vertices, $I$ is the number of internal lines, and $L$ is the number of loops. What is the formula if the graph has $n$ connected components?

\medskip

\noindent {\bf SOLUTION:}

Consider a connected graph $G$ where $V$ is the number of vertices, $I$ is the number of internal lines, and $L$ is the number of loops. Since the identity does not depend on the external lines, erase them. Take an internal line (edge) that can be deleted without making the graph disconnected, and remove it. Then clearly $I \to I-1$. If the edge joins vertices $v_1$ and $v_2$ and removing the edge leaves the graph connected, there must be a second path through the graph between $v_1$ and $v_2$, i.e.\ there is a loop which is removed when the edge is removed, so $L \to L-1$. The operation leaves $V-I+L$ invariant. Proceed this way till removing an edge makes the graph disconnected, i.e.\ there are no loops, and we are left with a tree graph.

Now pick any vertex $v$ in the graph, and move through the graph without retracing your path. There are no loops, so the path does not end back at $v$. Since the number of vertices is finite, eventually the path must end at a vertex (a ``leaf node'' of the tree graph). Remove this last node and edge. Then $V \to V-1$ and $I \to I-1$ keeping $V-I+L$. Keep repeating this process until $I=0$. Then we have a graph with one vertex $V=1$, $I=0$, $L=0$, $V-I+L=1$. Since our operators preserved $V-I+L=1$, this completes the proof. If there are $n$ connected components, the formula holds for each component, so the total is
$V-I+L=n$ since the components have no vertices, edges or loops in common.

\end{Lexercise}
\begin{Lexercisenb}{1}{2}

Work out the transformation of fermion bilinears $\overline \psi(\mathbf{x},t)\, \Gamma\, \chi(\mathbf{x},t)$ under $C$, $P$, $T$, where $\Gamma=P_L, P_R,\gamma^\mu P_L, \gamma^\mu P_R ,\sigma^{\mu \nu} P_L, \sigma^{\mu \nu}P_R$. Use your results to find the transformations under $CP$, $CT$, $PT$ and $CPT$.

\medskip

\noindent {\bf SOLUTION:}

Under parity $\mathscr{P}$, charge conjugation $\mathscr{C}$ and time-reversal $\mathscr{T}$,
\begin{align}
\mathscr{P} \psi(\mathbf{x},t) \mathscr{P}^{-1} &= \gamma^0 \psi(-\mathbf{x},t), \nonumber \\
\mathscr{C} \psi(\mathbf{x},t) \mathscr{C}^{-1} &= i \gamma^2 \psi(\mathbf{x},t)^{\dagger T}, \nonumber \\
\mathscr{T} \psi(\mathbf{x},t) \mathscr{T}^{-1} &= i \gamma^1 \gamma^3 \psi(\mathbf{x},-t) \,.
\end{align}
Note that $\mathscr{C}$ is a \emph{unitary} operator, and $\mathscr{T}$ is \emph{antiunitary} despite the $\dagger$ for the $\mathscr{C}$ transformation and not for the $\mathscr{T}$ transformation. $\mathscr{P}, \mathscr{C}$ are operators and commute with $\gamma$ matrices; $\mathscr{T}$ complex conjugates the $\gamma$ matrices. Transformations of $\overline \psi$ are given by taking conjugates of the above.

Under parity,
\begin{align}
\mathscr{P} \overline \psi(\mathbf{x},t)\Gamma \chi(\mathbf{x},t) \mathscr{P}^{-1} &= \mathscr{P} \overline \psi(\mathbf{x},t) \mathscr{P}^{-1} \Gamma \mathscr{P} \chi(\mathbf{x},t) \mathscr{P}^{-1} =
\overline \psi(-\mathbf{x},t) \gamma^0 \Gamma \gamma^0 \chi(-\mathbf{x},t).
\end{align}

Under charge conjugation,
\begin{align}
\mathscr{C} \overline \psi(\mathbf{x},t)\Gamma \chi(\mathbf{x},t) \mathscr{C}^{-1} &= \mathscr{C} \overline \psi(\mathbf{x},t) \mathscr{C}^{-1} \Gamma \mathscr{C} \chi(\mathbf{x},t) \mathscr{C}^{-1} =
\psi(\mathbf{x},t)^T i\gamma^2 \gamma^0 \Gamma i \gamma^2 \chi(\mathbf{x},t)^{\dagger T}.
\end{align}
Taking the transpose (and including the Fermi minus sign for the exchange of operators),
\begin{align}
\mathscr{C} \overline \psi(\mathbf{x},t)\Gamma \chi(\mathbf{x},t) \mathscr{C}^{-1}&=
-\chi(\mathbf{x},t)^\dagger (i\gamma^2)^T \Gamma^T (\gamma^0)^T (i \gamma^2)^T \psi(\mathbf{x},t) \nonumber \\
&=
-\overline \chi(\mathbf{x},t) (i\gamma^0 \gamma^2) \Gamma^T (i\gamma^0 \gamma^2) \psi(\mathbf{x},t) \nonumber \\
&=
\overline \chi(\mathbf{x},t)\, C\, \Gamma^T C^{-1} \psi(\mathbf{x},t), & C &= i \gamma^0 \gamma^2
\end{align}

Under time-reversal,
\begin{align}
\mathscr{T} \overline \psi(\mathbf{x},t)\Gamma \chi(\mathbf{x},t) \mathscr{T}^{-1} &= \mathscr{T} \overline \psi(\mathbf{x},t) \mathscr{T}^{-1} \mathscr{T} \Gamma \mathscr{T}^{-1} \mathscr{T} \chi(\mathbf{x},t) \mathscr{T}^{-1} \nonumber \\
&=
\overline \psi(\mathbf{x},-t) \gamma^0 i \gamma^1 \gamma^3 \gamma^0 \Gamma^* i \gamma^1 \gamma^3 \chi(\mathbf{x},-t) \nonumber \\
&=
\overline \psi(\mathbf{x},-t) (i \gamma^1 \gamma^3) \Gamma^* (i \gamma^1 \gamma^3) \chi(\mathbf{x},-t) \nonumber \\
&=
\overline \psi(\mathbf{x},-t) \, T\, \Gamma^* T^{-1} \chi(\mathbf{x},-t), & T &= i \gamma^1 \gamma^3
\end{align}

The transformations can then be determined by computing $\gamma^0 \Gamma \gamma^0$, $(i\gamma^0 \gamma^2) \Gamma^T (i\gamma^0 \gamma^2)$ and $ (i \gamma^1 \gamma^3) \Gamma^* (i \gamma^1 \gamma^3)$ to give the results in the table. The second line gives the coordinate arguments, and
\begin{align*}
\hat \mu = \left\{ \begin{array}{ll}
\mu & \mbox{if $\mu=0$} \\
-\mu & \mbox{if $\mu=1,2,3$} \end{array} \right.
\end{align*}

\begin{align*}
\setlength{\tabcolsep}{3mm}
\renewcommand{\arraystretch}{2}
\begin{array}{|c|ccc|}
\hline
& \mathscr{C} & \mathscr{P} & \mathscr{T} \\ \hline
(\mathbf{x},t) & (\mathbf{x},t) & (-\mathbf{x},t) & (\mathbf{x},-t) \\ \hline
\bar\chi P_L \psi & \bar\psi P_L \chi & \bar\chi P_R \psi & \bar\chi P_L \psi
\\
\bar \chi P_R \psi & \bar \psi P_R \chi
& \bar \chi P_L \psi & \bar \chi P_R \psi \\
\bar \chi \gamma^\mu P_L \psi & -\bar \psi \gamma^\mu P_R
\chi & \bar \chi \gamma^{\hat \mu} P_R \psi & \bar \chi \gamma^{\hat \mu} P_L \psi \\
\bar \chi \gamma^\mu P_R\psi &
-\bar \psi \gamma^\mu P_L \chi & \bar \chi \gamma^{\hat \mu}P_L \psi
& \bar \chi \gamma^{\hat \mu}P_R \psi \\
\bar \chi \sigma^{\mu \nu} P_L \psi & - \bar \psi \sigma^{\mu \nu}P_L \chi & \bar \chi
\sigma^{\hat \mu \hat \nu} P_R \psi & -\bar \chi \sigma^{\hat \mu \hat \nu} P_L \psi \\
\bar \chi \sigma^{\mu \nu} P_R\psi & -\bar \psi \sigma^{\mu \nu} P_R \chi & \bar \chi \sigma^{\hat \mu
\hat \nu} P_L \psi & -\bar \chi \sigma^{\hat \mu \hat \nu} P_R \psi \\
\hline \end{array}
\end{align*}
Combining the above gives
\begin{align*}
\setlength{\tabcolsep}{3mm}
\renewcommand{\arraystretch}{2}
\begin{array}{|c|cccc|}
\hline
& \mathscr{CP} & \mathscr{PT} & \mathscr{CT} & \mathscr{CPT} \\ \hline
(\mathbf{x},t) & (-\mathbf{x},t) & (-\mathbf{x},-t) & (\mathbf{x},-t) & (-\mathbf{x},-t) \\ \hline
\bar\chi P_L \psi & \bar\psi P_R \chi & \bar\chi P_R \psi & \bar\psi P_L \chi & \bar\psi P_R \chi
\\
\bar \chi P_R \psi & \bar \psi P_L \chi
& \bar \chi P_L \psi & \bar\psi P_R \chi & \bar\psi P_L \chi \\
\bar \chi \gamma^\mu P_L \psi & -\bar \psi \gamma^{\hat \mu} P_L
\chi & \bar \chi \gamma^{\mu} P_R \psi & -\bar \psi \gamma^{\hat \mu} P_R \chi & - \bar \psi \gamma^\mu P_L \chi \\
\bar \chi \gamma^\mu P_R \psi & -\bar \psi \gamma^{\hat \mu} P_R
\chi & \bar \chi \gamma^{\mu} P_L \psi & -\bar \psi \gamma^{\hat \mu} P_L \chi & - \bar \psi \gamma^\mu P_R \chi \\
\bar \chi \sigma^{\mu \nu} P_L \psi & - \bar \psi \sigma^{\hat \mu \hat \nu}P_R \chi & -\bar \chi
\sigma^{ \mu \nu} P_R \psi & \bar \psi \sigma^{\hat \mu \hat \nu} P_L \chi & \bar \psi
\sigma^{ \mu \nu} P_R \chi \\
\bar \chi \sigma^{\mu \nu} P_R \psi & - \bar \psi \sigma^{\hat \mu \hat \nu}P_L \chi & -\bar \chi
\sigma^{ \mu \nu} P_L \psi & \bar \psi \sigma^{\hat \mu \hat \nu} P_R \chi & \bar \psi
\sigma^{ \mu \nu} P_L \chi \\
\hline \end{array}
\end{align*}
Note that for any operator $O(\mathbf{x},t)$, $\mathscr{CPT}$ transforms it to $(-1)^nO^\dagger(-\mathbf{x},-t)$ where $n$ is the number of Lorentz indices.
Thus the Lagrange density transforms as $\mathcal{L}(\mathbf{x},t) \to \mathcal{L}^\dagger(-\mathbf{x},-t)$, and a Hermitian action is $\mathscr{CPT}$ invariant.

\end{Lexercisenb}
\begin{Lexercisenb}{1}{3}
\label{ex:nfierz}

Show that for $SU(N)$,
\begin{align}\label{sun}
[T^A]^\alpha_{\ \beta}\, [T^A]^{\lambda}_{\ \sigma} &= \frac12 \delta^\alpha_\sigma\, \delta^\lambda_\beta - \frac{1}{2N} \delta^\alpha_\beta \, \delta^\lambda_\sigma,
\end{align}
where the $SU(N)$ generators are normalized to $\text{Tr}\, T^A T^B=\delta^{AB}/2$. From this, show that
\begin{align}\
\delta^\alpha_{\ \beta}\, \delta^{\lambda}_{\ \sigma} &= \frac{1}{N} \delta^\alpha_\sigma\, \delta^\lambda_\beta + 2 [T^A]^\alpha_{\ \sigma} \, [T^A]^\lambda_{\ \beta}, \nn
[T^A]^\alpha_{\ \beta}\, [T^A]^{\lambda}_{\ \sigma} &= \frac{N^2-1}{2N^2} \delta^\alpha_\sigma\, \delta^\lambda_\beta - \frac{1}{N} [T^A]^\alpha_{\ \sigma}\, [T^A]^\lambda_{\ \beta}.
\end{align}

\medskip

\noindent {\bf SOLUTION:}

The identity and the $T^a$ ($a=1, ..., N^2 -1 $) provide a basis for $N \times N$ matrices, so any such matrix can be decomposed as
\begin{equation}
A = c_{0} \mathbbm{1} + c_a T^a \,.
\end{equation}
If $A$ is Hermitian, the coefficients are real.
Now, assume that the Killing form is normalized as,
\begin{equation}
\text{Tr} \left( T^a T^b \right) = \frac{1}{2} \delta^{ab} \,
\end{equation}
and therefore
\begin{equation}
\text{Tr} \, A = N c_0 , \;\;\;\; \text{Tr} \, T^a A = \frac{1}{2} c_a\,.
\end{equation}
With explicit indices, this tells us that,
\begin{equation}
A_{ij} = \left( \frac{ A_{k \ell} \delta_{k \ell} }{N} \right) \delta_{ij} + \left( 2 T^a_{k \ell} A_{\ell k} \right) T^a_{ij}\,,
\end{equation}
$$ \implies \;\; A_{\ell k} \left[ \delta_{i \ell} \delta_{jk} - \frac{1}{N} \delta_{k \ell} \delta_{ij} - 2 T^a_{ij} T^a_{k \ell} \right] = 0 \,, $$
on picking out the coefficient of $A_{\ell k}$.
As this identity holds for all elements $A_{ij}$ of all Hermitian matrices, we conclude that the square bracket vanishes identically, and thus arrive at the desired Fierz identity,
\begin{equation}
T^a_{ij} T^a_{k \ell} = \frac{1}{2} \delta_{i \ell} \delta_{jk} - \frac{1}{2 N} \delta_{ij} \delta_{k \ell} \,,
\end{equation}
which is eqn~(L1.2).
Rewriting this equation as
\begin{equation}
\frac{1}{2 N} \delta_{ij} \delta_{k \ell} + T^a_{ij} T^a_{k \ell} = \frac{1}{2} \delta_{i \ell} \delta_{jk} \,,
\end{equation}
multiplying by $2$, and renaming the indices gives the first of eqn~(1.3). Adding eqn~(1.1) and the corresponding equation with $\beta \leftrightarrow \sigma$ multiplied by $1/N$ gives the second of eqn~(1.3).

\end{Lexercisenb}
\begin{Lexercisenb}{1}{4}
\label{ex:spinfierz}

Spinor Fierz identities are relations of the form
\begin{align*}
(\overline A\, \Gamma_1\, B)(\overline C\, \Gamma_2\, D) = \sum_{ij} c_{ij} (\overline C\, \Gamma_i\, B)(\overline A\, \Gamma_j\, D)
\end{align*}
where $A,B,C,D$ are fermion fields, and $c_{ij}$ are numbers. They are much simpler if written in terms of chiral fields using $\Gamma_i=P_L, P_R,\gamma^\mu P_L, \gamma^\mu P_R ,\sigma^{\mu \nu} P_L, \sigma^{\mu \nu}P_R$, rather than Dirac fields. Work out the Fierz relations for
\begin{align*}
& (\overline A P_L B)(\overline C P_L D), &&
(\overline A \gamma^\mu P_L B)(\overline C \gamma_\mu P_L D), &&
(\overline A \sigma^{\mu \nu} P_L B)(\overline C \sigma_{\mu \nu} P_L D), \nn
& (\overline A P_L B)(\overline C P_R D),&&
(\overline A \gamma^\mu P_L B)(\overline C \gamma_\mu P_R D), &&
(\overline A \sigma^{\mu \nu} P_L B)(\overline C \sigma_{\mu \nu} P_R D).
\end{align*}
Do not forget the Fermi minus sign. The $P_R \otimes P_R$ identities are obtained from the $P_L \otimes P_L$ identities by using $L \leftrightarrow R$.

\medskip

\noindent {\bf SOLUTION:}

Define
\begin{equation}
\gamma^5 = \gamma_5 = i \gamma^0 \gamma^1 \gamma^2 \gamma^3 , \;\;\;\; \sigma^{\mu \nu} = \frac{i}{2} [ \gamma^\mu, \gamma^\nu ]
\end{equation}
and raise/lower spacetime indices with $\eta_{\mu \nu} = (+1, -1,-1,-1)$.

Use a chiral basis,
$$ \left\{ \Gamma^A \right\} = \left\{ P_R, P_L, \gamma^\mu P_R , \gamma^\mu P_L, \sigma^{\mu \nu} \right\} \,.$$
$\gamma^\mu \gamma_5$ are four elements of the basis for $\mu=\{0,1,2,3\}$, and $\sigma^{\mu \nu}$ are six elements of the basis for $\mu \nu=\{01,02,03,12,23,31\}$. The 16 elements of $\Gamma^A$ are linearly independent and any $4 \times 4$ matrix can be written as a linear combination of $\Gamma^A$.
Define a dual basis,
$$\left\{ \Gamma_A \right\}= \left\{ P_R, P_L , \gamma_\mu P_L, \gamma_\mu P_R, \frac{1}{2} \sigma_{\mu \nu} \right\} $$
such that,
$$ \text{Tr} [ \Gamma_A \Gamma^B ] = 2 \delta^B_A. $$
Any $4\times4$ matrix can be written as
\begin{align}
X &= c_A \Gamma^A & c_A = \frac12 \text{Tr} \left[ X \Gamma_A \right]
\end{align}

Then it is straightforward to project the bilinears onto this basis,
\begin{equation}\label{sol1}
\left( \Gamma^A \right) \left[ \Gamma^B \right] = - \frac{1}{4} \text{Tr} \left[ \Gamma^A \Gamma_C \Gamma^B \Gamma_D \right] \;\; \left( \Gamma^D \right] \left[ \Gamma^C \right)
\end{equation}
where the overall minus sign accounts for the anti-commutativity of the fermion fields, and the type of parenthesis indicates whether the $\Gamma$ matrix is contracted with $\overline A$, $B$, $\overline C$ or $D$.

Evaluating eqn~(\ref{sol1}) for the required cases:
\begin{align}\label{sol2}
\left( P_L \right) \left[ P_L \right] &= - \frac{1}{4} \text{Tr} \left[ P_L \Gamma_C P_L \Gamma_D \right] \left( \Gamma^D \right] \left[ \Gamma^C \right) \nonumber \\
&= - \frac{1}{2} \left( P_L \right] \left[ P_L \right) - \frac{1}{16} \left( \sigma^{\mu \nu} \right] \left( \sigma_{\mu \nu} \right] - \frac{ i }{ 16 } \epsilon_{\mu \nu \alpha \beta} \left( \sigma^{\mu \nu} \right] \left( \sigma_{\alpha \beta} \right]
\end{align}
[Note that $ \left( \sigma^{\mu \nu} \right] \left( \sigma_{\mu \nu} \right]$ sums over both $\mu < \nu$ and $\mu > \nu$ and so is twice the sum over $\sigma^{\mu \nu}$ in the basis set $\Gamma^A$.]

\begin{align}\label{sol3}
\left( P_L \right) \left[ P_R \right] &= - \frac{1}{2} \left( \gamma^\mu P_R \right] \left( \gamma_\mu P_L \right]
\end{align}

\begin{align}\label{sol4}
\left( \gamma^\mu P_L \right) \left[ \gamma_\mu P_L \right] &= \left( \gamma^\mu P_L \right] \left( \gamma_\mu P_L\right]
\end{align}

\begin{align}\label{sol5}
\left( \gamma^\mu P_L \right) \left[ \gamma_\mu P_R \right] &= - \frac{1}{4} \text{Tr} \left[ P_R \gamma^\mu \Gamma_C P_L \gamma_\mu \Gamma_D \right] \; \left( \Gamma^D \right] \left[ \Gamma^C \right) \nonumber \\
&= - \frac{1}{4} \text{Tr} \left[ P_R \gamma^\mu P_L^2 \gamma_\mu P_R \right] \; \left( R \right] \left[ L \right) \nonumber \\
&= - 2 \left( R \right] \left[ L \right) \;\;\;\; \text{as} \;\;\;\; \gamma^\mu \gamma_\mu = 4 \mathbbm{1} , \;\;\;\; \text{Tr} \left[ P_R^4 \right] = \text{Tr} \left[ P_R \right] = 2
\end{align}

\begin{align}\label{sol6}
\left( \sigma^{\mu \nu} P_L \right) \left[ \sigma_{\mu \nu} P_L \right] &=
- 6 \left( P_L \right] \left( P_L \right]
- \frac{1}{4} \left( \sigma^{\mu \nu} \right] \left( \sigma_{\mu \nu} \right] - \frac{ i}{4} \epsilon^{\mu \nu \alpha \beta} \left( \sigma_{\mu \nu} \right] \left( \sigma_{\alpha \beta} \right]
\end{align}

\begin{align}\label{sol7}
\left( \sigma^{\mu \nu} P_L \right) \left[ \sigma_{\mu \nu} P_R \right] &= 0
\end{align}
Note that
$$ \frac{i}{2} \epsilon^{\mu \nu \alpha \beta} \sigma_{\alpha \beta} = -\gamma^5 \sigma^{\mu \nu} $$
($\epsilon_{0123}=+1$)
which can be used to replace the $\epsilon$ tensors by $\gamma_5$. eqn~(\ref{sol2}) becomes
\begin{align}\label{sol8}
\left( P_L \right) \left[ P_L \right] &= - \frac{1}{2} \left( P_L \right] \left[ P_L \right) - \frac{1}{16} \left( \sigma^{\mu \nu} \right] \left( \sigma_{\mu \nu} \right] + \frac{ 1 }{ 16 }\left( \sigma^{\mu \nu} \right] \left( \gamma_5 \sigma_{\mu \nu} \right] \nonumber \\
&=- \frac{1}{2} \left( P_L \right] \left[ P_L \right) - \frac{1}{8} \left( \sigma^{\mu \nu} P_L \right] \left( \sigma_{\mu \nu} P_L \right]
\end{align}
In the last line, we have put back the projectors on $\sigma^{\mu \nu}$, since they are contracted with left-handed fields, and used $\gamma_5 P_L=-P_L$. Similarly, eqn~(\ref{sol6}) becomes
\begin{align}\label{sol9}
\left( \sigma^{\mu \nu} P_L \right) \left[ \sigma_{\mu \nu} P_L \right] &=
- 6 \left( P_L \right] \left( P_L \right]
+\frac12 \left( \sigma^{\mu \nu} P_L \right] \left( \sigma_{\mu \nu} P_L \right]
\end{align}

To summarize, the identities (including the Fermi minus sign) are:
\begin{align}\label{sol10}
\left( P_L \right) \left[ P_L \right]
&=- \frac{1}{2} \left( P_L \right] \left[ P_L \right) - \frac{1}{8} \left( \sigma^{\mu \nu} P_L \right] \left( \sigma_{\mu \nu} P_L \right] \nonumber \\
\left( P_L \right) \left[ P_R \right] &= - \frac{1}{2} \left( \gamma^\mu P_R \right] \left( \gamma_\mu P_L \right] \nonumber \\
\left( \gamma^\mu P_L \right) \left[ \gamma_\mu P_L \right] &= \left( \gamma^\mu P_L \right] \left( \gamma_\mu P_L\right] \nonumber \\
\left( \gamma^\mu P_L \right) \left[ \gamma_\mu P_R \right]
&= - 2 \left( R \right] \left[ L \right) \nonumber \\
\left( \sigma^{\mu \nu} P_L \right) \left[ \sigma_{\mu \nu} P_L \right] &=
- 6 \left( P_L \right] \left( P_L \right]
+\frac12 \left( \sigma^{\mu \nu} P_L \right] \left( \sigma_{\mu \nu} P_L \right] \nonumber \\
\left( \sigma^{\mu \nu} P_L \right) \left[ \sigma_{\mu \nu} P_R \right] &=0
\end{align}

\end{Lexercisenb}

\begin{Lexercisenb}{3}{1}

Compute the mass renormalization factor $Z_m$ in QCD at one loop. Use this to determine the one-loop mass anomalous dimension $\gamma_m$,
\begin{align}
\mu \frac{\rd m}{\rd \mu} &= \gamma_m m,
\end{align}
by differentiating $m_0=Z_m m$, and noting that $m_0$ is $\mu$-independent.

\medskip

\noindent {\bf SOLUTION:}

The Lagrangian for massless QCD is
\beq
\mathcal{L}_{m=0} = \bar q_0 \slashed D q_0 - \frac{1}{4} G_{\mu\nu,0}^A G^{\mu\nu}_{A,0}\,,
\eeq
where $q_0$ is the bare quark field, $G_{\mu\nu,0}$ is the bare gluon field strength tensor and $\slashed D$ is the covariant derivative (we are only interested in quark-gluon interactions, so other gauge bosons are neglected here)
\beq
D_\mu = \partial_\mu - i g_{0,s} t_A G^A_{\mu,0}\,,
\eeq
where $t_A$ are the $SU(3)$ generators and $g_{0,s}$ the respective bare gauge coupling; $G^A_{\mu,0}$ is the bare gluon field. The insertion of the quark mass operator $\mathcal{O}_f = m_0 \bar q_0 q_0$, where $m_0$ is the bare quark mass, modifies the Lagrangian
\beq
\mathcal{L} = \mathcal{L}_{m=0} - \mathcal{O}_f\,.
\eeq

\begin{figure}
\begin{center}
\includegraphics[width=4cm]{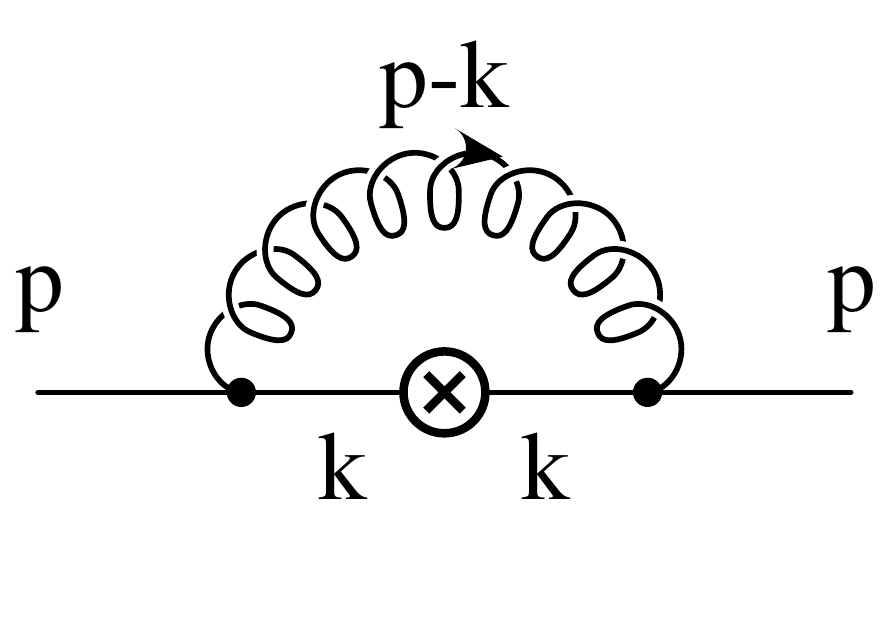}
\end{center}\caption{One loop correction to the mass operator.}\label{fig:Ofvertex}
\end{figure}

To renormalize the theory we define the renormalization constants
\beq
q_0 = Z_q^{1/2} q\,, \quad m_0 = Z_m m\,,
\eeq
where the fields without subscripts are renormalized fields. These constants will absorb the divergences. The other renormalization constants (for the gauge field and coupling) are not needed for this problem. In the ${\overline{\rm MS}}$ scheme we impose that only the $1/\epsilon$ pole and the dim-reg constants are absorbed, thus giving
\beq\label{eq:renconstexp}
Z_i = 1 + z_{i,1}\frac{g^2_s}{(4\pi)^2}\frac{1}{\hat \epsilon} + \dots\,,
\eeq
where the $z_{i,1}$ are $\mu$-independent constants (all the $\mu$ dependence is in the coupling constant).

To solve this problem we are interested in the Lagrangian and operator counterterms
\begin{align} \label{eq:counterterm}
\mathcal{L}_{\text{ct}} &= \left[(Z_q -1)\slashed p \right]\,, & O_{f,\text{ct}} &= (Z_q Z_m -1) m \bar q q
\end{align}
which make the one-loop corrections finite.

The loop integral with the insertion of the operator $\mathcal{O}_f$ is showed in Fig. \ref{fig:Ofvertex}
\beq
I_{\bar q q}=m(ig_s)^2 \mu^{2 \epsilon}\int \frac{d^d k}{(2\pi)^d} \gamma_\alpha t_A \frac{i \slashed k}{k^2} \frac{i \slashed k}{k^2} \gamma_\beta t_B \left( \frac{-ig^{\alpha\beta}\delta^{AB}}{(k-p)^2} \right)\,.
\eeq
Notice that we are only keeping the large $k$ region, since we are only interested in the divergent part. We can simplify this expression noting that $\delta^{AB} t_A t_B = C_F = 4/3$ is the eigenvalue of the Casimir operator and that, in $d$ dimensions, $g^{\alpha\beta} \gamma_\alpha \gamma_\beta = d$. Also $\slashed k \slashed k = k^2$, thus we have
\beq
I_{\bar q q}=-img_s^2 C_F \mu^{2 \epsilon}\int \frac{d^d k}{(2\pi)^d} \frac{d}{k^2(k-p)^2}\,.
\eeq
Introducing the Feynman parameters we have
\beq
I_{\bar q q}=-ig_s^2m C_F \mu^{2 \epsilon}\int \frac{d^d k}{(2\pi)^d} \int_0^1 dx \frac{d}{\left[ (1-x)k^2 + x(k-p)^2 \right]^2}\,.
\eeq
Define $\ell = k - xp$ and $\Delta = -x (1 - x) p^2$, so we can write
\beq
I_{\bar q q}=-ig_s^2 mC_F \mu^{2 \epsilon} \int \frac{d^d \ell}{(2\pi)^d} \int_0^1 dx \frac{d}{\left[ \ell - \Delta \right]^2} = g_s^2 C_F \int_0^1 dx \frac{d}{(4\pi)^\frac{d}{2}} \frac{\Gamma(2 - \frac{d}{2})}{\Gamma(2)} \left( \frac{\mu^2}{\Delta} \right)^{2 - \frac{d}{2}} \,.
\eeq
Expanding and integrating, we get
\beq
I_{\bar q q}= \frac{g_s^2 mC_F}{(4\pi)^2} 4 \left( \frac{1}{\hat\epsilon} - {\rm ln} \frac{-p^2}{\mu^2} + \frac{3}{2}\right) \,,
\eeq
so that the divergent part is
\beq \label{eq:Iqqdiv}
I_{\bar q q,\text{div}}= \frac{4mg_s^2 C_F}{(4\pi)^2} \frac{1}{\hat \epsilon}\,.
\eeq
The counterterm $Z_q Z_m m$ cancels the divergence,
\beq \label{eq:zm}
Z_q Z_m = 1-\frac{4g_s^2 C_F}{(4\pi)^2} \frac{1}{\hat \epsilon}\,.
\eeq

\begin{figure}
\begin{center}
\includegraphics[width=4cm]{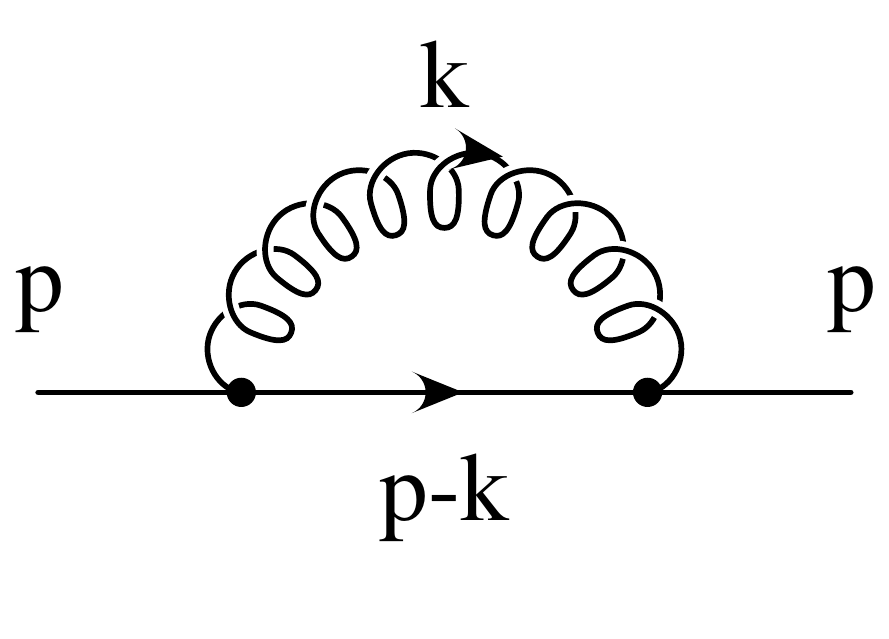}
\end{center}\caption{One-loop diagram for the quark self energy.\label{fig:counter}}
\end{figure}

We need now to calculate the one-loop contribution to the quark propagator, shown in Fig. \ref{fig:counter}. This gives
\beq
\begin{split}
I_2 &= (ig_s)^2\mu^{2 \epsilon}\int \frac{d^d k}{(2\pi)^d} \gamma_\alpha t_A \frac{i \left(\slashed p - \slashed k \right)}{\left( p - k \right)^2} \gamma_\beta t_B \left( \frac{-ig^{\alpha\beta}\delta^{AB}}{k^2} \right) \\
&= - g_s^2 C_F\mu^{2 \epsilon} \int \frac{d^d k}{(2\pi)^d} \frac{\gamma_\alpha\left(\slashed p - \slashed k \right)\gamma^\alpha}{\left( p - k \right)^2k^2}\,,
\end{split}
\eeq
where we used again the Casimir operator eigenvalue. To simplify the Dirac structure in the numerator, we use the $d$-dimensional identity
\beq
\gamma_\mu \gamma^\nu \gamma^\mu = (2 - d) \gamma^\nu\,,
\eeq
thus
\beq
I_2 = -g_s^2 C_F\mu^{2 \epsilon} \int \frac{d^d k}{(2\pi)^d}\frac{(2 - d)\left(\slashed p - \slashed k \right)}{\left( p - k \right)^2k^2} \,.
\eeq
We introduce now the Feynman parameter and get
\beq
\begin{split}
I_2 &= -g_s^2 C_F\mu^{2 \epsilon} \int_0^1 dx \int \frac{d^d k}{(2\pi)^d} \frac{(2 - d)\left(\slashed p - \slashed k \right)}{\left[x \left( p - k \right)^2 + (1-x)k^2\right]^2} \\
&= -g_s^2 C_F \mu^{2 \epsilon}\int_0^1 dx \int \frac{d^d \ell}{(2\pi)^d} \frac{(2 - d)\left((1-x)\slashed p - \slashed \ell \right)}{\left[\ell^2 -\Delta\right]^2}\,,
\end{split}
\eeq
where we defined $\ell = k - xp$ and $\Delta = - (1-x) xp^2$. The term linear in $\ell$ vanishes, so the only contribution is
\beq
\begin{split}
I_2 &= -g_s^2 C_F \slashed p \int_0^1 dx (1-x)\frac{(2-d)}{(4\pi)^\frac{d}{2}} \frac{\Gamma\left( 2 - \frac{d}{2}\right)}{\Gamma(2)}\left(\frac{\mu^2}{\Delta}\right)^{2 - \frac{d}{2}} \\
&= g_s^2 C_F \slashed p \int_0^1 dx (1-x)\frac{2(1 + \epsilon)}{(4\pi)^2}\left[ \frac{1}{\hat\epsilon} - {\rm ln}\left(\frac{-(x(1-x)p^2}{\mu^2}\right) \right]\,,
\end{split}
\eeq
where we used $d = 4 - 2\epsilon$, from which we have $2-d = -2(1+\epsilon)$. The divergent part is then
\beq\label{eq:I2div}
I_{2,\text{div}} = \frac{g_s^2 C_F \slashed p}{(4\pi)^2} \frac{1}{\hat \epsilon}\,.
\eeq
The counterterm $Z_q$ cancels the divergence,
\beq\label{eq:zq}
Z_q = 1-\frac{g_s^2 C_F}{(4\pi)^2} \frac{1}{\hat \epsilon}\,.
\eeq

We can now find the anomalous dimension. This can be computed from the equation
\beq \label{eq:andim}
\gamma_{\mathcal{ O} }= \frac{1}{Z_m}\frac{d\,Z_m}{d\ln \mu} = -2g_s^2 \frac{dz_{m,1}}{dg_s^2}\,,
\eeq
where the last step can be deduced from eqn~\eqref{eq:renconstexp} (see Neubert's lectures). This means that $\gamma_{\mathcal{ O}}$ can be directly obtained from the $1/\hat \epsilon$ pole constant. From eqns~(\ref{eq:zq}) and (\ref{eq:zm}),
\beq
Z_m =\frac{1 - 4\frac{g_s^2 C_F}{(4\pi)^2} \frac{1}{\hat \epsilon}}{1 - \frac{g_s^2 C_F p}{(4\pi)^2} \frac{1}{\hat \epsilon}}= 1 - 3\frac{g_s^2 C_F}{(4\pi)^2} \frac{1}{\hat \epsilon} + {\mathcal{ O}}(g_s^4) \,.
\eeq
Finally, we can now plug this result into \eqref{eq:andim} and get
\beq
\gamma_{\mathcal{ O}} = 6\frac{g_s^2 C_F}{(4\pi)^2} = C_F \frac{3 \alpha_s}{2\pi}=\frac{2 \alpha_s}{\pi}\,.
\eeq
The anomalous dimension of $m$ is the negative of that for $O$,
\beq
\gamma_m = - C_F \frac{3 \alpha_s}{2\pi}=-\frac{2 \alpha_s}{\pi}\,.
\eeq

The individual graphs, and thus $Z_q$ and $Z_m Z_q$ depend on the choice of gauge, but $Z_m$ and $\gamma_m$ are gauge independent.

\end{Lexercisenb}

\begin{Lexercisenb}{3}{2}

Verify eqn~(L3.16) and eqn~(L3.17).

\medskip

\noindent {\bf SOLUTION:}

The one-loop QCD running,
\begin{equation}
\frac{1}{\alpha_s ( \mu) } = \frac{1}{\alpha_s (Q)} + \frac{33 - 2n_f}{6 \pi} \text{log} \left( \frac{\mu}{Q} \right)
\end{equation}
where $n_f$ is the number of active flavors between $\mu$ and $Q$, allows us to relate a coupling fixed at 1$\,$TeV to the hadronization scale $\Lambda_{\rm QCD}$ (at which $1/ \alpha_s ( \Lambda_{\rm QCD} ) = 0$) by running down through the heavy quark thresholds,
\begin{equation}\label{3.52}
\frac{6 \pi}{\alpha_s ( 1 \text{TeV} ) } = 21\, \text{log} \frac{1 \text{TeV}}{m_t} + 23 \, \text{log} \frac{m_t}{m_b} + 25\, \text{log} \frac{m_b}{m_c} + 27\, \text{log} \frac{m_c}{\Lambda_{\rm QCD}}
\end{equation}
$$ \implies \;\;\;\; \Lambda_{\rm QCD} = e^{- \frac{ 6\pi }{ 27 \alpha_s (1 \text{TeV} ) } } (1\, \text{TeV} )^{21/27} \left( m_t \, m_b \, m_c \right)^{2/27} $$
The proton mass is proportional to $\Lambda_{\rm QCD}$, and so scales as $m_t^{2/27}$ when the coupling is fixed at an energy $\mu > m_t$ (providing other quark masses are held fixed). Applying $m_t \rd/\rd m_t$ to
\begin{equation}\label{3.53}
\frac{6 \pi}{\alpha_s ( 1 \text{TeV} ) } = 21\, \text{log} \frac{1 \text{TeV}}{m_t} + 23 \, \text{log} \frac{m_t}{m_b} + 25\, \text{log} \frac{m_b}{m_c} + 27\, \text{log} \frac{m_c}{\mu_L} + \frac{6 \pi}{\alpha_s(\mu_L)}
\end{equation}

keeping $\alpha_s ( 1 \text{TeV} )$ fixed gives
\begin{align}
0 &= -21 + 23 + m_t \frac{\rd}{\rd m_t} \frac{6\pi}{\alpha_s(\mu_L)} \implies m_t \frac{\rd}{\rd m_t} \frac{1}{\alpha_s(\mu_L)} = - \frac{1}{3\pi}
\end{align}

\end{Lexercisenb}

\begin{Lexercisenb}{4}{1}

In $\d=4$ spacetime dimensions, work out the field content of Lorentz-invariant operators with dimension $\opdim$ for $\opdim=1,\ldots,6$. At this point, do not try and work out which operators are independent, just the possible structure of allowed operators. Use the notation $\phi$ for a scalar, $\psi$ for a fermion, $X_{\mu\nu}$ for a field strength, and $D$ for a derivative. For example, an operator of type $\phi^2 D$ such as $\phi D_\mu \phi$ is not allowed because it is not Lorentz-invariant. An operator of type $\phi^2 D^2$ could be either $D_\mu \phi D^\mu \phi$ or $\phi D^2 \phi$, so a $\phi^2 D^2$ operator is allowed, and we will worry later about how many independent $\phi^2 D^2$ operators can be constructed.

\medskip

\noindent {\bf SOLUTION:}

We use the notation $[A] = n$ for the mass dimension of the operator A. The dimensionality of gauge boson, fermion and scalar fields in $d$ spacetime dimensions can be determined from the kinetic terms in the Lagrangian imposing the condition
$$
\left[ \mathcal{L} \right] = d \,.
$$
The three kinetic terms are (numerical constants are irrelevant for this discussion)
\beq
\partial_\mu \phi \partial^\mu\phi\,,\qquad \bar\psi i \slashed \partial \psi\,,\qquad X_{\mu\nu} X^{\mu\nu}\,,
\eeq
where $\phi$, $\psi$ and $X$ are the scalar field, the fermion field and the gauge boson field strength respectively. It follows immediately
\beq
\left[ X X \right] = d\,\, \Rightarrow \,\, \left[X\right] = \frac{d}{2}\,,
\eeq
for the field strength. For the scalar field we have
\beq
[\partial_\mu \phi \partial^\mu\phi] = 2 [\partial] + 2 [\phi] = d \,\, \Rightarrow \,\, [\phi] = \frac{d-2}{2}\,,
\eeq
while for the fermion field we have
\beq
[\bar\psi i \slashed \partial \psi] = [\partial] + 2 [\psi] = d \,\, \Rightarrow \,\, [\psi] = \frac{d-1}{2}\,.
\eeq
In $d=4$ we then have
\beq
[\phi] = 1\,, \quad [D] = 1\,,\quad [\psi] = \frac{3}{2}\,, \quad [X] = 2\,.
\eeq
The Lorentz-invariant operator structures of mass dimension $n$ are:
\begin{itemize}
\item $n = 1$: $\phi$;
\item $n = 2$: $\phi^2$,
\item $n = 3$: $\psi^2$, $\phi^3$;
\item $n = 4$: $\phi^2 D^2$, $\phi^4$, $X^2$, $\phi \psi^2$, $\psi^2 D$;
\item $n = 5$: $\psi^2 D^2$, $\psi^2\phi^2$, $\psi^2 X$, $\phi X^2$, $\phi^5$, $\phi^3 D^2$, $\psi^2 \phi D$;
\item $n = 6$: $\phi^6$, $\phi^2X^2$, $\phi^2D^4$, $D^2X^2$, $X^3$, $\psi^4$, $\psi^2 D^3$, $X \phi^2 D^2$, $\phi^4 D^2$, $\psi^2 \phi D^2$, $\psi^2 \phi^2 D$, $\psi^2 X\phi$, $\psi^2 \phi^3$.
\end{itemize}
$D^n$ for $n=2,4,6$ are not allowed, because there must be at least one field for $D$ to act on. For $n=0$, there is always the operator $\mathbbm{1}$. $D^2 \phi$ for $n=3$ is a total derivative and integrates to zero. $D^2 X$ for $n=4$ is $D_\mu D_\nu X^{\mu \nu}$,
and antisymmetry of $X$ converts this to $\left[D_\mu, D_\nu \right] X^{\mu \nu} \propto X^2$. Several operators, such as $\phi^2 D^4$ can be eliminated by field redefinitions. Similar simplifications occur in the next exercise.

\end{Lexercisenb}

\begin{Lexercisenb}{4}{2}

For $\d=2,3,4,5,6$ dimensions, work out the field content of operators with dimension $\opdim \le \d$, i.e. the ``renormalizable'' operators.

\medskip

\noindent {\bf SOLUTION:}

Using the general result for mass dimensionality in the previous exercise, we can now work out Lorentz-invariant renormalizable operators in different spacetime dimensions., \emph{i.e.} those with dimension $n \leq d$. Note that at $n=d$ there will be the kinetic term operators and these will always appear as $\phi^2D^2$, $\psi^2 D$ and $X^2$ for scalar, fermion and gauge boson respectively. Except in the particular case of $n=2$, we will omit these operators in the lists. There is also the operator $\mathbbm{1}$ (cosmological constant) which is also omitted in the lists.

\begin{itemize}

\item $d = 2$
\beq
[\phi] = 0\,, \quad [D] = 1\,,\quad [\psi] = \frac{1}{2}\,, \quad [X] = 1\,.
\eeq
In this particular case, the scalar field can enter with an arbitrary power $p$ by itself at $n=0$ or in other operators at $n=1,2$ since it does not add any mass dimension to $n$ and is Lorentz-invariant. For each class of operator we then have an infinite set of operators when we change the value of $p$. So we have
\begin{itemize}
\item $n = 0$: $\phi^p$;
\item $n = 1$: $\phi^p \psi^2$;
\item $n = 2$: $\phi^p \psi^4$, $\phi^p D^2$, $\phi^p X^2$, $\phi^p \psi^2 D$, $\psi^2 X \phi^p$.
\end{itemize}
\item $d = 3$
\beq
[\phi] = \frac{1}{2}\,, \quad [D] = 1\,,\quad [\psi] = 1\,, \quad [X] = \frac{3}{2}\,.
\eeq
From now on there are no more non-trivial $n=0$ operators.
\begin{itemize}
\item $n=1/2$: $\phi$;
\item $n = 1$: $\phi^2$;
\item $n=3/2$: $\phi^3$;
\item $n = 2$: $\phi^4$, $\psi^2$;
\item $n=5/2$: $\phi^5$, $\psi^2 \phi$;
\item $n = 3$: $\phi^6$, $\phi^2\psi^2$;
\end{itemize}
\item $d = 4$
\beq
[\phi] = 1\,, \quad [D] = 1\,,\quad [\psi] = \frac{3}{2}\,, \quad [X] = 2\,.
\eeq
This case has already been analyzed in the previous exercise, so we can skip it.
\item $d = 5$
\beq
[\phi] = \frac{3}{2}\,, \quad [D] = 1\,,\quad [\psi] = 2\,, \quad [X] = \frac{5}{2}\,.
\eeq
There are no more operator for $n=1$ and from now on the only $n=2$ operator is $D^2$, so we don't write it anymore.
\begin{itemize}
\item $n=3/2$: $\phi$;
\item $n=2$: \text{none};
\item $n=5/2$: \text{none};
\item $n = 3$: $\phi^2$;
\item $n=3/2$: \text{none};
\item $n = 4$: $\psi^2$.
\item $n=9/2$: $\phi^3$;
\end{itemize}
In this case, at $n=5$ there are only the kinetic terms.
\item $d = 6$
\beq
[\phi] = 2\,, \quad [D] = 1\,,\quad [\psi] = \frac{5}{2}\,, \quad [X] = 3\,.
\eeq
\begin{itemize}
\item $n=1$: \text{none};
\item $n=2$: $\phi$;
\item $n=3$: \text{none};
\item $n = 4$: $\phi^2$;
\item $n = 5$: $\psi^2$;
\item $n = 6$: $\phi^3$.
\end{itemize}

\end{itemize}

\end{Lexercisenb}

\begin{Lexercisenb}{4}{3}

Compute the decay rate $\Gamma( b \to c e^- \overline \nu_e)$ with the interaction Lagrangian
\begin{align*}
L &= -\frac{4 G_F}{\sqrt 2}V_{cb} ( \overline c \gamma^\mu P_L b)(\overline \nu_e \gamma_\mu P_L e)
\end{align*}
with $m_e \to 0$, $m_\nu \to 0$, but retaining the dependence on $\rho = m_c^2/m_b^2$. It is convenient to write the three-body phase space in terms of the variables $x_1=2E_e/m_b$ and $x_2=2 E_\nu/m_b$.

\medskip

\noindent {\bf SOLUTION:}

At tree-level, the amplitude for the decay $b \to c e^- \bar{\nu}_e$ is given by,
\begin{equation}
i \mathcal{A} = \frac{4 G_F}{\sqrt{2}} V_{cb} \left( \bar c \gamma^\mu b_L \right) \left(e_L \gamma_\mu \bar{\nu}_e \right)
\end{equation}
where we use the same symbol for the spinor fields and their polarization spinors $u(p,s)$.

To compute the square of the amplitude, first consider
\begin{align}
\left( \bar c_L \gamma^\mu b_L \right) \left( \bar c_L \gamma^\nu b_L \right)^\dagger
&= \left( \bar c_L \gamma^\mu b_L \right) \left( \bar b_L \gamma^{\nu} c_L \right) \\
&= \frac14 \text{Tr} \left[ \gamma^\mu ( 1 - \gamma_5 ) b \bar b \gamma^\nu ( 1- \gamma_5) c \bar c \right]
\end{align}
Then performing a sum over the different available spin states,
\begin{equation}
\text{Tr} \left[ \gamma^\mu ( 1 - \gamma_5 ) ( \slashed{p}_b + m_b ) \gamma^\nu ( 1- \gamma_5) (\slashed{p}_c + m_c ) \right] = T^{\mu \nu}_{\alpha \beta} p_b^\alpha p_c^{\beta}
\end{equation}
\begin{align}
T^{\mu \nu}_{\alpha \beta} = 8 \left( \eta^{\mu \alpha} \eta^{\nu \beta} - \eta^{\mu \nu} \eta^{\alpha \beta} + \eta^{\mu \beta} \eta^{\nu \alpha} + i \epsilon^{\mu \alpha \nu \beta} \right) \,.
\end{align}

The probability is then,
\begin{equation}
\vev{\abs{\mathcal{A}}^2}=
\frac12 \sum_{\text{spins}} | i \mathcal{A} |^2 = \frac{G_F^2 \abs{V_{cb}}^2}{4} T^{\mu \nu}_{\alpha \beta} T^{\rho \sigma}_{\mu \nu} \, p_b^\alpha p_c^\beta p_{\nu, \rho} p_{e, \sigma} = 64\, G_F^2 \abs{V_{cb}}^2 \; (p_b \cdot p_{\nu}) \; (p_c \cdot p_e)
\end{equation}
summing over all spins, and including a $1/2$ for spin-averaging over the initial $b$-quark spin.

The three body phase space is,
\begin{align}\label{4.16}
d\Pi_3 &= \int \frac{d^3 \mathbf{p}_c}{(2\pi)^3 \, 2E_c} \int \frac{d^3 \mathbf{p}_e }{(2\pi)^3 \, 2 | \mathbf{p}_e |} \int \frac{d^3 \vec{p}_\nu}{ (2\pi)^3 \, 2 | \vec{p}_\nu | }
( 2\pi)^4 \delta^{(4)} ( {p}_c + {p}_e + {p}_\nu-p_b )\,.
\end{align}
Note that when we integrate over the spin-summed probability, we are going to need the integral,
\begin{equation}
I^{\alpha \beta} (q) = \int \frac{d^3 \vec{p}_e }{(2\pi)^3 \, 2 | \vec{p}_e |} \int \frac{d^3 \vec{p}_\nu}{ (2\pi)^3 \, 2 | \vec{p}_\nu | } ( 2\pi)^4 \delta^{(4)} ( {p}_e + {p}_\nu-q ) \; p_e^\alpha p_\nu^\beta
\end{equation}
The simplest way to evaluate this integral is to exploit Lorentz invariance to write it as,
\begin{equation}
I^{\alpha \beta} (q) = \eta^{\alpha \beta} I_1 + q^{\alpha} q^{\beta} I_2
\end{equation}
and then to calculate the two independent Lorentz scalars $I_{1,2}$ in the frame $\bm{q}=0$, and use
\begin{align}\label{4.19}
p_e \cdot p_\nu &= \frac12 q^2 & q \cdot p_\nu &= \frac 12 q^2 & q \cdot p_e &= \frac 12 q^2
\end{align}
which follow from $q=p_e+p_\nu$, $p_e^2=p_\nu^2=0$.
Explicitly,
\begin{align}
\eta_{\alpha \beta} I^{\alpha \beta} &= \int \frac{d^3 \mathbf{p}_e }{(2\pi)^3 \, 2 | \mathbf{p}_e |} \int \frac{d^3 \mathbf{p}_\nu}{ (2\pi)^3 \, 2 | \mathbf{p}_\nu | } ( 2\pi)^4 \delta^{(4)} ( {p}_e + {p}_\nu-q ) \; (p_e \cdot p_\nu)
\end{align}
The space $\delta$-function fixes $\mathbf{p}_\nu=-\mathbf{p}_e$, and $E_\nu=E_e=\abs{\mathbf{p}_e}$, giving
\begin{align}
&= \frac{q^2}{2\pi} \int \frac{ \abs{\mathbf{p}_e}^2 d \abs{\mathbf{p}_e} }{ (2 \abs{\mathbf{p}_e}) (2 \abs{\mathbf{p}_e}) } \delta ( q^0 -2 \abs{\mathbf{p}_e}) = \frac{1}{16 \pi} q^2
\end{align}
and
\begin{align}
q_{\alpha}q_{ \beta} I^{\alpha \beta} &= \int \frac{d^3 \mathbf{p}_e }{(2\pi)^3 \, 2 | \mathbf{p}_e |} \int \frac{d^3 \mathbf{p}_\nu}{ (2\pi)^3 \, 2 | \mathbf{p}_\nu | } ( 2\pi)^4 \delta^{(4)} ( {p}_e + {p}_\nu-q ) \; (q \cdot p_e)(q \cdot p_\nu) \nonumber \\
&= \frac{1}{32 \pi} q^4
\end{align}
from eqn~(\ref{4.19}).
These give
\begin{equation}
I^{\alpha \beta} (q) = \frac{1}{96 \pi} \left( \eta^{\alpha \beta} q^2 + 2 q^\alpha q^\beta \right) .
\end{equation}

The differential decay rate is
\begin{equation}
d\Gamma ( b \to c e^- \bar \nu_e ) = \frac{1}{2 m_b} d \Pi_3 \; \vev{\abs{\mathcal{A}}^2}
\end{equation}
and the $\delta$-function in the three-body phase space eqn~(\ref{4.16}) can be written as
\begin{align}
\delta^{(4)} ( {p}_c + {p}_e + {p}_\nu-p_b ) &= \int \rd^4 q \ \delta^{(4)} ( {p}_e + {p}_\nu -q ) \delta^{(4)} ( {p}_c + q-p_b ) \,.
\end{align}
This gives
\begin{align}
d\Gamma ( b \to c e^- \bar \nu_e ) &= \frac{1}{2 m_b} \frac{64 G_F^2 \abs{V_{cb}}^2}{\pi} \int \frac{d^3 \mathbf{p}_c}{(2\pi)^3 2 E_c} \int \rd^4 q \ \delta^{(4)} ( {p}_c + q-p_b) I^{\alpha \beta}(q) p_{c \alpha} p_{b \beta} \nonumber \\
&= \frac{G_F^2 \abs{V_{cb}}^2}{ 3 \pi^2 }\int \frac{d^3 \mathbf{p}_c}{(2\pi)^3 2 E_c} \left[ - 4 m_b E_c^2 + 3 E_c ( m_b^2 + m_c^2) - 2 m_b m_c^2 \right] \end{align}
The $\mathbf{p}_c$ integral is
\begin{align}
\frac{d^3 \mathbf{p}_c}{(2\pi)^3 2 E_c} &= \frac{1}{4\pi^2} \abs{\mathbf{p}_c} \rd E_c
\end{align}

Using the above results, we find,
\begin{equation}
\frac{d \Gamma ( b \to c e^- \bar \nu_e ) }{d E_c} = | V_{cb} |^2 \frac{G_F^2}{12 \pi^3} \sqrt{E_c^2 - m_c^2} \left[ - 4 m_b E_c^2 + 3 E_c ( m_b^2 + m_c^2) - 2 m_b m_c^2 \right]
\end{equation}

The total decay rate is given by integrating between $E_c = m_c$ and $E_c = (m_c^2+m_b^2)/2m_c$, the maximum energy kinematically allowed,
\begin{align}
\Gamma ( b \to c e^- \bar \nu_e ) &= | V_{cb} |^2 \frac{G_F^2 m_b^5}{192 \pi^3} \left( 1 - 8 \rho + 8 \rho^3 - \rho^4 - 12 \rho^2 \log \, \rho \right) & \rho & = \frac{m_c^2}{m_b^2}
\end{align}

\end{Lexercisenb}

\begin{Lexercisenb}{5}{1}

Compute the one-loop scalar graph Fig.~\ref{fig:quad} with a scalar of mass $m$ and interaction vertex $-\lambda \phi^4/4!$ in the \MSbar\ scheme. Verify the answer is of the form eqn~(L5.15). The overall normalization will be different, because this exercise uses a real scalar field, and $H$ in the SM is a complex scalar field.

\begin{figure}
\begin{center}
\includegraphics[width=2cm]{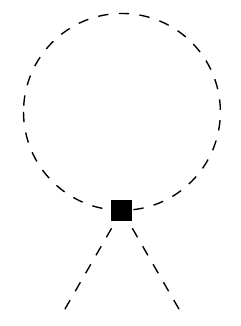}
\caption{\label{fig:quad}One loop correction to the scalar mass from the $-\lambda \phi^4/4!$ interaction.}
\end{center}
\end{figure}

\medskip

\noindent {\bf SOLUTION:}

The Lagrangian is
\beq
\mathcal{L} = \frac{1}{2}\partial_\mu\phi\partial^\mu\phi - \frac12m^2 \phi^2 - \frac{\lambda}{4!}\phi^4\,.
\eeq
The Feynman vertex for the self-interaction is $-i\lambda$ (the $4!$ factor is canceled by the $4!$ different ways of combining the four $\phi$ lines in the diagram).

Calling $k$ the momentum running in the loop, we have (the graph has a symmetry factor of $1/2$)
\beq
I = -i \frac12 \lambda\mu^{2\epsilon} \int \frac{d^dk}{(2\pi)^d} \frac{1}{\left[ k^2 - m^2\right]} = \frac{\lambda m^2}{2(4\pi)^\frac{d}{2}} \frac{\Gamma(1 - \frac{d}{2})}{\Gamma(1)} \left( \frac{\mu^2}{m^2}\right)^{2 - \frac{d}{2}}\,.
\eeq
Expanding it for $\epsilon \to 0$ we have
\beq
I = \frac{\lambda m^2}{32 \pi^2}\left( \frac{1}{\hat \epsilon} - \ln \frac{m^2}{\mu^2} \right)\,.
\eeq
In the $\overline{\rm MS}$ scheme the counterterm will cancel the $1/\hat\epsilon$ pole, leaving
\beq
I + I_{\text{c.t.}} =- \frac{\lambda m^2}{32 \pi^2} \ln \frac{m^2}{\mu^2} \,.
\eeq
The Higgs mass correction has a slightly different prefactor because $H$ is a complex field, and because of a different normalization convention for the $(H^\dagger H)^2$ vertex.

\end{Lexercisenb}

\begin{Lexercisenb}{5}{2}
\label{ex5.2}

Compute $I_F$ and $I_{\text{EFT}}$ given in eqns~(L5.19,L5.21) in dimensional regularization in $\d=4-2\epsilon$ dimensions. Both integrals have UV divergences, and the $1/\epsilon$ pieces are canceled by counterterms. Determine the counterterm contributions $I_{F,\text{ct}}$, $I_{\text{EFT},\text{ct}}$ to the two integrals.

\medskip

\noindent {\bf SOLUTION:}

The full integral is given by,
\begin{align}
I_{\text{F}} &= - i \mu^{2 \epsilon} \int_0^1 dx \, \int \frac{d^d k}{ (2 \pi)^d } \, \frac{1}{ (k^2 - x m^2 - (1-x) M^2 )^2 } \nonumber \\
&= \frac{1}{16 \pi^2} \Gamma ( \epsilon ) \int_0^1 dx \, \left( \frac{4 \pi \mu^2 }{ x m^2 + (1-x) M^2} \right)^\epsilon \nonumber \\
&= \frac{1}{16 \pi^2} \left[ \frac{1}{ \epsilon} + \int_0^1 dx \, \log \left( \frac{\bar \mu^2}{ xm^2 + (1-x) M^2 } \right) \right] \nonumber \\
&=\frac{1}{16 \pi^2} \left[ \frac{1}{ \epsilon} - \frac{1}{M^2 - m^2} \left( M^2 \log \frac{M^2}{\mu^2} - m^2 \log \frac{m^2}{\bar \mu^2} \right) + 1 \right]
\end{align}
The full theory counterterm cancels the divergence,
\begin{align}
I_{\text{F,c.t.}} &=-\frac{1}{16 \pi^2} \frac{1}{ \epsilon}
\end{align}

In an EFT expansion, we try to capture this answer by adding higher dimension operators, each suppressed by the scale $M$. This looks like,
\begin{align}
I_{\text{EFT}} &= i \mu^{2 \epsilon} \int \frac{d^d k}{ (2 \pi)^d } \frac{1}{ k^2 - m^2} \, \frac{1}{M^2} \left[ 1 + \frac{k^2}{M^2} + \frac{k^4}{M^4} + ... \right] \nonumber \\
&= - \frac{1}{16 \pi^2} \sum_{a=0}^\infty \left( \frac{m^2}{M^2} \right)^{a+1} \, (-1)^{a+1} \frac{ \Gamma ( \epsilon - a - 1 ) \Gamma ( 2 + a - \epsilon) }{ \Gamma (2 - \epsilon) } \, \left( \frac{4 \pi \mu^2}{m^2} \right)^\epsilon \nonumber \\
&= - \frac{1}{16 \pi^2} \sum_{a=0}^\infty \left( \frac{m^2}{M^2} \right)^{a+1} \, \left[ \frac{1}{\epsilon} - \gamma_E + 1 + \mathcal{O} ( \epsilon ) \right] \, \left[ 1 + \epsilon\, \log \left( \frac{ 4 \pi \mu^2}{m^2} \right) + \mathcal{O} ( \epsilon^2 ) \right] \nonumber \\
&= - \frac{1}{16 \pi^2} \left[ \frac{1}{ \epsilon} + \log \frac{ \bar \mu^2}{ m^2 } \right] \sum_{a=0}^\infty \left( \frac{m^2}{M^2} \right)^{a+1} \nonumber \\
&= - \frac{1}{16 \pi^2} \left[ \frac{1}{ \epsilon} - \log \frac{m^2}{\bar \mu^2} + 1 \right] \frac{m^2}{M^2 - m^2}
\end{align}
Each term in the EFT expansion is UV divergent. Note that in the EFT, each of the operators has its own counterterm,
\begin{equation}
I_{\text{EFT,c.t.}}^{(n)} = \frac{1}{16 \pi^2 \epsilon} \left( \frac{m^2}{M^2} \right)^n
\end{equation}
and so the anomalous dimensions in the two theories are different,
$$ \gamma_{\text{Full}} \neq \gamma_{\text{EFT}} $$
which has important consequences for resumming large logarithms. The sum of all the counterterms (needed for Exercise~5.6) is
\begin{equation}\label{5.5}
I_{\text{EFT,c.t.}} = \frac{1}{16 \pi^2 \epsilon} \frac{m^2}{M^2-m^2}
\end{equation}

The resummed answer captures the correct analyticity structure in the IR scale $m$ (i.e. the branch cut from the logarithm), but not the UV scale $M$. Note that the series only formally converges for the soft part of the integral, $\int_0^M d k$, which is the regime of validity of this EFT. In order to successfully match the full integral, we need to add an additional term, given by the ``matching condition",
\begin{equation}
I_{\text{Match}} = I_{\text{Full}} - I_{\text{EFT}} .
\end{equation}
Mathematically, this corresponds to the remainder of the integration $\int_M^\infty dk$, and physically reflects the high energy modes which we have integrated out in going to the EFT description (see the next problem).

\end{Lexercisenb}

\begin{Lexercisenb}{5}{3}
\label{ex5.4}

Compute $I_M \equiv \left( I_F + I_{F,\text{ct}} \right) - \left( I_{\text{EFT}} + I_{\text{EFT},\text{ct}} \right)$ and show that it is analytic in $m$.

\medskip

\noindent {\bf SOLUTION:}

Using $I_F$ and $I_{\text{EFT}}$ from the previous solution, with the $1/\epsilon$ pieces dropped,
\begin{align}\label{5.7}
I_M &= \frac{1}{16 \pi^2} \left[ - \frac{1}{M^2 - m^2} \left( M^2 \log \frac{M^2}{\mu^2} - m^2\log \frac{m^2}{\bar \mu^2} \right) + 1 \right] + \frac{1}{16 \pi^2} \left[ - \log \frac{m^2}{\bar \mu^2} + 1 \right] \frac{m^2}{M^2 - m^2} \nonumber \\
&= \frac{1}{16 \pi^2}\frac{M^2}{m^2 - M^2}\left[ \log \frac{M^2}{\mu^2} - 1 \right] \,.
\end{align}
The non-analytic $\log m^2$ term has canceled, and $I_M$ is analytic in $m$.

\end{Lexercisenb}

\begin{Lexercisenb}{5}{4}

Compute $I_F^{(\text{exp})}$, i.e.\ $I_F$ with the IR $m$ scale expanded out
\begin{align*}
I_F^{(\text{exp})} &= -i \mu^{2\epsilon} \int \frac{\rd^d k}{(2\pi)^d} \frac{1}{(k^2-M^2)}\left[\frac{1}{k^2} + \frac{m^2}{k^4} + \ldots \right] \,.
\end{align*}
Note that the first term in the expansion has a $1/\epsilon$ UV divergence, and the remaining terms have $1/\epsilon$ IR divergences.

\medskip

\noindent {\bf SOLUTION:}

\begin{align}
I_F^{(\text{exp})} &= - i \mu^{2\epsilon} \int \frac{d^d k}{ (2\pi)^d} \frac{1}{k^2 - M^2} \frac{1}{k^2} \left[ 1 + \frac{m^2}{k^2} + \frac{m^4}{k^4} + ... \right] \\
&= \frac{1}{16 \pi^2} \left[ \frac{1}{ \epsilon} - \log \frac{M^2}{\bar \mu^2} + 1 \right] \frac{M^2}{M^2 - m^2}
\end{align}
on integrating term-by-term and adding. The finite part agrees with eqn~(\ref{5.7}).

\end{Lexercisenb}

\begin{Lexercisenb}{5}{5}

Compute $I_F^{(\text{exp})} + I_{F,\text{ct}}$ using $I_{F,\text{ct}}$ determined in Exercise~\ref{ex5.2}. Show that the UV divergence cancels, and the remaining $1/\epsilon$ IR divergence is the same as the UV counterterm $I_{\text{EFT},ct}$ in the EFT.

\medskip

\noindent {\bf SOLUTION:}

\begin{align}
I_F^{(\text{exp})} &= \frac{1}{16 \pi^2} \left[ \frac{1}{ \epsilon} - \log \frac{M^2}{\bar \mu^2} + 1 \right] \frac{M^2}{M^2 - m^2}
\end{align}
from Exercise~5.5 and
\begin{align}
I_{\text{F,c.t.}} &=-\frac{1}{16 \pi^2} \frac{1}{ \epsilon}
\end{align}
from Exercise~5.3, so
\begin{align}\label{5.12}
I_F^{(\text{exp})} + I_{F,\text{ct}} &= \frac{1}{16 \pi^2}\left\{ \frac{1}{ \epsilon}\frac{m^2}{M^2-m^2}+ \left[ - \log \frac{M^2}{\bar \mu^2} + 1 \right] \frac{M^2}{M^2 - m^2}
\right\}
\end{align}
and the infinite part is the same as eqn~(\ref{5.5}).

\end{Lexercisenb}

\begin{Lexercisenb}{5}{6}

Compute $I_{\text{EFT}}^{(\text{exp})}$, i.e.\ $I_{\text{EFT}}$ with the IR $m$ scale expanded out. Show that it is a scaleless integral which vanishes. Using the known UV divergence from Exercise~\ref{ex5.2}, write it in the form
\begin{align*}
I_{\text{EFT}}^{(\text{exp})} &=-B \frac{1}{16\pi^2} \left[ \frac{1}{\eUV}-\frac{1}{\eIR} \right]\,,
\end{align*}
and show that the IR divergence agrees with that in $I_F^{(\text{exp})} + I_{F,\text{ct}}$.

\medskip

\noindent {\bf SOLUTION:}

\begin{align}
I_\text{EFT}^{(\text{exp})} &=-i \mu^{2\epsilon}\int \frac{{\rm d}^\d k}{(2\pi)^\d}\ \left[\frac{1}{k^2} + \frac{m^2}{k^4} + \ldots \right] \left[-\frac{1}{M^2}-\frac{k^2}{M^4}-\ldots\right].
\end{align}
Multiplying out gives integrals of the form
\begin{align}
-i \mu^{2\epsilon}\int \frac{{\rm d}^\d k}{(2\pi)^\d}\ \frac{ (m^2)^r}{(M^2)^s} (k^2)^{2+s-r}
\end{align}
which are scaleless and vanish. From Exercise~5.3,
\begin{align}
B &=\frac{m^2}{M^2-m^2}.
\end{align}
and the $1/\eIR$ divergent term agrees with eqn~(\ref{5.12}). [Note the various signs.]
\begin{align}\label{5.16}
I_\text{EFT}^{(\text{exp})} + I_\text{EFT,c.t.}^{(\text{exp})} &= \frac{1}{16 \pi^2 } \frac{1}{ \eIR} B
\end{align}

We can also compute the integral directly. The log divergent terms are
\begin{align}
I_\text{EFT}^{(\text{exp})} &=-i \mu^{2\epsilon}\int \frac{{\rm d}^\d k}{(2\pi)^\d}\ \left[-\frac{m^2}{M^2} - \frac{m^4}{M^4} - \ldots
\right] \frac{1}{k^4}
\end{align}
Using eqn~(L5.34) gives
\begin{align}
I_\text{EFT}^{(\text{exp})} &= \left[-\frac{m^2}{M^2} - \frac{m^4}{M^4} - \ldots
\right] \left[\frac{1}{\eUV}-\frac{1}{\eIR} \right]
\end{align}
so that
\begin{align}
B &= \left[\frac{m^2}{M^2} + \frac{m^4}{M^4} +\ldots \right] = \frac{m^2}{M^2-m^2}.
\end{align}
The $1/\eUV$ terms are canceled by the counterterm eqn~(\ref{5.5}), and the remaining terms, which are IR divergent, agree with eqn~(\ref{5.12}).

\end{Lexercisenb}

\begin{Lexercisenb}{5}{7}

Compute $\left(I_F^{(\text{exp})} + I_{F,\text{ct}}\right)-\left(I_{\text{EFT}}^{(\text{exp})} + I_{\text{EFT},\text{ct}} \right)$ and show that all the $1/\epsilon$ divergences (both UV and IR) cancel, and the result is equal to $I_M$ found in Exercise~\ref{ex5.4}.

\medskip

\noindent {\bf SOLUTION:}

This follows immediately using eqn~(\ref{5.12}) and eqn~(\ref{5.16}).
\end{Lexercisenb}

\begin{Lexercisenb}{5}{8}

Make sure you understand why you can compute $I_M$ simply by taking $I_F^{(\text{exp})} $ and dropping all $1/\epsilon$ terms (both UV and IR).

\medskip

\noindent {\bf SOLUTION:}

No written solution needed.

\end{Lexercisenb}

\begin{Lexercisenb}{5}{9}
\label{exerciseManohar15}

Compute the QED on-shell electron form factors $F_1(q^2)$ and $F_2(q^2)$ expanded to first order in $q^2/m^2$ using dimensional regularization to regulate the IR and UV divergences. This gives the one-loop matching to heavy-electron EFT. Note that it is much simpler to \emph{first} expand and then do the Feynman parameter integrals. A more difficult version of the problem is to compute the on-shell quark form factors in QCD, which gives the one-loop matching to the HQET Lagrangian. For help with the computation, see Ref.~\cite{Manohar:1997qy}. Note that in the non-Abelian case, using background field gauge is helpful because the amplitude respects gauge invariance on the external gluon fields.

\medskip

\noindent {\bf SOLUTION:}

\begin{figure}
\begin{center}
\includegraphics[width=3cm]{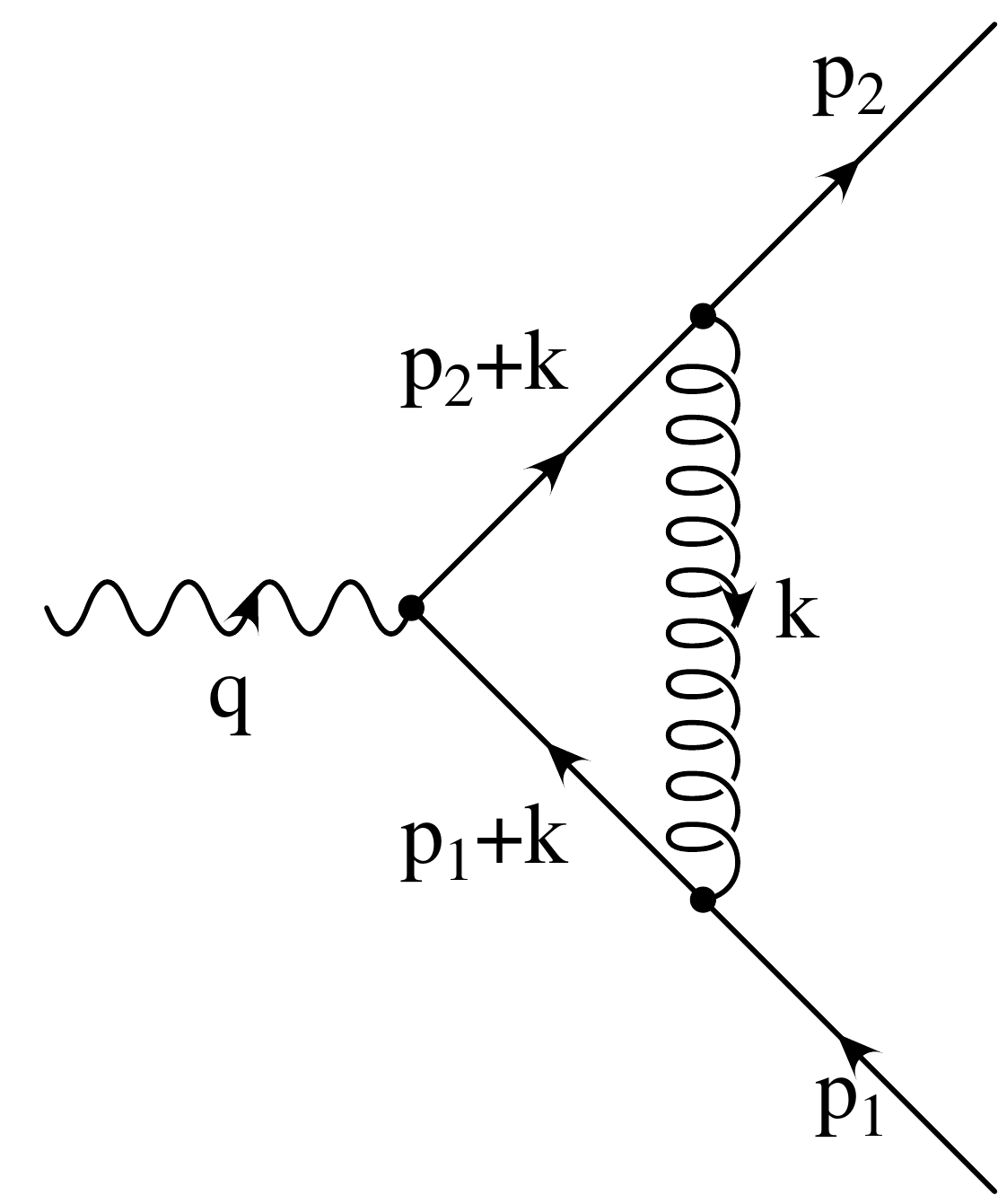}
\end{center}
\caption{\label{fig:vertex_manohar15} Feynman diagram of 1-loop QED vertex correction, as calculated in Exercise \ref{exerciseManohar15}.}
\end{figure}

To calculate the QED electron on-shell form factors $F_1 (q^2)$ and $F_2 (q^2)$ up to one loop, we consider the vertex as pictured in Figure \ref{fig:vertex_manohar15}. The form factors $F_i(q^2)$ are defined as
\begin{equation}
\bar{u}(p_2)\Gamma^\mu(q^2)u(p_1)=\bar{u}(p_2)\left(F_1(q^2)\gamma^\mu+F_2(q^2)\frac{i\sigma^{\mu\nu}q_\nu}{2m}\right)u(p_1) \label{eq:defFormfactorsManohar15}
\end{equation}
with $q=p_2-p_1$, where we have obviously $F_1(0)=1$ and $F_2(0)=0$ at tree level. We define $\tilde{F}_i(q^2)$ as the one-loop contribution to the full form factors $F_i(q^2)$.

We will calculate the $\tilde{F}_i(q^2)$ by writing down the Feynman rules of Figure \ref{fig:vertex_manohar15}, identify the integrals that will give us $\tilde{F}_1(q^2)$ and $\tilde{F}_2(q^2)$, do the integrals over the loop momentum in $d$ dimensions, expand in $\hat{q}^2=q^2/m^2$ to first order and evaluate the leftover Feynman parameter integrals.

Applying the Feynman rules gives
\begin{align}
-ie\Gamma^\mu(q^2)&=\int\frac{d^4 k}{(2\pi )^4} \left(-ie\gamma^\nu \right) \frac{i\left(\slashed{k}+\slashed{p}_2+m\right)}{\left(k+p_2\right)^2-m^2+i\epsilon} \left(-ie\gamma^\mu \right)\frac{i\left(\slashed{k}+\slashed{p}_1+m\right)}{\left(k+p_1\right)^2-m^2+i\epsilon} \left(-ie\gamma^\rho \right) \frac{-ig_{\nu\rho}}{k^2+i\epsilon}\nonumber\\
&=-e^3\int\frac{d^4 k}{(2\pi )^4} \frac{\gamma^\nu \left(\slashed{k}+\slashed{p}_2+m\right)\gamma^\mu \left(\slashed{k}+\slashed{p}_1+m\right) \gamma_\nu }{\left[\left(k+p_2\right)^2-m^2+i\epsilon\right]\left[\left(k+p_1\right)^2-m^2+i\epsilon\right]\left[k^2+i\epsilon\right]}. \label{eq:feynmanrules15}
\end{align}
First, we want to massage the denominator. By introducing Feynman parameters, we get
\begin{align}
\frac{1}{[\dots]}=\int_0^1 dx \int_0^1 dy \int_0^1 dz \frac{2\,\delta(1-x-y-z)}{\left(y\left[\left(k+p_2\right)^2-m^2\right]+x\left[\left(k+p_1\right)^2-m^2\right]+zk^2\right)^3}.
\end{align}
The denominator can be simplified further (using $x+y+z=1$, overall momentum conservation $q+p_1=p_2$ and on-shellness of the real electrons $p_1^2=p_2^2=m^2$) to
\begin{align}
\left(\dots\right)^3&=\left(\left[k+xp_1+yp_2\right]^2-\left[xp_1+yp_2\right]^2+x(p_1^2-m^2)+y(p_2^2-m^2)\right)^3\nonumber\\
&=\left(\left[k+xp_1+yp_2\right]^2-x^2p_1^2+xy\left((p_2-p_1)^2-p_1^2-p_2^2\right)-y^2p_2^2\right)^3\nonumber\\
&=\left(\left[k+xp_1+yp_2\right]^2+xyq^2-x(x+y)p_1^2-y(x+y)p_2^2\right)^3\nonumber\\
&=\left(\left[k+xp_1+yp_2\right]^2+xyq^2-(x+y)^2m^2\right)^3\nonumber\\
&=\left(l^2+xyq^2-(1-z)^2m^2\right)^3\label{eq:denom15},
\end{align}
with shifted momenta $l=k+xp_1+yp_2$, which replaces the integrations over loop momentum $k$.

The numerator of \eqref{eq:feynmanrules15} can be reduced using the following identities of the Dirac $\gamma$ matrices in $d$ dimensions
\begin{align}
\gamma^\mu\gamma_\mu &= d\\
\gamma^\mu\gamma^\alpha\gamma_\mu &= (2-d)\gamma^\alpha\\
\gamma^\mu\gamma^\alpha\gamma^\beta\gamma_\mu &= 4g^{\alpha\beta}-(4-d)\gamma^\alpha\gamma^\beta\\
\gamma^\mu\gamma^\alpha\gamma^\beta\gamma^\delta\gamma_\mu &= -2\gamma^\delta\gamma^\beta\gamma^\alpha+(4-d)\gamma^\alpha\gamma^\beta\gamma^\delta.
\end{align}
We get
\begin{align}
\gamma^\nu\dots\gamma_\nu&=-2m^2\gamma^\mu+4m(2k^\mu+p_1^\mu+p_2^\mu)-2(\slashed{k}+\slashed{p}_1)\gamma^\mu(\slashed{k}+\slashed{p}_2)\\
&\hspace{2cm}+(4-d)\left(\slashed{k}+\slashed{p}_2-m\right)\gamma^\mu \left(\slashed{k}+\slashed{p}_1-m\right).\nonumber
\end{align}
By replacing $k$ with parameter $l$ as above ($k\rightarrow l-xp_1-yp_2$), and erasing the terms linear in $l$ (note that we will eventually integrate over $l$ and everything else is even in $l$), we get
\begin{align}
\gamma^\nu\dots\gamma_\nu&=-2m^2\gamma^\mu+4m(p_1^\mu(1-2x)+p_2^\mu(1-2y))+(2-d)\slashed{l}\gamma^\mu\slashed{l} \\
&\hspace{2cm}-2(-y\slashed{p}_2+(1-x)\slashed{p}_1)\gamma^\mu(-x\slashed{p}_1+(1-y)\slashed{p}_2)\nonumber\\
&\hspace{2cm}+(4-d)\left(-x\slashed{p}_1+(1-y)\slashed{p}_2-m\right)\gamma^\mu \left((1-x)\slashed{p}_1-y\slashed{p}_2-m\right) \nonumber\\
&\hspace{2cm}+\text{terms linear in $l$}.\nonumber
\end{align}
Using relations $q+p_1=p_2$ and $1=x+y+z$, one gets
\begin{align}
\gamma^\nu\dots\gamma_\nu&=-2m^2\gamma^\mu+4mz(p_1^\mu+p_2^\mu)+4m(x-y)q^\mu+(2-d)\slashed{l}\gamma^\mu\slashed{l} \\
&\hspace{2cm}-2(z\slashed{p}_2-(1-x)\slashed{q})\gamma^\mu(z\slashed{p}_1+(1-y)\slashed{q})\nonumber\\
&\hspace{2cm}+(4-d)\left(x\slashed{q}+z\slashed{p}_2-m\right)\gamma^\mu \left(z\slashed{p}_1-y\slashed{q}-m\right). \nonumber
\end{align}
Using on-shellness and completeness relations (we always have $\bar{u}(p_2)\Gamma^\mu(q^2)u(p_1)$ and therefore can write $\slashed{p}_2\,A=m\,A$ as well as $A\,\slashed{p}_1=A\, m$ for any $A$), we get for the numerator (after some rather uninspiring and tedious work)
\begin{align}
\gamma^\nu\dots\gamma_\nu&=(2-d)\slashed{l}\gamma^\mu\slashed{l}+4mz(p_1^\mu+p_2^\mu)\\
&\hspace{1cm}+mq^\mu(x-y)\left(4-2z-(4-d)(1-z)\right)\nonumber\\
&\hspace{1cm}+m^2\gamma^\mu\left(-2-2z^2+(4-d)(1-z)^2\right)\nonumber\\
&\hspace{1cm}-q^2\gamma^\mu\left(2(z+xy)-(4-d)xy\right)\nonumber\\
&\hspace{1cm}+im\sigma^{\mu\nu}q_\nu\left(2z(1+z)-(4-d)(1-z)^2\right).\nonumber
\end{align}
In the first line, we can apply the Gordon identity $p_1^\mu+p_2^\mu=2m\gamma^\mu-i\sigma^{\mu\nu}q_\nu$ and make use of the relation $\slashed{l}\gamma^\mu\slashed{l}=\gamma^\alpha\gamma^\mu\gamma^\beta g_{\alpha\beta}\frac{l^2}{d}=(2-d)\gamma^\mu \frac{l^2}{d}$ to get
\begin{align}
\gamma^\nu\dots\gamma_\nu&=(2-d)^2\gamma^\mu \frac{l^2}{d} \label{eq:numeratorManohar15}\\
&\hspace{1cm}+mq^\mu(x-y)\left(4-2z-(4-d)(1-z)\right)\nonumber\\
&\hspace{1cm}+m^2\gamma^\mu\left(8z-2-2z^2+(4-d)(1-z)^2\right)\nonumber\\
&\hspace{1cm}-q^2\gamma^\mu\left(2(z+xy)-(4-d)xy\right)\nonumber\\
&\hspace{1cm}-im\sigma^{\mu\nu}q_\nu(1-z)\left(2z+(4-d)(1-z)\right).\nonumber
\end{align}

Obviously, the denominator \eqref{eq:denom15} and the delta function is invariant under parameter exchange $x\leftrightarrow y$. Therefore, the second line of the numerator \eqref{eq:numeratorManohar15} gives zero contribution, as it changes sign under the same exchange.

Comparing to \eqref{eq:defFormfactorsManohar15}, the only contribution to $\tilde{F}_2(q^2)$ comes from the last line in \eqref{eq:numeratorManohar15}, while lines 1, 3 and 4 give contributions to $\tilde{F}_1(q^2)$. We can finally write down the one-loop form factors as
\begin{align}\label{eq:FiManohar15}
\tilde{F}_i(q^2)&=-2ie^2\mu^{4-d}\int_0^1 dx \int_0^1 dy \int_0^1 dz \int\frac{d^d l}{(2\pi)^d}\delta(1-x-y-z)\frac{A_i}{\left(l^2+xyq^2-(1-z)^2m^2\right)^3}
\end{align}
with
\begin{align}
A_1&= \frac{(2-d)^2l^2}{d}-(d-2)\left((1-z)^2m^2+xyq^2\right)+2z\left(2m^2-q^2\right)\label{eq:A1toFiManohar15}\\
A_2&=-2m^2(1-z)\left(2z+(4-d)(1-z)\right).\label{eq:A2toFiManohar15}
\end{align}

We encounter two types of phase space integrals in $d$ dimensions, namely
\begin{align}
\mathcal{I}_{1}&=\int\frac{d^d l}{(2\pi)^d} \frac{l^2}{\left(l^2-\Delta \right)^3}\text{, and}\\
\mathcal{I}_{2}&=\int\frac{d^d l}{(2\pi)^d} \frac{1}{\left(l^2-\Delta \right)^3}.
\end{align}
It is worth to point out, that integrals of the form $\mathcal{I}_2$ are not divergent in $d=4$ dimensions. To be consistent, we will still keep the dependency on $d$ explicit. This will help us to evaluate the Feynman parameter integrals, some of which diverge in $d=4$ dimensions. We will not go through the calculation of the integrals $\mathcal{I}_1$ and $\mathcal{I}_2$ step by step, as it is basically following the standard recipe of Wick rotating, parametrizing the loop momentum $l$ in $d$ dimensional polar coordinates, identifying the $d-1$ angular integrations as the surface of the $d$ dimensional unit sphere (as we have only dependence on the radius of $l$) and computing the radial integration. We get the integrals to be
\begin{align}
\mathcal{I}_1=\frac{d}{4}\frac{i}{(4\pi)^{d/2}}\frac{1}{\Delta^{2-d/2}}\Gamma\left(\frac{4-d}{2}\right) \label{eq:integral1manohar15}\\
\mathcal{I}_2=\frac{-i}{2(4\pi)^{d/2}}\frac{1}{\Delta^{3-d/2}}\Gamma\left(\frac{6-d}{2}\right).\label{eq:integral2manohar15}
\end{align}

Let us now calculate $\tilde{F}_1(q^2)$. From \eqref{eq:integral1manohar15} and \eqref{eq:integral2manohar15} as well as using the $\delta$-function for the integration over $z$, we get \eqref{eq:FiManohar15} with \eqref{eq:A1toFiManohar15} to become
\begin{align}
\tilde{F}_1(q^2)=&-2ie^2\mu^{4-d}\int_0^1 dx \int_0^{1-x} dy \frac{(2-d)^2 i}{4(4\pi)^{d/2}}\frac{1}{\left((x+y)^2m^2-xyq^2\right)^{2-d/2}}\Gamma\left(\frac{4-d}{2}\right)\nonumber\\
&-2ie^2\mu^{4-d}\int_0^1 dx \int_0^{1-x} dy \frac{-i}{2(4\pi)^{d/2}}\frac{(2-d)\left((x+y)^2m^2+xyq^2\right)}{\left((x+y)^2m^2-xyq^2\right)^{3-d/2}}\Gamma\left(\frac{6-d}{2}\right)\nonumber\\
&-2ie^2\mu^{4-d}\int_0^1 dx \int_0^{1-x} dy \frac{-i}{2(4\pi)^{d/2}}\frac{2(1-x-y)\left(2m^2-q^2\right)}{\left((x+y)^2m^2-xyq^2\right)^{3-d/2}}\Gamma\left(\frac{6-d}{2}\right)\nonumber\\
=&2e^2\frac{(2-d)^2 }{4(4\pi)^{d/2}}\frac{\mu^{4-d}}{m^{4-d}}\Gamma\left(\frac{4-d}{2}\right)\int_0^1 dx \int_0^{1-x} dy \frac{1}{\left((x+y)^2-xy\hat{q}^2\right)^{2-d/2}}\nonumber\\
&-2e^2\frac{(2-d)}{2(4\pi)^{d/2}}\frac{\mu^{4-d}}{m^{4-d}}\Gamma\left(\frac{6-d}{2}\right)\int_0^1 dx \int_0^{1-x} dy \frac{\left((x+y)^2+xy\hat{q}^2\right)}{\left((x+y)^2-xy\hat{q}^2\right)^{3-d/2}}\nonumber\\
&-2e^2\frac{1}{(4\pi)^{d/2}}\frac{\mu^{4-d}}{m^{4-d}}\Gamma\left(\frac{6-d}{2}\right)\int_0^1 dx \int_0^{1-x} dy \frac{(1-x-y)\left(2-\hat{q}^2\right)}{\left((x+y)^2-xy\hat{q}^2\right)^{3-d/2}}.\label{eq:f1masterformulaManohar15}
\end{align}
Along the way, we introduced $\hat{q}^2=q^2/m^2$. Before we calculate the leftover integrals over $x$ and $y$, we now want to expand the integrands in the heavy-electron-limit $\hat{q}^2\approx0$ to order $\hat{q}^2$. These expansions read
\begin{align}
\frac{1}{(a-b\hat{q}^2)^{2-d/2}}&=\frac{1}{a^{2-d/2}}-\frac{1}{2}\frac{b(d-4)}{a^{3-d/2}}\hat{q}^2+\mathcal{O}(\hat{q}^4)\\
\frac{a+b\hat{q}^2}{(a-b\hat{q}^2)^{3-d/2}}&=\frac{1}{a^{2-d/2}}-\frac{1}{2}\frac{b(d-8)}{a^{3-d/2}}\hat{q}^2+\mathcal{O}(\hat{q}^4)\\
\frac{c-\hat{q}^2}{(a-b\hat{q}^2)^{3-d/2}}&=\frac{c}{a^{3-d/2}}-\frac{1}{2}\frac{bc(d-6)+2a}{a^{4-d/2}}\hat{q}^2+\mathcal{O}(\hat{q}^4).
\end{align}

Therefore, we can solve the Feynman parameter integrals of the first line in \eqref{eq:f1masterformulaManohar15}
\begin{align}
\int_0^1 dx \int_0^{1-x} dy \frac{1}{\left(\dots\right)^{2-d/2}}\approx&\int_0^1 dx \int_0^{1-x} dy \frac{1}{(x+y)^{4-d}} - \frac{xy(d-4)}{2(x+y)^{6-d}}\hat{q}^2\nonumber\\
&=\frac{1}{d-2}-\frac{d-4}{12(d-2)}\hat{q}^2 \label{eq:f1firstintegralManohar15},
\end{align}
the one in the second line as
\begin{align}
\int_0^1 dx \int_0^{1-x} dy \frac{(\dots)}{\left(\dots\right)^{3-d/2}}\approx&\int_0^1 dx \int_0^{1-x} dy \frac{1}{(x+y)^{4-d}}-\frac{xy(d-8)}{2(x+y)^{6-d}}\hat{q}^2\nonumber\\
&=\frac{1}{d-2}-\frac{d-8}{12(d-2)}\hat{q}^2 \label{eq:f1secondintegralManohar15},
\end{align}
while the Feynman parameter integrals in the last line of \eqref{eq:f1masterformulaManohar15} reads
\begin{align}
\int_0^1 dx \int_0^{1-x} dy \frac{(\dots)(\dots)}{(\dots)^{3-d/2}}\approx&\int_0^1 dx \int_0^{1-x} dy\Bigg[\frac{2(1-x-y)}{(x+y)^{6-d}}\nonumber\\
&\hspace{2.8cm}-\frac{(1-x-y)(xy(d-6)+(x+y)^2)}{(x+y)^{8-d}}\hat{q}^2\Bigg]\nonumber\\
&=\frac{2-\hat{q}^2}{(d-3)(d-4)}-\frac{d-6}{6(d-3)(d-4)}\hat{q}^2 .\label{eq:f1thirdintegralManohar15}
\end{align}
Note, that the last integral is divergent in $d=4$ dimensions, which is the reason why we did the phase-space integral in $d$ dimensions.

We will now put the solutions \eqref{eq:f1firstintegralManohar15}, \eqref{eq:f1secondintegralManohar15} and \eqref{eq:f1thirdintegralManohar15} in \eqref{eq:f1masterformulaManohar15}. In the same step, we set the dimensions to $d=4-2\epsilon$ and expand to order $\epsilon^0$. The result (with $\bar{\mu}^2=4\pi e^{-\gamma_E}\mu^2$) turns out to be
\begin{align}
\tilde{F}_1(\hat{q}^2)=\frac{e^2}{16\pi^2}\left(\frac{9-2\hat{q}^2}{3\epsilon} + \frac{8-\hat{q}^2}{2}-\frac{9-2\hat{q}^2}{3}\ln\left(\frac{m^2}{\bar\mu^2}\right)\right)+\mathcal{O}(\epsilon)+\mathcal{O}(\hat{q}^4).
\end{align}
The wave function graph Fig.~\ref{fig:counter} gives
\begin{align}
\delta Z &= \frac{\alpha}{4\pi}\left( \frac{3}{\epsilon}+4 - 3 \log \frac{m^2}{\bar\mu^2}\right)
\end{align}
Adding the tree-level contribution, and subtracting $\delta Z$, we have
\begin{align}
F_1(\hat{q}^2)=1+\frac{\alpha}{4\pi}\left(-\frac{2\hat{q}^2}{3\epsilon} - \frac{\hat{q}^2}{2}+\frac{2\hat{q}^2}{3}\ln\left(\frac{m^2}{\bar\mu^2}\right)\right)+\mathcal{O}(\hat{q}^4,\epsilon).
\end{align}

Now, we will calculate $\tilde{F}_2(q^2)$ in the same manner. This will be very straight-forward as the procedure is exactly the same as in the calculation of $\tilde{F}_1(q^2)$. From \eqref{eq:FiManohar15} with \eqref{eq:A2toFiManohar15}, we have
\begin{align}
\tilde{F}_2(q^2)&=-2ie^2\mu^{4-d}\int_0^1 dx \int_0^1 dy \int_0^1 dz \int\frac{d^d l}{(2\pi)^d}\\ &\hspace{2cm}\times\delta(1-x-y-z)\frac{-2m^2(1-z)(2z+(4-d)(1-z))}{\left(l^2+xyq^2-(1-z)^2m^2\right)^3}.\nonumber
\end{align}
We can use \eqref{eq:integral2manohar15} for the phase-space integration and use the $\delta$-function for the integration over $z$
\begin{align}
\tilde{F}_2(q^2)&=-2ie^2\mu^{4-d}\int_0^1 dx \int_0^{1-x} dy \frac{-i}{2(4\pi)^{d/2}}\frac{-2m^2(x+y)(2(1+x+y)-d(x+y))}{((x+y)^2m^2-xyq^2)^{3-d/2}}\Gamma\left(\frac{6-d}{2}\right)\nonumber\\
&=2e^2\frac{1}{(4\pi)^{d/2}}\frac{\mu^{4-d}}{m^{4-d}}\Gamma\left(\frac{6-d}{2}\right)\int_0^1 dx \int_0^{1-x}dy \frac{2(x+y)+(2-d)(x+y)^2}{((x+y)^2-xy\hat{q}^2)^{3-d/2}}\label{eq:f2masterformulaManohar15},
\end{align}
where we again introduced $\hat{q}^2=q^2/m^2$. The expansion of the integrand reads
\begin{align}
\frac{1}{(a-b\hat{q}^2)^{3-d/2}}&=\frac{1}{a^{3-d/2}}-\frac{1}{2}\frac{b(d-6)}{a^{4-d/2}}\hat{q}^2+\mathcal{O}(\hat{q}^4),
\end{align}
and therefore we may write the leftover Feynman parameter integrations of $\tilde{F}_2(q^2)$ as follows
\begin{align}
\int_0^1 dx \int_0^{1-x} dy \frac{\dots}{(\dots)^{3-d/2}}\approx& \int_0^1 dx \int_0^{1-x} dy \Bigg[\frac{2}{(x+y)^{5-d}}+ \frac{(2-d)}{(x+y)^{4-d}}\nonumber\\
&\hspace{1cm}-\left(\frac{xy(d-6)}{(x+y)^{7-d}}+\frac{(2-d)xy(d-6)}{2(x+y)^{6-d}}\right)\hat{q}^2\Bigg]\nonumber\\
&=\frac{2}{d-3}-1-\frac{d-6}{6(d-3)}\hat{q}^2+\frac{d-6}{12}\hat{q}^2.\label{eq:f2integralManohar15}
\end{align}
We end the calculation by injecting \eqref{eq:f2integralManohar15} into \eqref{eq:f2masterformulaManohar15}, set $d=4-2\epsilon$ and expand to order $\epsilon^0$:
\begin{align}
F_2(\hat{q}^2)
&=\frac{e^2}{16\pi^2}\left(2+\frac{1}{3}\hat{q}^2\right)+\mathcal{O}(\epsilon)+\mathcal{O}(\hat{q}^4)\nonumber\\
&=\frac{\alpha}{4\pi}\left(2+\frac{1}{3}\hat{q}^2\right)+\mathcal{O}(\hat{q}^4, \epsilon)
\end{align}
which is finite.

\end{Lexercisenb}

\begin{Lexercisenb}{5}{10}

\item The SCET matching for the vector current $\overline \psi \gamma^\mu \psi$ for the Sudakov form factor is a variant of the previous problem. Compute $F_1(q^2)$ for on-shell massless quarks, in pure dimensional regularization with $Q^2=-q^2 \not =0$. Here $Q^2$ is the big scale, whereas in the previous problem $q^2$ was the small scale. The spacelike calculation $Q^2>0$ avoids having to deal with the $+i 0^+$ terms in the Feynman propagator which lead to imaginary parts. The timelike result can then be obtained by analytic continuation.

\medskip

\noindent {\bf SOLUTION:}

\begin{figure}
\center
\includegraphics[width=.2\textwidth]{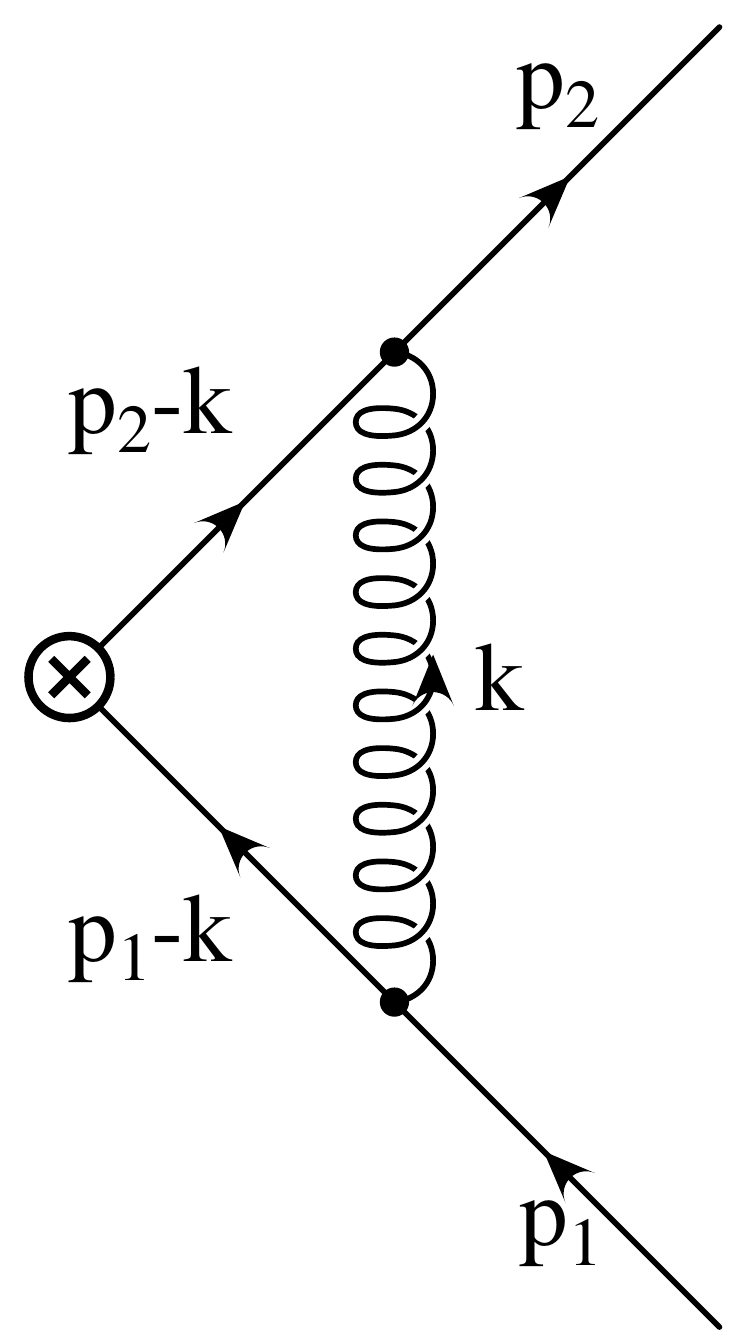}
\caption{\label{fig:virt} The Feynman diagram encoding the virtual 1-loop correction to the vector current $\overline\psi\gamma^\mu\psi$.}
\end{figure}

In SCET, the IR scales in which the full theory integral has to be expanded in are $m^2\sim\lambda^2$ and $p^2\sim\lambda^2$, where $\lambda$ is the small power counting parameter in SCET. Consequently, expanding the integrand in the IR scales leads to the massless version of the full calculation.

Using standard QCD Feynman rules for the diagram in Fig.~\ref{fig:virt}, describing the one-loop correction of the vector current $\overline{\psi}\gamma^\mu\psi$, we get
\begin{equation}
\mathcal A^\mu=-i\,(4\pi)\,\alpha_s\, C_F\, \bar u(p_2,s_2)\,\tilde\mu^{2\varepsilon}\int\frac{\dd^d k}{(2\pi)^d}\frac{\gamma^\alpha(\slashed{p}_2-\slashed{k})\gamma^\mu(\slashed{p}_1-\slashed{k})\gamma_\alpha}{(k^2-2\,p_2\cdot k)(k^2-2\,p_1\cdot k)k^2}\,v(-p_1,s_2)\,,
\end{equation}
where $p_i^2=0$ was used when multiplying out the propagators, and the usual $+i0^+$ prescription is implied.

After introducing standard Feynman parameters for the three propagators, i.e.
\begin{align}
&\frac{1}{(k^2-2\,p_2\cdot k)(k^2-2\,p_1\cdot k)k^2}\\
={}&2\int_0^1\dd x_1\int_0^{1-x_1}\dd x_2\,\frac{1}{(x_2 k^2-2\,x_2 p_2\cdot k+x_1 k^2-2\,x_1p_1\cdot k+(1-x_1-x_2)k^2)^3}\\
={}&2\int_0^1\dd x_1\int_0^{1-x_1}\dd x_2\,\frac{1}{(k^2-2\,k\cdot(x_1 p_1+x_2 p_2))^3}\\
={}&2\int_0^1\dd x_1\int_0^{1-x_1}\dd x_2\,\frac{1}{((k-(x_1 p_1+x_2 p_2))^2-x_1x_2Q^2)^3}\,,
\end{align}
and using $p_1\cdot p_2=1/2\,Q^2$, we apply a shift in the loop momentum $k\equiv \ell+x_1 p_1+x_2 p_2$. Consequently the denominator reduces to $(\ell^2-x_1 x_2 Q^2)^3$, with $\ell$ the new loop momentum.

Next we simplify the Dirac structure in the numerator. Applying the standard identity $\gamma^\alpha\gamma^\beta\gamma^\mu\gamma^\delta\gamma_\alpha=-2\gamma^\delta\gamma^\mu\gamma^\beta+(4-d)\gamma^\beta\gamma^\mu\gamma^\delta$, and using that $\bar u(p_2,s_2)\slashed{p}_2=0$ and $\slashed{p}_1v(-p_1,s_1)=0$ in the massless case, we get
\begin{equation}
\mathcal N^\mu\equiv\gamma^\alpha(\slashed p_2-\slashed k)\gamma^\mu(\slashed p_1-\slashed k)\gamma_\alpha\cong-2(\slashed p_1 - \slashed k)\gamma^\mu(\slashed p_2 - \slashed k)+(4-d)\slashed k\gamma^\mu\slashed k\,,
\end{equation}
where $\cong$ means that the equality is valid only when sandwiched between the spinors and under the loop integral. The loop momentum shift from above leads to (again using the spinors on the left and right)
\begin{equation}
\mathcal N^\mu\cong-2((1-x_1)\slashed p_1-\slashed \ell)\gamma^\mu((1-x_2)\slashed p_2-\slashed \ell)+(4-d)(\slashed \ell+x_1 \slashed p_1)\gamma^\mu(\slashed \ell+x_2\slashed p_2)\,.
\end{equation}
Reconsidering the denominator of the loop integral, we notice that the dependence is only quadratic in $\ell$, such that in the numerator all terms linear in $\ell$ evaluate to zero when integrated over due to symmetry. Consequently
\begin{equation}
\mathcal N^\mu\cong(2-d)\slashed \ell\gamma^\mu\slashed \ell-2(1-x_1)(1-x_2)\slashed p_1\gamma^\mu\slashed p_2+(4-d)x_1 x_2\slashed p_1\gamma^\mu\slashed p_2\,.
\end{equation}
The two remaining Dirac structures can furthermore be rewritten as
\begin{equation}
\slashed p_1\gamma^\mu\slashed p_2\cong-\gamma^\mu\slashed p_1\slashed p_2\cong -2\,\gamma^\mu p_1\cdot p_2=-\gamma^\mu Q^2\,,
\end{equation}
and
\begin{equation}
\slashed \ell\gamma^\mu\slashed \ell=\gamma_\alpha\gamma^\mu\gamma_\beta \ell^\alpha \ell^\beta\cong\frac{1}{d}\,\gamma_\alpha\gamma^\mu\gamma_\beta \ell^2 g^{\alpha\beta}=\gamma^\mu\frac{2-d}{d}\ell^2\,,
\end{equation}
where we used the identity $\gamma_\alpha\gamma^\mu\gamma^\alpha=(2-d)\gamma^\mu$ and the fact that due to the Lorentz structure of the loop integral we can rewrite $\ell^\alpha \ell^\beta\cong1/d\, \ell^2 g^{\alpha\beta}$ under the integral sign.

Ultimately, we arrive at
\begin{equation}
\mathcal N^\mu\cong\gamma^\mu\left[\frac{(2-d)^2}{d}\ell^2+Q^2\left(2(1-x_1)(1-x_2)-(4-d)x_1x_2\right)\right]\,.
\end{equation}

The loop integral can now be solved by using the standard formulas
\begin{align}
\tilde\mu^{2\varepsilon}\int\frac{\dd^d\ell}{(2\pi)^d}\frac{1}{(\ell^2-x_1x_2Q^2)^3}&=-\frac{i\tilde\mu^{2\varepsilon}}{(4\pi)^{d/2}}\frac{\Gamma(3-d/2)}{\Gamma(3)}(x_1x_2Q^2)^{d/2-3}\,,\\
\tilde\mu^{2\varepsilon}\int\frac{\dd^d\ell}{(2\pi)^d}\frac{\ell^2}{(\ell^2-x_1x_2Q^2)^3}&=\frac{i\tilde\mu^{2\varepsilon}}{(4\pi)^{d/2}}\frac{\Gamma(1+d/2)\Gamma(2-d/2)}{\Gamma(d/2)\Gamma(3)}(x_1x_2Q^2)^{d/2-2}\,,
\end{align}
and subsequently the Feynman parameter integrals by using
\begin{equation}
\int_0^1\dd x_1\int_0^{1-x_1}\dd x_2\,x_1^a x_2^b=\frac{\Gamma(1+a)\Gamma(1+b)}{\Gamma(3+a+b)}\,.
\end{equation}

After putting everything together and a few simplifications one can write the final result as
\begin{align}
\mathcal A^\mu&=\bar u(p_1,s_1)\gamma^\mu v(p_2,s_2)\frac{\alpha_s C_F}{4\pi}\frac{2^{-1+4\varepsilon}\,\pi^{1/2+\varepsilon}\,(\varepsilon(1-2\varepsilon)-2)\,\Gamma(1-\varepsilon)\Gamma(\varepsilon)}{\varepsilon\,\Gamma(3/2-\varepsilon)}\left(\frac{\tilde\mu^2}{Q^2}\right)^\varepsilon\\
&=\bar u(p_1,s_1)\gamma^\mu v(p_2,s_2)\frac{\alpha_s C_F}{4\pi}\bigg[-\frac{2}{\varepsilon^2}+\frac{1}{\varepsilon}\left(-2\log\frac{\mu^2}{Q^2}-3\right)-8+\frac{\pi^2}{6}-3\log\frac{\mu^2}{Q^2}\nonumber\\
&\qquad\qquad\qquad\qquad\qquad\qquad\quad-\log^2\frac{\mu^2}{Q^2}+\mathcal O(\varepsilon)\bigg]\,,
\end{align}
where $d=4-2\varepsilon$ and $\tilde\mu^{2\varepsilon}=\mathrm e^{\varepsilon\gamma_E}(4\pi)^{-\varepsilon}\mu^{2\varepsilon}$.

Dropping the $1/\varepsilon^n$ terms (and the spinor structure) we obtain the SCET matching coefficient for the vector current
\begin{equation}
C(\mu^2,Q^2)=1+\frac{\alpha_sC_F}{4\pi}\left[-8+\frac{\pi^2}{6}-3\log\frac{\mu^2}{Q^2}-\log^2\frac{\mu^2}{Q^2}\right]+\mathcal O(\alpha_s^2)\,.
\end{equation}

\end{Lexercisenb}

\begin{Lexercisenb}{5}{11}

\item Compute the SCET matching for timelike $q^2$, by analytically continuing the previous result. Be careful about the sign of the imaginary parts.

\medskip

\noindent {\bf SOLUTION:}

Remembering that $Q^2\equiv -q^2-i0^+$ and $\lim_{\epsilon\to0}\log(-1+i \epsilon)=\mathrm{sgn}\,(\epsilon)i\pi$ we can rewrite
\begin{equation}
\log\frac{\mu^2}{Q^2}=\log\frac{\mu^2}{-q^2-i0^+}=\log\frac{\mu^2}{q^2}+i\pi\,,
\end{equation}
resulting in
\begin{equation}
C(\mu^2,q^2)=1+\frac{\alpha_sC_F}{4\pi}\left[-8+\frac{7\pi^2}{6}-3\log\frac{\mu^2}{q^2}-\log^2\frac{\mu^2}{q^2}-3\,i\pi-2\,i\pi\log\frac{\mu^2}{q^2}\right]\,,
\end{equation}
for the SCET matching coefficient and
\begin{equation}
H(\mu^2,q^2)=|C(\mu^2,q^2)|^2=1+\frac{\alpha_sC_F}{4\pi}\left[-16+\frac{7\pi^2}{3}-6\log\frac{\mu^2}{q^2}-2\log^2\frac{\mu^2}{q^2}\right]\,,
\end{equation}
for the ``hard function'' appearing in SCET factorization theorems.

\end{Lexercisenb}

\begin{Lexercisenb}{5}{12}

Compute the anomalous dimension mixing matrix in eqn~(L5.56),
\begin{align*}
O_1 &= (\overline b^\alpha \gamma^\mu P_L c_\alpha)(\overline u^\alpha \gamma^\mu P_L d_\alpha) &
O_2 &= (\overline b^\alpha \gamma^\mu P_L c_\beta)(\overline u^\beta \gamma^\mu P_L d_\alpha)
\end{align*}
\begin{align*}
\mu \frac{\rd}{\rd \mu} \left[ \begin{array}{c} c_1 \\ c_2 \end{array}\right] &= \left[ \begin{array}{cc} \gamma_{11} & \gamma_{12} \\
\gamma_{21} & \gamma_{22} \end{array}\right] \left[ \begin{array}{c} c_1 \\ c_2 \end{array}\right] \,.
\end{align*}

Two other often used bases are
\begin{align*}
Q_1 &= (\overline b \gamma^\mu P_L c)(\overline u \gamma^\mu P_L d) &
Q_2 &= (\overline b \gamma^\mu P_L T^A c)(\overline u \gamma^\mu P_L T^A d)
\end{align*}
and
\begin{align*}
O_{\pm} &= O_1 \pm O_2
\end{align*}

So let
\begin{align*}
\mathcal{L} &= c_1 O_1 + c_2 O_2 = d_1 Q_1 + d_2 Q_2 = c_+ O_+ + c_- O_-
\end{align*}
and work out the transformation between the anomalous dimensions for $d_{1,2}$ and $c_{+,-}$ in terms of those for $c_{1,2}$,

\medskip

\noindent {\bf SOLUTION:}

In the EFT, the following diagrams contribute to the renormalization of the four-fermion operators,
\begin{figure}[H]
\begin{center}
\begin{tikzpicture}[
decoration={
markings,
mark=at position 0.55 with {\arrow[scale=1.5]{stealth'}};
},scale=0.7]

\draw[decorate,decoration={snake}] (-1,1) -- (-1,-1) ;
\filldraw (-0.1,-0.1) rectangle (0.1,0.1);

\filldraw (-1,1) circle (0.05);
\filldraw (-1,-1) circle (0.05);

\draw[postaction=decorate] (-1.5,1.5) -- (0,0) ;
\draw[postaction=decorate] (0,0) -- (1.5,1.5) ;
\draw[postaction=decorate] (-1.5,-1.5) -- (0,0) ;
\draw[postaction=decorate] (0,0) -- (1.5,-1.5) ;

\end{tikzpicture}\hspace{1cm}
\begin{tikzpicture}[
decoration={
markings,
mark=at position 0.55 with {\arrow[scale=1.5]{stealth'}};
},scale=0.7]

\draw[decorate,decoration={snake}] (-1,1) -- (1,1) ;
\filldraw (-0.1,-0.1) rectangle (0.1,0.1);

\filldraw (-1,1) circle (0.05);
\filldraw (1,1) circle (0.05);

\draw[postaction=decorate] (-1.5,1.5) -- (0,0) ;
\draw[postaction=decorate] (0,0) -- (1.5,1.5) ;
\draw[postaction=decorate] (-1.5,-1.5) -- (0,0) ;
\draw[postaction=decorate] (0,0) -- (1.5,-1.5) ;

\end{tikzpicture}\hspace{1cm}
\begin{tikzpicture}[
decoration={
markings,
mark=at position 0.55 with {\arrow[scale=1.5]{stealth'}};
},scale=0.7]

\draw[decorate,decoration={snake}] (135:0.75) arc (135:-45:0.75) ;
\filldraw (-0.1,-0.1) rectangle (0.1,0.1);

\filldraw (135:0.75) circle (0.05);
\filldraw (-45:0.75) circle (0.05);

\draw[postaction=decorate] (-1.5,1.5) -- (0,0) ;
\draw[postaction=decorate] (0,0) -- (1.5,1.5) ;
\draw[postaction=decorate] (-1.5,-1.5) -- (0,0) ;
\draw[postaction=decorate] (0,0) -- (1.5,-1.5) ;

\end{tikzpicture}
\end{center}
\end{figure}
we compute these in the Feynman gauge, and set the external quark masses (IR scales) to zero.

We find,
\begin{equation}
\langle \mathcal{O}_1 \rangle_{1-\text{loop}} = \left[ 1 + \frac{\alpha_s}{4 \pi} \left( 2 C_F + \frac{3}{N_c} \right) \left( \frac{1}{\epsilon} + \text{log} \frac{ \mu^2 }{ -p^2 } \right) \right] \langle \mathcal{O}_1 \rangle_{\text{tree}} - \frac{3 \alpha_s}{4 \pi} \left( \frac{1}{\epsilon} + \text{log} \frac{\mu^2}{-p^2} \right) \langle \mathcal{O}_2 \rangle_{\text{tree}}
\end{equation}
where the angled brakcets denote the matrix element $\langle \bar u d c | \mathcal{O} | b \rangle$. The same expression holds for $\langle \mathcal{O}_2 \rangle_{1-\text{loop}} $, with $\mathcal{O}_1 \leftrightarrow \mathcal{O}_2$. If $\mathcal{L}$ has
$c_1 O_1 + c_2 O_2$, one needs to add a counterterm
\begin{align}
\mathcal{L}_{\text{ct}} &= \frac{\alpha_s}{4 \pi \epsilon} c_1 \left[- \left( 2 C_F + \frac{3}{N_c} \right) O_1 +3 O_2 \right]
+ 1 \leftrightarrow 2 \,.
\end{align}

The wave function renormalization in $\MSbar$ is (see eqn~\ref{eq:zq})
\begin{equation}
Z_{\psi} = 1 - \frac{ \alpha_s C_F}{4 \pi \epsilon} .
\end{equation}
and hence
\begin{align}
c_i^{\text{bare}}&= Z_{ij} c_j & Z_{ij} &= \mathbbm{1}_{ij} - \frac{\alpha_s}{4 \pi \epsilon} \left(
\begin{array}{c c}
3/N_c & - 3 \\
-3 & 3/N_c
\end{array} \right)
\end{align}

Then using the fact that,
\begin{equation}
\mu \frac{d}{d \mu} \alpha_s ( \mu ) = - 2 \epsilon \alpha_s + \beta_{\alpha_s}
\end{equation}
where $\beta_{\alpha_s}$ is finite as $\epsilon \to 0$, we find the anomalous dimension,
\begin{align}
\mu \frac{\rd }{\rd \mu} c_i &= \gamma_{ij} c_j & \gamma &= - Z^{-1} \mu \frac{\rd }{\rd \mu} Z
\end{align}
so that
\begin{equation}
\gamma = \frac{\alpha_s}{4 \pi} \left( \begin{array}{c c}
-6 / N_c & 6 \\
6 & -6/N_c
\end{array} \right)
\end{equation}

Transforming to the basis $\mathcal{O}_{\pm} = \mathcal{O}_1 \pm \mathcal{O}_2$ diagonalizes this matrix,
\begin{equation}
\gamma^{(+,-)} =
\frac{\alpha_s}{4 \pi} \; 6 \left( \begin{array}{c c}
1 - 1/N_c & 0 \\
0 & -1 - 1/N_c
\end{array} \right)
\end{equation}

For the operators $Q_1, Q_2$, they are related to the previous operators by color and spin Fierz identities (see Exercise~\ref{ex:nfierz})
\begin{align}
Q_1 &= O_1, & Q_2 &= \left( \bar b^\alpha \gamma^\mu P_L T^A_{\alpha \beta} c^\beta \right) \left( \bar u^\mu \gamma^\mu P_L T^A_{\mu \nu} d^\nu \right) = \frac{1}{2} \mathcal{O}_2 - \frac{1}{2N_c} \mathcal{O}_1
\end{align}
and so we have,
\begin{equation}
\left( \begin{array}{c}
Q_1 \\
Q_2
\end{array} \right) =
\left( \begin{array}{c c}
1 & 0 \\
-1/2N_c & 1/2
\end{array} \right) \left( \begin{array}{c}
O_1 \\
O_2
\end{array} \right) := M_{ij} O_j
\end{equation}
and so the anomalous dimension matrix becomes,
\begin{align}
\gamma^{(d)} &= (M^T)^{-1} \gamma^{(c)} M^{T} \nonumber \\
&= \frac{\alpha_s}{4 \pi} \left( \begin{array}{c c}
0 & 3 -\frac{3}{N_c^2} \\
12 & - \frac{12}{N_c}
\end{array} \right)
\end{align}

\end{Lexercisenb}

\begin{Lexercisenb}{6}{1}

The classical equation of motion for $\lambda \phi^4$ theory,
\begin{align*}
L &= \frac12 (\partial_\mu \phi)^2 - \frac 12 m^2 \phi^2 - \frac{\lambda}{4!} \phi^4\,,
\end{align*}
is
\begin{align*}
E[\phi] &= (-\partial^2-m^2)\phi - \frac{\lambda}{3!} \phi^3\,.
\end{align*}
The EOM Ward identity for $\theta = F[\phi] E$ is eqn~(L6.32). Integrate both sides
with
\begin{align*}
\int \rd x\ e^{-i q \cdot x} \prod_i \int \rd x_i\ e^{-i p_i \cdot x_i}
\end{align*}
to get the momentum space version of the Ward identity
\bea
&&\label{ward}
\braket{0 | T \left\{ \widetilde \phi(p_1) \ldots \widetilde \phi(p_n) \widetilde \theta(q) \right \} | 0} \\
&&= i \sum_{r=1}^n
\braket{0 | T \left\{ \widetilde \phi(p_1) \ldots \cancel{\widetilde \phi(p_r)} \ldots \widetilde \phi(p_n) \widetilde F(q+p_r) \right\} | 0}\,.
\nonumber
\eea
(a) Consider the equation of motion operator
\begin{align*}
\theta_1 &= \phi \, E[\phi] = \phi (-\partial^2-m^2)\phi - \frac{\lambda}{3!} \phi^4\,,
\end{align*}
and verify the Ward identity by explicit calculation at order $\lambda$ (i.e.\ tree level) for $\phi \phi$ scattering, i.e. for $\phi \phi \to \phi \phi$. \\
(b) Take the on-shell limit $p_r^2 \to m^2$ at fixed $q \not = 0$ of
\begin{align*}
\prod_r (-i) (p_r^2-m^2) \times \text{Ward identity}\,,
\end{align*}
and verify that both sides of the Ward identity vanish. Note that both sides do not vanish if one first takes $q=0$ and then takes the on-shell limit. \\
(c) Check the Ward identity to one loop for the equation of motion operator
\begin{align*}
\theta_2 &= \phi^3 \, E[\phi] = \phi^3 (-\partial^2-m^2)\phi - \frac{\lambda}{3!} \phi^6\,.
\end{align*}

\medskip

\noindent {\bf SOLUTION:}

(a) Here $F[\phi]=\phi$ and $\theta_1=\phi E[\phi]$ in the Ward identity eqn~(\ref{ward}). The l.h.s.\ of the Ward identity is given by
the matrix element
\begin{align}
\vev{ \widetilde \phi(p_1) \, \widetilde \phi(p_2)\, \widetilde \phi(p_3) \, \widetilde \phi(p_4) \, \theta_1 (q) }
\end{align}
with all momenta incoming so that $\sum_i p_i + q=0$.

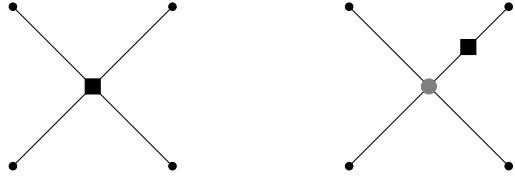
\begin{figure}
\begin{center}
\begin{tikzpicture}

\draw (0,0) -- +(45:1.5);
\draw (0,0) -- +(135:1.5);
\draw (0,0) -- +(-45:1.5);
\draw (0,0) -- +(-135:1.5);

\filldraw (0,0)+(45:1.5) circle (0.05);
\filldraw (0,0)+(135:1.5) circle (0.05);
\filldraw (0,0)+(-45:1.5) circle (0.05);
\filldraw (0,0)+(-135:1.5) circle (0.05);

\filldraw (-0.1,-0.1) -- +(0.2,0) -- +(0.2,0.2) -- +(0,0.2) -- +(0,0);

\end{tikzpicture}
\hspace{2cm}
\begin{tikzpicture}

\draw (0,0) -- +(45:1.5);
\draw (0,0) -- +(135:1.5);
\draw (0,0) -- +(-45:1.5);
\draw (0,0) -- +(-135:1.5);

\filldraw (0,0)+(45:1.5) circle (0.05);
\filldraw (0,0)+(135:1.5) circle (0.05);
\filldraw (0,0)+(-45:1.5) circle (0.05);
\filldraw (0,0)+(-135:1.5) circle (0.05);

\filldraw (45:0.6) -- +(0.2,0) -- +(0.2,0.2) -- +(0,0.2) -- +(0,0);

\filldraw[black!50] (0,0) circle (0.1);

\end{tikzpicture}
\end{center}
\caption{\label{fig6.1} Tree-level matrix element of $\theta_1$. The square box is the insertion of $\theta_1$, and the grey circle is the $-\lambda\phi^4/4!$ vertex. One has to sum over all possible insertions of $\theta_1$.}
\end{figure}
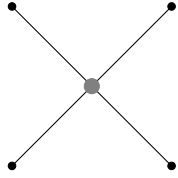
\begin{figure}
\begin{center}
\begin{tikzpicture}

\draw (0,0) -- +(45:1.5);
\draw (0,0) -- +(135:1.5);
\draw (0,0) -- +(-45:1.5);
\draw (0,0) -- +(-135:1.5);

\filldraw (0,0)+(45:1.5) circle (0.05);
\filldraw (0,0)+(135:1.5) circle (0.05);
\filldraw (0,0)+(-45:1.5) circle (0.05);
\filldraw (0,0)+(-135:1.5) circle (0.05);

\filldraw[black!50] (0,0) circle (0.1);

\end{tikzpicture}
\end{center}
\caption{\label{fig6.1a} Tree-level matrix element of $ \widetilde \phi(p_2)\, \widetilde \phi(p_3) \, \widetilde \phi(p_4) \, \widetilde \phi(p_1+q) $}
\end{figure}

The graphs in Fig.~\ref{fig6.1} give
\begin{align}\label{6.174}
& \prod_i \Delta(p_i) \left[ -4 \lambda + \sum_k [p_k^2-m^2+(p_k+q)^2-m^2] \frac{i}{(p_k+q)^2-m^2} (-i \lambda) \right] \nn
&= \prod_i \Delta(p_i) \lambda \left[ \sum_k \frac{p_k^2-m^2}{(p_k+q)^2-m^2} \right]
\end{align}
since the $-\lambda \phi^3/3!$ and $\phi(-\partial^2-m^2)\phi)$ vertices with momentum $q$ incoming have Feynman rules $-4\lambda$ and $p^2-m^2+(p+q)^2-m^2$.
Here
\begin{align}
\Delta(p) &= \frac{i}{p^2-m^2}
\end{align}
is the scalar propagator. Note that part of the $\phi(-\partial^2-m^2)\phi)$ insertion has canceled the $-\lambda \phi^3/3!$ insertion.

The r.h.s.\ of eqn~(\ref{ward}) is given by Fig.~\ref{fig6.1a}
\begin{align}
& i\vev{ \widetilde \phi(p_2)\, \widetilde \phi(p_3) \, \widetilde \phi(p_4) \widetilde \phi(p_1+q) } \, + \ldots \nn
&= i \Delta(p_2)\Delta(p_3) \Delta(p_4) \frac{i}{(p_1+q)^2-m^2} (-i \lambda) \nn
&=\lambda \prod_i \Delta(p_i) \left[ \frac{(p_1^2-m^2)}{(p_1+q)^2-m^2} + \ldots \right]=\lambda \prod_i \Delta(p_i) \sum_k \left[\frac{(p_k^2-m^2)}{(p_k+q)^2-m^2} \right]
\end{align}
which agrees with eqn~(\ref{6.174}), so the Ward identity is satisfied.

(b) Truncating the external legs, and taking the limit $p_i \to m^2$ at fixed $q$ gives zero, since the numerators vanish. However, if one first sets $q=0$, then the numerator and denominator cancel, and one gets $4\lambda$.

(c) Here $F[\phi]=\phi^3$ in the Ward identity eqn~(\ref{ward}). The tree-level contribution to the $\phi\phi\phi\phi \theta_2$ amplitude is show in Fig.~\ref{fig7.1}.

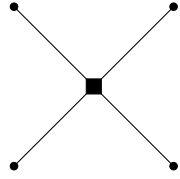
\begin{figure}
\begin{center}
\begin{tikzpicture}

\draw (0,0) -- +(45:1.5);
\draw (0,0) -- +(135:1.5);
\draw (0,0) -- +(-45:1.5);
\draw (0,0) -- +(-135:1.5);

\filldraw (0,0)+(45:1.5) circle (0.05);
\filldraw (0,0)+(135:1.5) circle (0.05);
\filldraw (0,0)+(-45:1.5) circle (0.05);
\filldraw (0,0)+(-135:1.5) circle (0.05);

\filldraw (-0.1,-0.1) -- +(0.2,0) -- +(0.2,0.2) -- +(0,0.2) -- +(0,0);

\end{tikzpicture}
\end{center}
\caption{\label{fig7.1} Tree-level matrix element of $\theta_2$.}
\end{figure}

\begin{align}\label{6.178}
I &= \sum_i 3!\, (p_i^2-m^2)
\end{align}
omitting the four external propagators.

The right hand side of the Ward identity is given by the matrix element in Fig.~\ref{fig7.2},
\begin{align}\label{6.179}
J &= i\sum_i 3!\, (-i)(p_i^2-m^2)
\end{align}
where the factor $(-i) (p_i^2-m^2)$ arises to cancel the missing propagator in the diagram, again dropping the four external propagators.

\begin{figure}
\begin{center}
\begin{tikzpicture}

\draw (0,0) -- +(45:1.5);
\draw (0,0) -- +(-45:1.5);
\draw (0,0) -- +(180:1.5);

\filldraw (0,0)+(45:1.5) circle (0.05);
\filldraw (0,0)+(-45:1.5) circle (0.05);
\filldraw (0,0)+(180:1.5) circle (0.05);

\filldraw (-0.1,-0.1) -- +(0.2,0) -- +(0.2,0.2) -- +(0,0.2) -- +(0,0);

\end{tikzpicture}
\end{center}
\caption{\label{fig7.2} Tree-level matrix element of the r.h.s.\ of the Ward identity eqn~(\ref{ward}) with $F=\phi^3$. }
\end{figure}
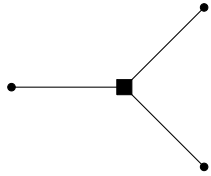

Eqn~(\ref{6.178},\,\ref{6.179}) agree, so the Ward identity is satisfied. Multiplying by $p_i^2\to m^2$ cancels the external propagators, and taking the limit $p_i^2 \to m^2$ gives zero on both sides.

The order $\lambda^2$ corrections to the l.h.s.\ of the Ward identity are shown in Fig.~\ref{fig7.3}.

\begin{figure}
\begin{center}
\begin{tikzpicture}[scale=0.9]

\draw (0,0) -- +(45:1.5);
\draw (0,0) -- +(135:1.5);
\draw (0,0) -- +(-45:1.5);
\draw (0,0) -- +(-135:1.5);

\filldraw (0,0)+(45:1.5) circle (0.05);
\filldraw (0,0)+(135:1.5) circle (0.05);
\filldraw (0,0)+(-45:1.5) circle (0.05);
\filldraw (0,0)+(-135:1.5) circle (0.05);

\draw (0.3536,0.707) circle (0.25);

\filldraw (-0.1,-0.1) -- +(0.2,0) -- +(0.2,0.2) -- +(0,0.2) -- +(0,0);

\filldraw[black!50] (0,0)+(45:0.75) circle (0.08);

\end{tikzpicture}\hspace{1cm}
\begin{tikzpicture}[scale=0.9]

\draw (0,0) -- +(45:1.5);
\draw (0,0) -- +(135:1.5);
\draw (0,0) -- +(-45:1.5);
\draw (0,0) -- +(-135:1.5);

\filldraw (0,0)+(45:1.5) circle (0.05);
\filldraw (0,0)+(135:1.5) circle (0.05);
\filldraw (0,0)+(-45:1.5) circle (0.05);
\filldraw (0,0)+(-135:1.5) circle (0.05);

\draw (0.3536,0.707) circle (0.25);

\filldraw (0.43033,0.43033) -- +(0.2,0) -- +(0.2,0.2) -- +(0,0.2) -- +(0,0);

\filldraw[black!50] (0,0) circle (0.08);

\end{tikzpicture}\hspace{1cm}
\begin{tikzpicture}[scale=0.6]

\draw (-1.5,0) -- +(135:1.5);
\draw (-1.5,0) -- +(-135:1.5);

\draw (1.5,0) -- +(45:1.5);
\draw (1.5,0) -- +(-45:1.5);

\draw (-1.5,0) .. controls (-0.5,1) and (0.5,1) .. (1.5,0);
\draw (-1.5,0) .. controls (-0.5,-1) and (0.5,-1) .. (1.5,0);

\filldraw (-1.5,0)+(135:1.5) circle (0.05);
\filldraw (-1.5,0)+(-135:1.5) circle (0.05);
\filldraw (1.5,0)+(-45:1.5) circle (0.05);
\filldraw (1.5,0)+(45:1.5) circle (0.05);

\filldraw[black!50] (1.5,0) circle (0.1);

\filldraw (-1.6,-0.1) -- +(0.2,0) -- +(0.2,0.2) -- +(0,0.2) -- +(0,0);

\end{tikzpicture}\hspace{1cm}
\begin{tikzpicture}

\draw (0,0) -- +(0:1);
\draw (0,0) -- +(180:1);
\draw (0,0) -- +(-45:1);
\draw (0,0) -- +(-135:1);

\filldraw (0,0)+(0:1) circle (0.05);
\filldraw (0,0)+(180:1) circle (0.05);
\filldraw (0,0)+(-45:1) circle (0.05);
\filldraw (0,0)+(-135:1) circle (0.05);

\draw (0,0.5) circle (0.5);

\filldraw (-0.1,-0.1) -- +(0.2,0) -- +(0.2,0.2) -- +(0,0.2) -- +(0,0);

\end{tikzpicture}
\end{center}
\caption{\label{fig7.3} One-loop matrix element of the l.h.s.\ of the Ward identity eqn~(\ref{ward}) with $F=\phi^3$. The square box is the insertion of $\theta_2$, and the grey circle is the $-\lambda\phi^4/4!$ vertex.}
\end{figure}
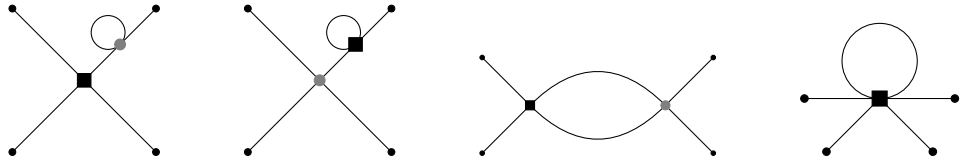

To evaluate the first graph, we need the one-loop graph from the bubble in the first graph of Fig.~\ref{fig7.3},
\begin{align}\label{m}
I_m &= \int \frac{i}{k^2-m^2} + \text{c.t.} = -\frac{ 1 }{16\pi^2} m^2 \left[1- \log \frac{m^2}{\bar \mu^2} \right]
\end{align}
which contributes to the one-loop mass shift
\begin{align}\label{m1}
\delta m^2 &= \frac{1}{2} \lambda I_m
\end{align}
of the scalar field.
The first graph has an insertion of $\phi^3(-\partial^2-m^2)\phi$. If the derivative acts on the internal line between the two vertices, it cancels the internal propagator. Otherwise, it produces a factor of $(p^2-m^2)4$ on one of the external lines. The total is (again dropping the external propagators $\prod_i\Delta(p_i)$)
\begin{align}
I_1 &= \frac{3!}2 \lambda I_m \times 4 + \frac{3!}2 \lambda I_m \left\{ (p_1^2-m^2) \left[ \frac{1}{p_2^2-m^2}+\frac{1}{p_3^2-m^2}+\frac{1}{p_4^2-m^2}\right] + \ldots\right\}
\end{align}
where $\ldots$ is the sum over $1 \leftrightarrow 2$, $1 \leftrightarrow 3$, $1 \leftrightarrow 4$.
The second graph has contributions where $p^2-m^2$ cancels the loop propagator, which gives zero, and where $p^2-m^2$ acts on one of the other lines.
The $\lambda$ from the $\lambda \phi^4$ vertex combines with the loop graph to give $\delta m^2$,
\begin{align}
I_2 &= \frac{3!}{2} \, \lambda I_m \left[ 1 + \frac{p_1^2-m^2}{(p_1+q)^2-m^2} \right] + \ldots = \frac{3!}{2} \lambda I_m\left[ 4 + \sum_i \frac{p_i^2-m^2}{(p_i+q)^2-m^2} \right]\,.
\end{align}
The third graph can have $(-\partial^2-m^2)$ acting on the internal lines, which do not satisfy $p^2=m^2$. One has to sum over the $s$, $t$, and $u$ channel diagrams. The basic loop integral for $\phi - \phi$ scattering is
\begin{align}\label{G}
G(p) &= -i \int \frac{1}{[k^2-m^2][(k+p)^2-m^2]} + \text{c.t.}
\end{align}
The $s$-channel graphs add to
\begin{align}
I_{3s} &= 3! \Bigl\{ \frac12 \lambda G(p_3+p_4) [p_1^2-m^2+p_2^2-m^2] + \frac12 \lambda G(p_1+p_2) [p_3^2-m^2+p_4^2-m^2] + 2\lambda I_m\Bigr\}
\end{align}
The last term arises from $(-\partial^2-m^2)$ acting on the internal lines, which converts the $G$ integral eqn~(\ref{G}) into the mass integral eqn~(\ref{m}).
The $t$ and $u$ channel results are given by $2 \leftrightarrow 3$ and $2 \leftrightarrow 4$.
The fourth graph gives
\begin{align}
I_{4} &= -\frac{6!}{3!} \frac12 \lambda I_m
\end{align}

The r.h.s.\ of the Ward identity is given by the $\phi \phi \phi \phi^3$ correlator, with diagrams shown in Fig.~\ref{fig7.4}.

\begin{figure}
\begin{center}
\begin{tikzpicture}[scale=0.8]

\draw (0,0) -- +(45:1.5);
\draw (0,0) -- +(180:1.5);
\draw (0,0) -- +(-45:1.5);

\filldraw (0,0)+(45:1.5) circle (0.05);
\filldraw (0,0)+(180:1.5) circle (0.05);
\filldraw (0,0)+(-45:1.5) circle (0.05);

\draw (0.3536,0.707) circle (0.25);

\filldraw (-0.1,-0.1) -- +(0.2,0) -- +(0.2,0.2) -- +(0,0.2) -- +(0,0);

\filldraw[black!50] (0,0)+(45:0.75) circle (0.09);

\end{tikzpicture}
\hspace{1cm}
\begin{tikzpicture}[scale=0.6]

\draw (-1.5,0) -- +(180:1.5);

\draw (1.5,0) -- +(45:1.5);
\draw (1.5,0) -- +(-45:1.5);

\draw (-1.5,0) .. controls (-0.5,1) and (0.5,1) .. (1.5,0);
\draw (-1.5,0) .. controls (-0.5,-1) and (0.5,-1) .. (1.5,0);

\filldraw (-1.5,0)+(180:1.5) circle (0.05);
\filldraw (1.5,0)+(-45:1.5) circle (0.05);
\filldraw (1.5,0)+(45:1.5) circle (0.05);

\filldraw[black!50] (1.5,0) circle (0.11);

\filldraw (-1.6,-0.1) -- +(0.2,0) -- +(0.2,0.2) -- +(0,0.2) -- +(0,0);

\end{tikzpicture}\hspace{1cm}
\begin{tikzpicture}[scale=0.6]

\draw (0,0) -- +(180:1.5);

\draw (0,0) -- +(0:1.5);
\draw (0,0) -- +(-90:1.5);
\draw (0,0) -- +(90:1);

\filldraw (0,0)+(180:1.5) circle (0.05);
\filldraw (0,0)+(-90:1.5) circle (0.05);
\filldraw[black!50] (0,0) circle (0.11);
\filldraw (0,0)+(0:1.5) circle (0.05);

\filldraw (1.5,0) circle (0.05);

\draw (0,1.5) circle (0.5);

\filldraw (-0.1,0.9) -- +(0.2,0) -- +(0.2,0.2) -- +(0,0.2) -- +(0,0);

\end{tikzpicture}
\end{center}
\caption{\label{fig7.4} One-loop matrix element of the r.h.s.\ of the Ward identity eqn~(\ref{ward}) with $F=\phi^3$. The square box is the insertion of $\phi^3$, and the grey circle is the $-\lambda\phi^4/4!$ vertex.}
\end{figure}

The graphs give
\begin{align}
J_1 &= \frac{3!}{2} \lambda I_m \left\{ \frac{1}{\Delta(p_1)} \left[ \Delta(p_2) + \Delta(p_3) + \Delta(p_4) \right] + \ldots \right\} \nn
J_{2s} &= \lambda \frac{3!}{2} \, \left\{G(p_3+p_4) \left[\frac{1}{\Delta(p_1)} + \frac{1}{\Delta(p_2)}\right] + G(p_1+p_2) \left[\frac{1}{\Delta(p_3)} + \frac{1}{\Delta(p_4)}\right] \right\} \nn
J_3 &= \frac{3!}{2} \lambda I_m \left\{ \frac{\Delta(p_1+q)}{\Delta(p_1)}+ \ldots \right\}
\end{align}

The Ward identity eqn~(\ref{ward}) is satisfied, since $I_1+I_2+I_{3s}+I_{3t}+I_{3u}=J_1+J_{2s}+J_{2t}+J_{2u}+J_3$. Note that there is a non-trivial cancellation of the $I_m$ terms between $I_1,I_2,I_3$.

\end{Lexercisenb}

\begin{Lexercisenb}{6}{2}

Write down all possible $C$-even dimension six terms in eqn~(L4.18), and show how they can be eliminated by field redefinitions.

\medskip

\noindent {\bf SOLUTION:}

The $C$-even terms have an even number of field strength tensors, so the dimensions six terms have two field-strength tensors and two derivatives. The possible terms are
\begin{align}
\partial_\alpha F_{\mu \nu}\, \partial^\alpha F^{\mu \nu}, \partial^2 F_{\mu \nu} F^{\mu \nu}, \partial^\alpha F_{\alpha \mu}\, \partial_\beta F^{\beta \mu}\,.
\end{align}
The second term reduces to the first one on integration by parts. Also, using
\begin{align}
\partial_\alpha F_{\mu \nu} + \partial_\mu F_{ \nu \alpha} + \partial_\nu F_{\alpha \mu } =0
\end{align}
This gives
\begin{align}
\partial_\alpha F_{\mu \nu} \partial^\alpha F^{\mu \nu} &= ( \partial_\mu F_{ \nu \alpha} + \partial_\nu F_{\alpha \mu } )^2
\implies \partial_\alpha F_{\mu \nu} \partial^\alpha F^{\mu \nu} = 2\, \partial^\alpha F_{\alpha \mu}\, \partial_\beta F^{\beta \mu}
\end{align}
Thus the Lagrangian to dimension six is
\begin{align}
\mathcal{L} &= - \frac{1}{4} F_{\mu \nu} F^{\mu nu} + \frac{c}{m_e^2} \partial^\alpha F_{\alpha \mu}\, \partial_\beta F^{\beta \mu}
\end{align}
Making the field redefinition
\begin{align}
A_\mu &\to A_\mu + \frac{1}{m_e^2} X_\mu
\end{align}
gives the Lagrangian
\begin{align}
\mathcal{L} &= - \frac{1}{4} F_{\mu \nu} F^{\mu nu} - \frac{1}{m_e^2} F_{\mu \nu} \partial_\mu X_\nu + \frac{c}{m_e^2} \partial^\alpha F_{\alpha \mu}\, \partial_\beta F^{\beta \mu} + \mathcal{O}\left( \frac{1}{m_e^4} \right)\,.
\end{align}
Choosing $X_\mu = -c \partial^\alpha F_{\alpha \mu}$ and integrating by parts eliminates the $1/m_e^2$ terms.

\end{Lexercisenb}

\begin{Lexercisenb}{6}{3}

Take the heavy quark Lagrangian
\begin{align*}
{\mathcal L}_v &= \bar Q_v \left\{ i v \cdot D + i {{\,\raise.15ex\hbox{/}\mkern-13.5mu D}}_\perp \frac{1}{ 2 m + i
v \cdot D} i {{\,\raise.15ex\hbox{/}\mkern-13.5mu D}}_\perp \right\} Q_v \nonumber \\
&= \bar Q_v \left\{ i v \cdot D - \frac{1}{2 m} {{\,\raise.15ex\hbox{/}\mkern-13.5mu D}}_\perp
{{\,\raise.15ex\hbox{/}\mkern-13.5mu D}}_\perp +\frac {1}{4 m^2} {{\,\raise.15ex\hbox{/}\mkern-13.5mu D}}_\perp \left(iv\cdot D\right)
{{\,\raise.15ex\hbox{/}\mkern-13.5mu D}}_\perp + \ldots \right\} Q_v \nonumber
\end{align*}
and use a sequence of field redefinitions to eliminate the $1/m^2$ suppressed $v \cdot D$ term. The equation of motion for the heavy quark field is $(i v \cdot D)Q_v=0$, so this example shows how to eliminate equation-of-motion operators in HQET. Here $v^\mu$ is the velocity vector of the heavy quark with $v \cdot v=1$, and
\begin{align*}
D_\perp^\mu \equiv D^\mu - (v \cdot D) v^\mu\,.
\end{align*}
If you prefer, you can work in the rest frame of the heavy quark, where $v^\mu=(1,0,0,0)$, $v \cdot D=D^0$ and $ D_\perp^\mu = (0,\mathbf{D})$. See Ref.~\cite{Manohar:1997qy} for help.

\medskip

\noindent {\bf SOLUTION:}

\noindent The aim of the exercise is to show how operators can be eliminated by using the equations of motion/field redefinitions
$(iv\cdot D)Q_v = 0$ and $\bar Q_v(iv\cdot D)Q_v = 0$. Indeed, these operators are not entirely eliminated, but sent to higher orders in the
$1/m$ expansion. So, we must think about the field redefinitions as $(iv\cdot D)Q_v = \mathcal{O}(1/m)$ and $\bar Q_v(iv\cdot D)Q_v = \mathcal{O}(1/m)$.
We will do the calculation in a general frame, even though we keep in mind that the field redefinition has to be applied in
such a way that in the rest frame temporal derivatives are removed of the Lagrangian. That is important because introducing
blindly the expression of $D_\perp$ in the $1/m$ term and using the equations of motion blindly, one could end up with a Lagrangian with
time derivatives in the rest frame. In that case the equation of motion at order $1/m$ should be used. Let's start with the local HQET
Lagrangian eqn~(2) and introduce eqn~(3) in it:

$$\mathcal{L}_v = \bar Q_v \bigg\{ i v\cdot D
- \frac{1}{2m}\slashed D_\perp \slashed D_\perp
+ \frac{1}{4m^2}\slashed D_\perp (iv\cdot D)\slashed D_\perp + \mathcal{O}(1/m^3) \bigg\}Q_v $$
we will not write $\mathcal{O}(1/m^3)$ in the following, since it is understood that we are only interested in the Lagrangian up to $\mathcal{O}(1/m^2)$.
Thus:

$$ = \bar Q_v \bigg\{ i v\cdot D
- \frac{1}{2m}\slashed D_\perp^2
+ \frac{i}{4m^2}\slashed D_\perp (v\cdot D)\slashed D_\perp \bigg\}Q_v $$

$$= \bar Q_v \bigg\{ i v\cdot D
- \frac{1}{4m}(\{\gamma^\mu, \gamma^\nu\} + [\gamma^\mu,\gamma^\nu]) D_{\perp,\mu} D_{\perp,\nu}
+ \frac{i}{8m^2}(\{\gamma^\mu, \gamma^\nu\} + [\gamma^\mu,\gamma^\nu]) D_{\perp\,\mu} (v\cdot D) D_{\perp\,\nu} \bigg\}Q_v$$

\begin{equation}
= \bar Q_v \bigg\{ i v\cdot D
- \frac{1}{2m} D_{\perp}^2
+ \frac{i}{2m} \sigma^{\mu\nu} D_{\perp,\mu} D_{\perp,\nu}
+ \frac{i}{4m^2} D_{\perp\,\mu} (v\cdot D) D_\perp^\mu
+ \frac{1}{4m^2}\sigma^{\mu\nu} D_{\perp\,\mu} (v\cdot D) D_{\perp\,\nu}\bigg\}Q_v
\end{equation}
where we used $\{\gamma^\mu,\gamma^\nu\} = 2g^{\mu\nu}I_4$ and $[\gamma^\mu,\gamma^\nu] = -2i\sigma^{\mu\nu}$.

\begin{equation}
= \bar Q_v \bigg\{ i v\cdot D
- \frac{1}{2m} D_{\perp}^2
+ \frac{i}{2m} \sigma^{\mu\nu} D_{\perp,\mu} D_{\perp,\nu}
+ \frac{i}{4m^2} D_{\perp\,\mu} (v\cdot D) D_\perp^\mu
+ \frac{1}{4m^2}\sigma^{\mu\nu} D_{\perp\,\mu} (v\cdot D) D_{\perp\,\nu}\bigg\}Q_v
\end{equation}\\
and

$$\mathcal{L}_v = \bar Q_v \bigg\{ i v\cdot D
- \frac{1}{2m} D_{\perp}^2
+ \frac{i}{4m} \sigma^{\mu\nu} [D_{\perp,\mu}, D_{\perp,\nu}]
+ \frac{i}{8m^2} [D_{\perp}^\mu, [v\cdot D, D_{\perp\,\mu}]]$$
\begin{equation}
+ \frac{1}{8m^2}\sigma^{\mu\nu} \{D_{\perp\,\mu}, [v\cdot D, D_{\perp\,\nu}]\}
\bigg\}Q_v
\end{equation}
where we added and subtracted $(v\cdot D)D_\perp^\mu$ or $D_\perp^\mu(v\cdot D)$ in order to make $v\cdot D$ to appear only inside commutators, and used the EOM conveniently. In this way, we ensure that,
in the rest frame, temporal derivatives never act over heavy quark fields, and terms involving these time derivatives can be rewritten in terms of
the chromoelectric and chromomagnetic fields.

In the rest frame, $v^\mu = (1,0,0,0)$, $v\cdot D = D^0$ and $D_\perp^\mu = (0,-{\bf D})$, the Lagrangian takes the form:

$$\mathcal{L}_v = \bar Q_v \bigg\{ i D^0
+ \frac{1}{2m} {\bf D}^2
+ \frac{i}{4m} \sigma^{ij} [{\bf D}^i, {\bf D}^j]
- \frac{i}{8m^2} [{\bf D}^i, [D^0, {\bf D}^i]]$$
\begin{equation}
+ \frac{1}{8m^2}\sigma^{ij} \{{\bf D}^i, [D^0, {\bf D}^j]\}
\bigg\}Q_v
\end{equation}
Where ${\bf D}$ is the covariant derivative in Euclidean space with the metric $g=\mbox{diag}(1,1,1)$, so we do not distinguish between up and down indices anymore.
Recall that $G^{\mu\nu} = -\frac{i}{g}[D^\mu,D^\nu]$, so the chromoelectric and chromomagnetic fields are given by
${\bf E}^{i}=G^{i0} = \frac{i}{g}[{\bf D}^i,D^0]$ and ${\bf B}^{i}=-\frac{1}{2}\epsilon^{ijk}G^{jk} = \frac{i}{2g}\epsilon^{ijk} [{\bf D}^j,{\bf D}^k]$, respectively, whereas
$\sigma^{ij}=\epsilon^{ijk}\sigma^k I_4$. Using all these relations the above Lagrangian can be written as:

\begin{equation}
\mathcal{L}_v = \bar Q_v \bigg\{ i D^0
+ \frac{1}{2m} {\bf D}^2
+ \frac{g}{2m}\sigma^i {\bf B}^{i}
+ \frac{g}{8m^2} [{\bf D}^i, {\bf E}^{i}]
+ \frac{ig}{8m^2}\epsilon^{ijk}\sigma^k \{{\bf D}^i, {\bf E}^{j}\}
\bigg\}Q_v
\end{equation}

\end{Lexercisenb}

\begin{Lexercisenb}{7}{1}

Verify that the first term in eqn~(L7.10) leads to the threshold correction in the gauge coupling given in eqn~(L7.11). If one matches at $\bar \mu=m$, then $e_L(\bar \mu)=e_H(\bar \mu)$, and the gauge coupling is continuous at the threshold. Continuity does not hold at higher loops, or when a heavy scalar is integrated out.

\medskip

{\bf SOLUTION:}

To simplify the algebra, absorb the coupling $e$ into the gauge field, so that the covariant derivative is $D_\mu
=\partial_\mu + i e A_\mu$. Then the gauge kinetic term above $m$ is
\begin{align}
\mathcal{L} &= - \frac1{4e_H^2} F_{\mu \nu}^2\,.
\end{align}
The one-loop fermion vacuum polarization graph is present in theory above $m$, and contributes as a correction to the Lagrangian below $m$.
The matrix element of $(-1/4) F_{\mu \nu}^2$ in a photon state of momentum $p$ is $-i (p^2 g_{\mu \nu}-p_\mu p_\nu)$, so
Eqn~(L7.11) gives a shift in the gauge kinetic term, and the Lagrangian below $m$ is
\begin{align}
\mathcal{L} &= - \frac1{4e_H^2} F_{\mu \nu}^2 - \frac{1}{2\pi^2} \frac16 \log \frac{m^2}{\bar \mu^2} \left( -\frac14 F_{\mu \nu}^2\right)
=- \frac1{4e_L^2} F_{\mu \nu}^2\,,
\end{align}
where the vacuum polarization graph no longer has the factor of $e^2$ because of rescaling the gauge field.
This gives
\begin{align}
\frac1{e_L^2(\bar \mu)}&= \frac1{e_H^2(\bar \mu)} - \frac1{12} \log \frac{m^2}{\bar \mu^2}
\end{align}

\end{Lexercisenb}

\begin{Lexercisenb}{7}{2}

Assume the threshold correction is of the form
\begin{align*}
\frac{1}{e_L^2(\overline\mu)} &= \frac{1}{e_H^2(\overline\mu)} + c \log \frac{m^2}{\overline\mu^2} \,.
\end{align*}
Find the relation between $c$ and the difference $\beta_H-\beta_L$ of the $\beta$-functions in the two theories, and check that this agrees with eqn~(L7.11).

\medskip

\noindent {\bf SOLUTION:}

Differentiating both sides w.r.t.\ $\mu$,
\begin{align}
- \frac{2}{e_L^3} \beta_L(e_L) &= -\frac{2}{e_H^3}\beta_H(e_H) -2c \implies
\beta_H-\beta_L = - c e^3
\end{align}
where $e = e_H =e_L$ to lowest order. In the example in the text, $\beta_H=e^3/(12\pi^2)$, $\beta_L=0$ and $c=-1/12$.

\end{Lexercisenb}

\begin{Lexercisenb}{8}{1}
\label{ex:nda}

\item Show that the power counting formula eqn~(L8.1) for an EFT Lagrangian is self-consistent, i.e.\ an arbitrary graph with insertions of vertices of this form generates an interaction which maintains
the same form. (See \cite{Gavela:2016bzc} and \cite{Manohar:1983md}). Show that eqn~(L8.1) is equivalent to
\begin{align*}
\widehat O &\sim \frac{\Lambda^4}{16 \pi^2 } \left[\frac{\partial}{\Lambda}\right]^{N_p} \left[\frac{ 4 \pi\, \phi}{ \Lambda} \right]^{N_\phi}
\left[\frac{ 4 \pi\, A}{ \Lambda } \right]^{N_A} \left[\frac{ 4 \pi \, \psi}{\Lambda^{3/2}}\right]^{N_\psi} \left[ \frac{g}{4 \pi } \right]^{N_g}
\left[\frac{y}{4 \pi } \right]^{N_y} \left[\frac{\lambda}{16 \pi^2 }\right]^{N_\lambda} .
\end{align*}

\medskip

\noindent {\bf SOLUTION:}

For consistency, each field and derivative should be scaled by the same scale, and so the most general vertex could take the form,
\begin{equation}
L_{abcd}^{ABC} = \Lambda_x^4 \; \lambda^{A} g^{B} y^{C} \; \left( \frac{ \partial }{ \Lambda_\partial } \right)^d \left( \frac{ \phi}{ \Lambda_\phi } \right)^a \left( \frac{\psi}{ \Lambda_\psi^{3/2} } \right)^b \left( \frac{A_\mu}{ \Lambda_A } \right)^c \,.
\end{equation}
Consider a diagram made of $V$ such vertices, with $L$ loops,
\begin{align}
&\left[ \int \frac{d^4 k}{(2\pi)^4} \frac{1}{k^2} \right]^{I_\phi + I_A} \left[ \int \frac{d^4 k}{(2\pi)^4} \frac{1}{k} \right]^{I_\psi} \left[ \Lambda_x^4 ( 2 \pi )^4 \delta^4 (p) \right]^{V-1} \nonumber \\
&\qquad \left( \frac{ k }{ \Lambda_\partial } \right)^{\sum_i d_i - d_E} \left( \frac{ 1 }{ \Lambda_\phi } \right)^{\sum_i a_i - a_E} \left( \frac{ 1 }{ \Lambda_\psi^{3/2} } \right)^{\sum_i b_i - b_E} \left( \frac{ 1 }{ \Lambda_A } \right)^{\sum_i c_i - c_E} L_{a_E b_E c_E d_E}^{\sum_i A_i , \sum_i B_i, \sum_i C_i}\,.
\end{align}
We replace all of the momenta with a factor of $\Lambda$, associated with the cutoff of the EFT, and perform the loop integrals over the delta functions---this gives a factor of $\Lambda^4 / 16 \pi^2$ per loop. By the conservation of ends, e.g. $\sum_i a_i - a_E = 2 I_\phi$, and so we are left with,
\begin{align}
&\left(\frac{\Lambda^4}{ 16 \pi^2 } \right)^{L} \left( \Lambda^4 \right)^{V -1 + I_\phi + I_A + I_\psi} \left( \frac{\Lambda_x}{\Lambda} \right)^{4V-4}
\left( \frac{\Lambda}{ \Lambda_\partial } \right)^{\sum_i d_i - d_E} \left( \frac{ \Lambda }{ \Lambda_\phi } \right)^{2 I_\phi} \left( \frac{\Lambda}{ \Lambda_\psi } \right)^{3 I_\psi} \left( \frac{ \Lambda }{ \Lambda_A } \right)^{2 I_A} \nonumber \\[5pt]
& \hspace{2cm} \times L_{a_E b_E c_E d_E}^{\sum_i A_i , \sum_i B_i, \sum_i C_i}\,.
\end{align}
Using $V - I + L = 1$ for a connected graph, we find that,
\begin{equation}
\left( 4 \pi \right)^{-2L} \left( \frac{\Lambda_x}{\Lambda} \right)^{4V-4} \left( \frac{\Lambda}{ \Lambda_\partial } \right)^{\sum_i d_i - d_E} \left( \frac{ \Lambda }{ \Lambda_\phi } \right)^{2 I_\phi} \left( \frac{\Lambda}{ \Lambda_\psi } \right)^{3 I_\psi} \left( \frac{ \Lambda }{ \Lambda_A } \right)^{2 I_A} L_{a_E b_E c_E d_E}^{\sum_i A_i , \sum_i B_i, \sum_i C_i}
\end{equation}
If every diagram is to give a correction to the tree-level $L$ which respects our various power counting rules, then we require,
\begin{equation}
\left( \frac{4 \pi \Lambda_x^2}{\Lambda^2} \right)^{2V-2} \left( \frac{\Lambda}{ \Lambda_\partial } \right)^{\sum_i d_i - d_E} \left( \frac{ \Lambda }{4 \pi \Lambda_\phi } \right)^{2 I_\phi} \left( \frac{\Lambda^{3/2} }{4 \pi \Lambda_\psi^{3/2} } \right)^{2 I_\psi} \left( \frac{ \Lambda }{ 4 \pi \Lambda_A } \right)^{2 I_A} \leq 1
\end{equation}
for every $V, I_\phi, I_{\psi}, I_A, d_E$.

Further, note that this counting gives kinetic terms,
\begin{equation}
\frac{\Lambda_x^4}{\Lambda^2 \Lambda_\phi^2} ( \partial_\mu \phi )^2 , \;\;\;\; \frac{ \Lambda_x^4 }{ \Lambda \Lambda_{\psi}^3 } \bar \psi \slashed{\partial} \psi , \;\;\; \frac{\Lambda_x^4 }{ \Lambda^2 \Lambda_A^2} ( F_{\mu \nu} )^2
\end{equation}
and so maintaining a canonical normalization requires,
$$ \frac{ \Lambda_x^4 }{ \Lambda^2 \Lambda_{\phi}^2 } , \; \frac{ \Lambda_x^4 }{ \Lambda \Lambda_{\psi}^3 }, \; \frac{ \Lambda_x^4 }{ \Lambda^2 \Lambda_A^2 } \leq 1 . $$
which is guaranteed by the above diagram condition.

Finally, note that for the renormalizable operators, saturating these bounds gives,
$$ \frac{\Lambda_x^4}{\Lambda_\phi^4} \phi^4 , \;\; \frac{\Lambda_x^4}{\Lambda_{\phi} \Lambda_\psi^3} \phi \bar \psi \psi , \;\; \frac{\Lambda_x^4}{\Lambda_A \Lambda_{\psi}^3 } i \bar \psi \slashed{A} \psi $$
and so to ensure order unity coefficients one must use the couplings $g' = g/4\pi$, $y' = y/4\pi$ and $\lambda' = \lambda/16 \pi^2$. Altogether then, we have a consistent power counting of,
\begin{equation}
L_{abcd}^{ABC} = \frac{ \mu^4 }{16 \pi^2} \; \left( \frac{\lambda}{16 \pi^2} \right)^{A} \left( \frac{g}{4\pi} \right)^{B} \left( \frac{y}{4 \pi} \right)^{C} \; \left( \frac{ \partial }{ \Lambda_\partial } \right)^d \left( \frac{4 \pi \phi}{ \Lambda_\phi } \right)^a \left( \frac{4 \pi \psi}{ \Lambda_\psi^{3/2} } \right)^b \left( \frac{4 \pi A_\mu}{ \Lambda_A } \right)^c
\end{equation}
where all the $\Lambda_i$ are greater than (or equal to) $\Lambda$, and the overall scale $\mu$ is less than (or equal to) $\Lambda$, where $\Lambda$ is the EFT cutoff (the scale up to which one can safely integrate loop momenta). In practice, given an EFT one can determine the cutoff in terms of $\Lambda_i$ by computing amplitudes and checking for the first breakdown of perturbative unitarity.

The most natural thing is to scale $\Lambda_x \Lambda_{\partial} = 1$, and $\Lambda_{\partial} = \Lambda_{\rm EFT}$.

\end{Lexercisenb}

\begin{Lexercisenb}{9}{1}

Show (by explicit calculation) for a general $2 \times 2$ matrix $A$ that
\begin{align*}
\frac 16 \langle A \rangle^3-\frac12\langle A \rangle \langle A^2 \rangle+\frac13\langle A^3 \rangle &=0 \,, \hspace{1em} \frac 12 \langle A \rangle^2- \frac12 \langle A^2 \rangle - \langle A \rangle A + A^2 = 0\,,\nn
\end{align*}
and for general $2 \times 2$ matrices $A,B,C$ that
\bea
0 &=& \langle A \rangle \langle B \rangle \langle C \rangle-\langle A \rangle\langle BC \rangle-\langle B \rangle \langle AC \rangle-\langle C \rangle \langle AB \rangle+\langle ABC \rangle+\langle ACB \rangle
\nonumber
\eea
Identities analogous to this for $3 \times 3$ matrices are used to remove $L_0$ and replace it by $L_{1,2,3}$ in $\chi$PT, as discussed by Pich in his lectures~\cite{Pich:cs}.

\medskip

\noindent {\bf SOLUTION:}

For a $2 \times 2$ matrix,
$$ A = \left( \begin{array}{c c}
a & b \\
c & d
\end{array} \right) $$
we have that,
\begin{align}
\langle A \rangle^3 &= (a+d)^3 , \;\;\; \langle A^2 \rangle = a^2 + d^2 + 2 bc , \;\;\; \langle A^3 \rangle = a^3 + d^3 + 3 bc ( a+d) \nonumber \\
\implies \;\; 0&= \langle A \rangle^3 - 3 \langle A \rangle \langle A^2 \rangle + 2 \langle A^3 \rangle \ \checkmark
\end{align}
and that,
\begin{align}
\langle A \rangle A &= \left(
\begin{array}{cc}
a (a+d) & b (a+d) \\
c (a+d) & d (a+d) \\
\end{array}
\right) , \;\;\;\;
A^2 = \left(
\begin{array}{cc}
a^2+b c & b (a+d) \\
c (a+d) & d^2+b c \\
\end{array}
\right) , \nonumber \\
\frac{1}{2} \left( \langle A \rangle^2 - \langle A^2 \rangle \right) \mathbbm{1} &= \left(
\begin{array}{cc}
a d-b c & 0 \\
0 & a d-b c \\
\end{array}
\right)
\end{align}
$$ \implies \;\;\;\; 0 = \frac{1}{2} \left( \langle A \rangle^2 - \langle A^2 \rangle \right)- \langle A \rangle A + A^2 \ \checkmark $$

\begin{align*}
\langle A \rangle \langle B \rangle \langle C \rangle &= \left(A_{11}+A_{22}\right) \left(B_{11}+B_{22}\right) \left(C_{11}+C_{22}\right) \\
\langle A B \rangle \langle C \rangle &= \left(C_{11}+C_{22}\right) \left(A_{11} B_{11}+A_{21} B_{12}+A_{12} B_{21}+A_{22}
B_{22}\right) \\
\langle A B C \rangle &= C_{12} \left(A_{21} B_{11}+A_{22} B_{21}\right)+A_{11} \left(B_{11} C_{11}+B_{12}
C_{21}\right) \\
&\quad +A_{12} \left(B_{21} C_{11}+B_{22} C_{21}\right)+C_{22}
\left(A_{21} B_{12}+A_{22} B_{22}\right)
\end{align*}
$$ \implies \;\;\;\; 0 = \langle A \rangle \langle B \rangle \langle C \rangle - 3 \langle A \rangle \langle BC \rangle + \langle ABC \rangle + \langle ACB \rangle + \left( A \leftrightarrow B \right) + \left( A \leftrightarrow C \right) \ \checkmark $$

\end{Lexercisenb}

\begin{Lexercisenb}{9}{2}

Show that the Jarlskog invariant
\begin{align*}
J &= \langle X_u^2 X_d^2 X_u X_d \rangle - \langle X_d^2 X_u^2 X_d X_u \rangle\,,
\end{align*}
is the lowest order $CP$-odd invariant made of the quark mass matrices. Here,
\begin{align*}
X_{u}\equiv M_{u}M_{u}^{\dagger},\hspace{1em}X_{d}\equiv M_{d}M_{d}^{\dagger},\hspace{1em}M_{u}\rightarrow LM_{u}R_{u}^{\dagger},\hspace{1em}M_{d}\rightarrow LM_{d}R_{d}^{\dagger}
\end{align*}
Show that $J$ can also be written in the form
\begin{align*}
J &= \frac13 \langle \left[X_u,X_d\right]^3 \rangle \,,
\end{align*}
and explicitly work out $J$ in the SM using the CKM matrix convention of the PDG~\cite{Patrignani:2016xqp}.

\medskip

\noindent {\bf SOLUTION:}

Under a change of basis in flavor space (bi-unitary transformation),
$M_{u}$ and $M_{d}$ transform as follows:

\begin{align*}
M_{u}\rightarrow LM_{u}R_{u}^{\dagger}\\
M_{d}\rightarrow LM_{d}R_{d}^{\dagger}
\end{align*}
where $L,R_{u}$ and $R_{d}$ are $3\times3$ unitary matrices. Let
us try to construct a basis-invariant quantity from the quarks mass
matrices. Because $R_{u}\neq R_{d}$, we must construct the Hermitian
matrices $X_{u}\equiv M_{u}M_{u}^{\dagger}$ and $X_{d}\equiv M_{d}M_{d}^{\dagger}$
which transform as $U(3)$ octets:

\begin{align*}
X_{u}\rightarrow LX_{u}L^{\dagger}\\
X_{d}\rightarrow LX_{d}L^{\dagger}
\end{align*}
Any polynomial $p(X_{u},X_{d})$ also transforms as an octet:
\[
p(X_{u},X_{d})\rightarrow Lp(X_{u},X_{d})L^{\dagger}.
\]
In order to eliminate the remaining $L$ and $L^{\dagger}$ to get
an invariant, we take the trace of $p(X_{u},X_{d})$. Indeed, because
the trace is cyclic $\langle AB\rangle=\langle BA\rangle$ and $L\in U(3)$
$(LL^{\dagger}=\mathbb{I}_{3})$, we have:
\[
\langle p(X_{u},X_{d})\rangle\rightarrow\langle Lp(X_{u},X_{d})L^{\dagger}\rangle=\langle p(X_{u},X_{d})\mathbb{I}_{3}\rangle=\langle p(X_{u},X_{d})\rangle.
\]
Therefore, the trace of any polynomial of the Hermitian matrices $X_{u}$
and $X_{d}$ is a flavor invariant. \newline \newline However, we
want a CP-odd invariant. So, we must select only purely imaginary
traces. In other words, we must find an invariant such that:
\begin{align*}
\text{Re}\langle p(X_{u},X_{d})\rangle=0\\
\text{Im}\langle p(X_{u},X_{d})\rangle\neq0.
\end{align*}
Let us examine the simplest monomials, the powers of $X_{u,d}$ of
the form $X_{u}^{n}$ or $X_{d}^{n}$, the products of two powers
of $X_{u,d}$ of the form $X_{u}^{n}X_{d}^{m}$, the products of three
powers of $X_{u,d}$ of the form $X_{u}^{n}X_{d}^{m}X_{u}^{p}$ and
so on... Where $n,m$ and $p$ are non-zero integers.

We have $(X_{u}^{n})^{\dagger}=X_{u}^{n}$ and $(X_{d}^{n})^{\dagger}=X_{d}^{n}$,
so $X_{u}^{n}$ and $X_{d}^{n}$ are Hermitian matrices and then their
diagonal entries are real and so is their traces. Thus, the corresponding
invariants $\langle X_{u}^{n}\rangle$ and $\langle X_{d}^{n}\rangle$
are real and then CP-even.

Regarding the monomials which are products of two powers of $X_{u,d}$,
of the form $X_{u,d}^{n}X_{d,u}^{m}$, let us calculate the imaginary
part of the corresponding invariants, by using $\text{Im}\langle M\rangle=\frac{1}{2i}\langle M-M^{\dagger}\rangle$:

\begin{align*}
2i\text{Im}\langle X_{u}^{n}X_{d}^{m}\rangle= & \langle X_{u}^{n}X_{d}^{m}-(X_{u}^{n}X_{d}^{m})^{\dagger}\rangle\\
= & \langle X_{u}^{n}X_{d}^{m}-X_{d}^{m}X_{u}^{n}\rangle\\
= & \langle X_{u}^{n}X_{d}^{m}\rangle-\langle X_{d}^{m}X_{u}^{n}\rangle\\
= & \langle X_{u}^{n}X_{d}^{m}\rangle-\langle X_{u}^{n}X_{d}^{m}\rangle=0.
\end{align*}
The same holds for $\langle X_{d}^{n}X_{u}^{m}\rangle$. Therefore,
the invariants of the form $\langle X_{u}^{n}X_{d}^{m}\rangle$ and
$\langle X_{d}^{n}X_{u}^{m}\rangle$ are real and then CP-even.

Let us now consider the monomials which are products of three powers
of $X_{u,d}$, of the form $\langle X_{u,d}^{n}X_{d,u}^{m}X_{u,d}^{p}\rangle$:

\begin{align*}
2i\text{Im}\langle X_{u}^{n}X_{d}^{m}X_{u}^{p}\rangle= & \langle X_{u}^{n}X_{d}^{m}X_{u}^{p}-(X_{u}^{n}X_{d}^{m}X_{u}^{p})^{\dagger}\rangle\\
= & \langle X_{u}^{n}X_{d}^{m}X_{u}^{p}-X_{u}^{p}X_{d}^{m}X_{u}^{n}\rangle\\
= & \langle X_{u}^{n}X_{d}^{m}X_{u}^{p}\rangle-\langle X_{u}^{p}X_{d}^{m}X_{u}^{n}\rangle\\
= & \langle X_{u}^{n+p}X_{d}^{m}\rangle-\langle X_{u}^{p+n}X_{d}^{m}\rangle=0
\end{align*}
The same holds for $\langle X_{d}^{n}X_{u}^{m}X_{d}^{p}\rangle$.
So, these invariants of the form $\langle X_{u}^{n}X_{d}^{m}X_{u}^{p}\rangle$
and $\langle X_{d}^{n}X_{u}^{m}X_{d}^{p}\rangle$ are also real and
then CP-even.

Next, we consider the monomials which are products of four powers
of $X_{u,d}$, of the form $X_{u,d}^{n}X_{d,u}^{m}X_{u,d}^{p}X_{d,u}^{q}$:
\begin{align*}
2i\text{Im}\langle X_{u}^{n}X_{d}^{m}X_{u}^{p}X_{d}^{q}\rangle= & \langle X_{u}^{n}X_{d}^{m}X_{u}^{p}X_{d}^{q}-(X_{u}^{n}X_{d}^{m}X_{u}^{p}X_{d}^{q})^{\dagger}\rangle\\
= & \langle X_{u}^{n}X_{d}^{m}X_{u}^{p}X_{d}^{q}-X_{d}^{q}X_{u}^{p}X_{d}^{m}X_{u}^{n}\rangle\\
= & \langle X_{u}^{n}X_{d}^{m}X_{u}^{p}X_{d}^{q}\rangle-\langle X_{d}^{q}X_{u}^{p}X_{d}^{m}X_{u}^{n}\rangle\\
= & \langle X_{u}^{n}X_{d}^{m}X_{u}^{p}X_{d}^{q}\rangle-\langle X_{u}^{n}X_{d}^{q}X_{u}^{p}X_{d}^{m}\rangle=0\text{ if \ensuremath{m=q}}\\
= & \langle X_{d}^{q}X_{u}^{n}X_{d}^{m}X_{u}^{p}\rangle-\langle X_{d}^{q}X_{u}^{p}X_{d}^{m}X_{u}^{n}\rangle=0\text{ if \ensuremath{n=p}}.
\end{align*}
As we want a non-zero imaginary part, we must choose $m\neq q$ and
$n\neq p$. The simplest choice (lowest order in quarks mass matrices)
is then $(n,m,p,q)=(2,2,1,1)$ and corresponds to the invariant: $\langle X_{u}^{2}X_{d}^{2}X_{u}X_{d}\rangle$.
This invariant is not CP-even (because it is complex) but is it CP-odd
(purely imaginary)? Let us compute its real part using $\text{Re}\langle M\rangle=\frac{1}{2}\langle M+M^{\dagger}\rangle$

\begin{align*}
2\text{Re}\langle X_{u}^{2}X_{d}^{2}X_{u}X_{d}\rangle= & \langle X_{u}^{2}X_{d}^{2}X_{u}X_{d}+(X_{u}^{2}X_{d}^{2}X_{u}X_{d})^{\dagger}\rangle\\
= & \langle X_{u}^{2}X_{d}^{2}X_{u}X_{d}+X_{d}X_{u}X_{d}^{2}X_{u}^{2}\rangle\\
= & \langle X_{u}^{2}X_{d}^{2}X_{u}X_{d}\rangle+\langle X_{d}X_{u}X_{d}^{2}X_{u}^{2}\rangle\\
= & \langle X_{u}^{2}X_{d}^{2}X_{u}X_{d}\rangle+\langle X_{u}^{2}X_{d}X_{u}X_{d}^{2}\rangle\neq0\text{ (a priori)}
\end{align*}
Thus, this invariant is not CP-odd. It is the sum of a CP-even and
a CP-odd invariant:
\[
\langle X_{u}^{2}X_{d}^{2}X_{u}X_{d}\rangle=\underset{\text{CP-even}}{\underbrace{\text{Re}\langle X_{u}^{2}X_{d}^{2}X_{u}X_{d}\rangle}}+\underset{\text{CP-odd}}{\underbrace{i\text{Im}\langle X_{u}^{2}X_{d}^{2}X_{u}X_{d}\rangle}},
\]
Therefore, the lowest order CP-odd invariant is:

\begin{align*}
i\text{Im}\langle X_{u}^{2}X_{d}^{2}X_{u}X_{d}\rangle= & \frac{1}{2}\langle X_{u}^{2}X_{d}^{2}X_{u}X_{d}-(X_{u}^{2}X_{d}^{2}X_{u}X_{d})^{\dagger}\rangle\\
= & \frac{1}{2}\langle X_{u}^{2}X_{d}^{2}X_{u}X_{d}-X_{d}X_{u}X_{d}^{2}X_{u}^{2}\rangle\\
= & \frac{1}{2}(\langle X_{u}^{2}X_{d}^{2}X_{u}X_{d}\rangle-\langle X_{d}^{2}X_{u}^{2}X_{d}X_{u}\rangle)\\
= & \frac{1}{2}J
\end{align*}

Let us calculate $\langle[X_{u},X_{d}]^{3}\rangle$:
\begin{align*}
\langle[X_{u},X_{d}]^{3}\rangle= & \langle(X_{u}X_{d}-X_{d}X_{u})^{3}\rangle\\
= & \langle X_{u}X_{d}X_{u}X_{d}X_{u}X_{d}-3X_{u}X_{d}X_{u}X_{d}^{2}X_{u}+3X_{u}X_{d}^{2}X_{u}X_{d}X_{u}-X_{d}X_{u}X_{d}X_{u}X_{d}X_{u}\rangle\\
= & \langle X_{u}X_{d}X_{u}X_{d}X_{u}X_{d}\rangle+3\langle X_{u}X_{d}^{2}X_{u}X_{d}X_{u}-X_{u}X_{d}X_{u}X_{d}^{2}X_{u}\rangle-\langle X_{d}X_{u}X_{d}X_{u}X_{d}X_{u}\rangle\\
= & 3\langle X_{u}X_{d}^{2}X_{u}X_{d}X_{u}-X_{u}X_{d}X_{u}X_{d}^{2}X_{u}\rangle\\
= & 3\langle X_{u}^{2}X_{d}^{2}X_{u}X_{d}-X_{d}^{2}X_{u}^{2}X_{d}X_{u}\rangle\\
= & 3J
\end{align*}
Then, $J$ can be written in the form $J=\frac{1}{3}\langle[X_{u},X_{d}]^{3}\rangle$.\newline\newline
Let us work out explicitly $J$. In the gauge basis in which all down-type
quarks are mass eigenstates, we have: $M_{u}=V_{\text{CKM}}^{\dagger}\left(\begin{array}{ccc}
m_{u} & 0 & 0\\
0 & m_{c} & 0\\
0 & 0 & m_{t}
\end{array}\right)$ and $M_{d}=\left(\begin{array}{ccc}
m_{d} & 0 & 0\\
0 & m_{s} & 0\\
0 & 0 & m_{b}
\end{array}\right)$.\newline
\begin{itemize}
\item In the standard parametrization (PDG), the CKM matrix is written as:
\[
V_{\text{CKM}}=\left(\begin{array}{ccc}
c_{12}c_{13} & s_{12}c_{13} & s_{13}e^{-i\delta}\\
-s_{12}c_{23}-c_{12}s_{23}s_{13}e^{i\delta} & c_{12}c_{23}-s_{12}s_{23}s_{13}e^{i\delta} & s_{23}c_{13}\\
s_{12}s_{23}-c_{12}c_{23}s_{13}e^{i\delta} & -c_{12}s_{23}-s_{12}c_{23}s_{13}e^{i\delta} & c_{23}c_{13}
\end{array}\right),
\]
where $c_{ij}=\cos\theta_{ij}$ and $s_{ij}=\sin\theta_{ij}$. The
explicit computation of $J=\frac{1}{3}\langle[M_{u}M_{u}^{\dagger},M_{d}M_{d}^{\dagger}]^{3}\rangle$
yields:
\bea
J&=&\frac{i}{2}\sin(2\theta_{12})\sin(2\theta_{13})\sin(\theta_{13})\cos^{2}(\theta_{13})\sin(\delta) \nonumber\\
&& \times (m_{b}^{2}-m_{d}^{2})(m_{b}^{2}-m_{s}^{2})(m_{s}^{2}-m_{d}^{2})(m_{t}^{2}-m_{u}^{2})(m_{t}^{2}-m_{c}^{2})(m_{c}^{2}-m_{u}^{2})
\nonumber
\eea
\item In the Wolfenstein parametrization (PDG), we express the CKM matrix
as:
\end{itemize}
\[
V_{\text{CKM}}=\left(\begin{array}{ccc}
1-\tfrac{\lambda^{2}}{2} & \lambda & A\lambda^{3}(\rho-i\eta)\\
-\lambda & 1-\tfrac{\lambda^{2}}{2} & A\lambda^{2}\\
A\lambda^{3}(1-\rho-i\eta) & -A\lambda^{2} & 1
\end{array}\right)+\mathcal{O}(\lambda^{4}),
\]
then we compute $J$ and obtain:
\[
J=2iA^{2}\eta\lambda^{6}(m_{b}^{2}-m_{d}^{2})(m_{b}^{2}-m_{s}^{2})(m_{s}^{2}-m_{d}^{2})(m_{t}^{2}-m_{u}^{2})(m_{t}^{2}-m_{c}^{2})(m_{c}^{2}-m_{u}^{2})+O(\lambda^{7})
\]

\end{Lexercisenb}

\begin{Lexercisenb}{9}{3}

Compute the Hilbert series for the ring of invariants generated by\\
(a) $x$, $y$ (each of dimension 1), and invariant under the transformation $(x,y)\to(-x,-y)$.\\
(b) $x$, $y$, $z$ (each of dimension 1), and invariant under the transformation $(x,y,z)\to(-x,-y,-z)$.

\medskip

\noindent {\bf SOLUTION:}

(a)

In a first step we count the number of invariants $N_n$ of degree $n$. Obviously, invariants can only be built for even $n$, namely of the form $x^a y^b$, with $a+b=n$ even. There are $n+1$ non-equivalent possibilities to build such an invariant of degree $n$, such that
\begin{equation}
N_n=\begin{cases}
n+1 & \text{for } n \text{ even}\\
0 & \text{for } n \text{ odd}
\end{cases}\,,
\end{equation}
and consequently the Hilbert series is
\begin{equation}
H(q)=\sum_{n=0}^\infty N_n q^n=\sum_{n=0}^\infty (2n+1)\, q^{2n}=\frac{1+q^2}{(1-q^2)^2}\,.
\end{equation}

This expressions can be easily interpreted: There are 2 invariant generators of degree $n=2$, $x^2$ and $y^2$, (indicated by the power of the denominator and the power of $q$ therein respectively) which are linearly independent regardless of to which power they are raised. Additionally, there is one invariant generator of degree $n=2$, $xy$, which satisfies the relation $(xy)^2=x^2y^2$ at degree $n=4$ such that expressions of this form can be written as a linear combination of the other two generators.

One can rewrite the given expression for $H(q)$ as
\begin{equation}
H(q)=\frac{1-q^4}{(1-q^2)^3}\,,
\end{equation}
allowing the same interpretation: there are three generators of degree $n=2$, with one relation between them at degree $n=4$.

(b)

Analogous to exercise (a), we count the number of invariants $N_n$ of degree $n$, which are now of the form $x^a y^b z^c$ with $a+b+c=n$, $n$ even. Writing these down in a systematic way one counts
\begin{equation}
\underbrace{x^0 y^0 z^n,\,x^0 y^1 z^{n-1},\,\dots,\,x^0 y^{n-1} z^1,\,x^0 y^n z^0}_{n+1\text{ terms}},\,\underbrace{x^{1} y^{0} z^{n-1},\,\dots,\,x^{1} y^{n-1} z^{0}}_{n \text{ terms}},\,\dots,\,\underbrace{x^n y^0 z^0}_{1 \text{ term}}\,,
\end{equation}
such that
\begin{equation}
N_n=
\begin{cases}
\sum_{j=1}^{n+1}j=\frac{1}{2}(1+n)(2+n) & \text{for } n \text{ even}\\
0 & \text{for } n \text{ odd}
\end{cases}\,.
\end{equation}
Therefore, the Hilbert series is
\begin{equation}
H(q)=\sum_{n=0}^\infty\frac{1}{2}(1+2n)(2+2n)q^{2n}=\frac{1+3\,q^2}{(1-q^2)^3}\,.
\end{equation}

The result can be interpreted analogous to part (a): There are 3 invariant generators of degree $n=2$, $x^2$, $y^2$ and $z^2$, where all powers of them remain linearly independent. This can be read off the power of the denominator and the power of $q$ therein respectively. Additionally, there are three generators of degree $n=2$, $xy$, $xz$ and $yz$, (indicated by the coefficient 3 in the numerator and the power of $q$ respectively) which fulfill the relations $(xy)^2=x^2 y^2$, $(xz)^2=x^2 z^2$ and $(yz)^2=y^2 z^2$ at degree $n=4$, such that expressions of this form can, analogously to part (a), be written as a linear combination of the three generaltors identified at the beginning of this paragraph.

\end{Lexercisenb}

\begin{Lexercisenb}{10}{1}
\label{ex:11.1}

Show that $(\psi_{Lr}^T C \psi_{Ls})$ is symmetric in $rs$ and $(\psi_{Lr}^T C \sigma^{\mu \nu} \psi_{Ls})$ is antisymmetric in $rs$.

\medskip

\noindent {\bf SOLUTION:}

{\bf Solution:}

\begin{align}
(\psi_{Lr})^{T} C \psi_{Ls} &= (\psi_{Lr \alpha}) C_{\alpha \beta} \psi_{Ls\beta} = - \psi_{Ls\beta} C_{\alpha \beta} (\psi_{Lr \alpha})
= - \psi_{Ls}^T C^T (\psi_{Lr})
\end{align}
where the minus sign is from anticommuting Fermi fields. Similarly
\begin{align}
(\psi_{Lr}^T C \sigma^{\mu \nu} \psi_{Ls}) &= - (\psi_{Ls}^T (C \sigma^{\mu \nu})^T \psi_{Lr})
\end{align}
Using two-component left-handed fields, $C_{\alpha \beta}= \epsilon_{\alpha \beta} = i \sigma^2$ so $C^T=-C$, so the scalar operator is symmetric in $rs$. $\sigma^{\mu \nu}$ is proportional to the Pauli matrices $\sigma^k$ ($\sigma^{ij} = \epsilon^{ijk} \sigma^k$, $\sigma^{0k} =- i \sigma^k$) and
\begin{align}
(i \sigma^2 \sigma^k)^T = (\sigma^k)^T (-i \sigma^2) = (\sigma^2)(-\sigma^k)(\sigma^2)(-i \sigma^2)
= -i \sigma^2 \sigma^k
\end{align}
so the tensor operator is symmetric.

\end{Lexercisenb}

\begin{Lexercisenb}{10}{2}

Prove the duality relations eqns~(L10.7,L10.8). The sign convention is $\gamma_5 = i \gamma^0 \gamma^1 \gamma^2 \gamma^3$ and $\epsilon_{0123}=+1$.

\medskip

\noindent {\bf SOLUTION:}

The identities are equivalent to showing that
\begin{align}
\frac{i}{2} \epsilon^{\alpha \beta \mu \nu} \sigma_{\mu \nu} \gamma_5 &= -\sigma^{\alpha \beta}.
\end{align}
The identity is true up to an overall normalization, since both sides are two-index antisymmetric tensors of the same parity. The normalization is fixed by looking at one term, e.g.\
\begin{align}
\frac{i}{2} \epsilon^{0 1 \mu \nu} \sigma_{\mu \nu} \gamma_5 =
i \epsilon^{0 1 2 3} \sigma_{2 3} \gamma_5 = i \epsilon^{0 1 2 3} (i \gamma_2 \gamma_3)(i \gamma^0 \gamma^1 \gamma^2
\gamma^3) = -i \gamma^0 \gamma^1
& \stackrel{?}{=} -\sigma^{0 1}.
\end{align}
so the normalization is correct.

\end{Lexercisenb}

\begin{Lexercisenb}{10}{3}

\item Show that eqn~(L10.9) is the unique dimension-five term in the SMEFT Lagrangian. How many independent operators are there for $n_g$ generations?

\medskip

\noindent {\bf SOLUTION:}

This discussion is taken from Buchm\"uller and Wyler~\cite{Buchmuller:1985jz}. A dimension-5 term in SMEFT cannot be built purely from fermions or scalars. Since fermions have mass-dimension 3/2 (in $d=4$ dimensions), they must appear in pairs in a Lagrangian term to have integer mass dimension. Consequently, a maximum of two fermions can appear in an operator with mass dimension less than six. The only scalar in the SM is the Higgs doublet, and it is not possible to construct an SU(2) singlet from five SU(2) doublets. Therefore, for reasons of dimensionality and the requirement of building an SU(2) singlet, it is not possible to have an odd number of fermions or scalars. The only possibility is to have two fermions and two scalars.

If the scalars are taken to be $H$ and $H^{*}$, then they have a total hypercharge of zero, and so the two fermions must also have total hypercharge zero so that the Lagrangian remains invariant under $\rm U(1)_{Y}$. This is only possible by taking a multiplet and its charge conjugate, but their product cannot then form a Lorentz scalar. Alternatively, the two scalars can both be taken to be $H$, in which case it is possible to write the operators \cite{Weinberg:1979sa}
\begin{equation}
\brackets{\ell^{T}_{i} C \ell_{j} }H_{k}H_{l}\epsilon^{ik}\epsilon^{jl} \hspace{10mm} {\rm and } \hspace{10mm} \brackets{\ell^{T}_{i} C \ell_{j} }H_{k}H_{l}\epsilon^{ij}\epsilon^{kl}.
\end{equation}
However, the second term is forbidden if there is only a single Higgs doublet (i.e. no extended Higgs sector), since $H^{T}\epsilon H$ is identically zero. Therefore the only dimension-5 term that can be written in SMEFT is
\begin{equation}
L_{5} = c_{5} \brackets{\ell^{T}_{i} C \ell_{j} }H_{k}H_{l}\epsilon^{ik}\epsilon^{jl}.
\end{equation}

For $n_{g}$ (fermion) generations, each lepton acquires a generation index that runs from $1,\dots,n_{g}$, and the Wilson coefficient becomes a matrix in generation space (analogous to the Yukawa matrices). From Exercise~\ref{ex:11.1}, the operator is symmetric in generation indices, so there are $n_g(n_g+1)/2$ independent operators.

\end{Lexercisenb}

\begin{Lexercisenb}{10}{4}

Show that eqn~(L10.9) generates a Majorana neutrino mass when $H$ gets a vacuum expectation value, and find the neutrino mass matrix $M_\nu$ in terms of $C_5$ and $v$.

\medskip

\noindent {\bf SOLUTION:}

{\bf Solution: } A Majorana mass term for a Dirac fermion $\psi$ may be written as
\begin{equation}
-\frac12 M \psi_{L}^{T} C \psi_{L},
\end{equation}
where $M$ is the Majorana mass.

Upon electroweak symmetry breaking, the Higgs acquires a VEV as
\begin{equation}
H \to \frac{1}{\sqrt{2}}
\begin{pmatrix}
0 \\
v
\end{pmatrix}.
\end{equation}
Then
\begin{align}
c_{5} \brackets{\ell^{T}_{i} C \ell_{j} }H_{k}H_{l}\epsilon^{ik}\epsilon^{jl} &\to \frac{c_{5}}{2}
\begin{pmatrix}
\nu_{L}^{T} & e_{L}^{T}
\end{pmatrix}
\begin{pmatrix}
0 & 1 \\
-1 & 0
\end{pmatrix}
\begin{pmatrix}
0 \\
v
\end{pmatrix}
\begin{pmatrix}
C \nu_{L} & Ce_{L}
\end{pmatrix}
\begin{pmatrix}
0 & 1 \\
-1 & 0
\end{pmatrix}
\begin{pmatrix}
0 \\
v
\end{pmatrix}
\\
&= \frac{c_{5} v^{2}}{2} \nu_{L}^{T} C \nu_{L}.
\end{align}

The Majorana mass can then be identified as
\begin{equation}
M= -c_{5} v^{2}.
\end{equation}

\end{Lexercisenb}

\begin{Lexercisenb}{10}{5}
Prove eqn~(L10.21).

\medskip

\noindent {\bf SOLUTION:}

Done in eqn~(\ref{sol10}).

\end{Lexercisenb}

\begin{Lexercisenb}{10}{6}

\item In the SMEFT for $n_g$ generations, how many operators are there of the following kind (in increasing order of difficulty): (a) $Q_{He}$ (b) $Q_{ledq}$ (c) $Q_{lq}^{(1)}$ (d) $Q_{qq}^{(1)}$ (e) $Q_{ll}$ (f) $Q_{uu}$ (g) $Q_{ee}$ \\ (h) show that there are a total of 2499 Hermitian dimension-six $\Delta B= \Delta L=0$ operators.

\medskip

\noindent {\bf SOLUTION:}

The enumeration of operators of each combination of elementary field is given by the procedure in Ref.~\cite{Henning:2015alf}, summarized by their master formula for the Hilbert series of the dim $6$ Standard Model in eqn~(3.16) for one generation. As the number of operators for each elementary field combination is so small, this is sufficient to deduce a complete set of independent operators, such as those tabulated in the present section, by simply guessing. To extend the enumeration to $n_g$ generations, as stated in Ref.~\cite{Henning:2015alf}, the Hilbert series may be derived by raising the plethystic exponential to a power of $n_g$ and then expanding and projecting onto the relevant components.

Alternatively, it is simple enough instead to inscribe flavor indices on the operators presented here and counting (see Appendix~A of Ref.~\cite{Alonso:2013hga}). Giving the fermions in each operator a flavor index, the number of independent flavor components may be simply counted by decomposing into representation of the flavor group, subject to symmetry constraints of the elementary fields.

The operator class $Q_{ledq}\sim (L_{i,I}\bar{e}_j)(Q^I_k\bar{d}_l)$, where $I$ denote isospin indices (included for clarity) and $i,j,k,l$ are flavor indices, consists of only distinct fields, so there are $n_g^4$ such operators, where each factor of $n_g$ is simply the number of possible flavor identities that each field can take. Then, as the conjugate operator is distinct, exactly the same enumeration holds for its conjugates, giving a total of $2n_g^4$ operators.

Similarly, for $Q_{He}\sim \bar{e}^{\dagger i}\bar{\sigma}^\mu\bar{e}_j(H^\dagger i D_\mu H - iD_\mu H^\dagger H)$, the $\bar{e}^\dagger$ and $\bar{e}$ fields are distinguished as conjugates, so are distinct, implying that there are $n_g\times n_g$ possible operators. The adjoint of these flavored operators just reverses the flavor indices, so there are no additional contributions to these $n_g^2$ operators from conjugates. Identical arguments apply to operators with the replacement of $\bar{e}$ with any of the other fermion fields (and is unaffected if the bilinear has a triplet isospin structure that is contracted with the Higgs bilinear, which can occur if the fermion is isospinning). There is one exceptional example $Q_{Hud}$, which is not Hermitian (because the fermions are distinct), so has twice the number of flavor components.

The operator $Q^{(1)}_{lq}\sim (L^{\dagger i}\bar{\sigma}^\mu L_j)(Q^{\dagger k}\bar{\sigma}_\mu Q_l)$ has similar structure, where all gauge indices are contracted within each bilinear. Each operator is distinct, so there are $n_g^4$ possible flavor components, that transform into each other under conjugation. Identical counting applies to other operators consisting of two different bilinears of fermions of the same type with their conjugates, including those where the bilinears are not gauge singlets.

Other four fermion operators involving identical fermions may be counted by giving them flavor indices and determining the number of non-zero entries. This may be done systematically beginning with the states of the most minimal internal structure.

Beginning generally with $(f^{\dagger i a}\bar{\sigma}^\mu f_{j a})(f^{\dagger k b}\bar{\sigma}_\mu f_{l b})\sim (f^{\dagger i a}f^{\dagger k b})(f_{j a}f_{l b})$ by a Fierz identity. Here $a$ and $b$ denote internal gauge indices that are contracted within each original vector bilinear. The two fermion bilinears of identical gauge species may be decomposed into symmetric and antisymmetric components in all indices. Fully antisymmetric combinations are $0$, because the fermion bilinear $(ff)$ is symmetric. For $f=\bar{e}$ (operators of type $Q_{ee}$), there are no gauge indices and so $\bar{e}_{[j}\bar{e}_{l]}=0$ by fermion statistics. This reduces the operators to $(\bar{e}^{\dagger (i}\bar{e}^{\dagger k)}(\bar{e}_{(j}\bar{e}_{l)})$, of which there are $\Big(\frac{1}{2}n_g(n_g+1)\Big)^2=\frac{1}{4}n_g^2(n_g+1)^2$.

For $f=L$ ($Q_{ll}$), the bilinear $L^I_jL^J_l$ is only non-zero if both flavor and isospin indices are symmetrized with the same parity. This gives an isospin triplet bilinear with symmetric flavor indices and an isospin singlet with antisymmetric flavor indices. These are then contracted in isospin indices with their conjugates. This requires that the isospin indices of each bilinear have the same parity in order to be non-zero, so results in $\Big(\frac{1}{2}n_g(n_g+1)\Big)^2$ terms $(\bar{L}^{\dagger (i}\bar{L}^{\dagger k)})(\bar{L}_{(j}\bar{L}_{l)})$ and $\Big(\frac{1}{2}n_g(n_g-1)\Big)^2$ terms $(\bar{L}^{\dagger [i}\bar{L}^{\dagger k]}(\bar{L}_{[j}\bar{L}_{l]})$, which gives a total of $\frac{1}{2}n_g^2(n_g^2+1)$.

For $f=\bar{u}$ ($Q_{uu}$), isospin indices in the above case are replaced with color indices, but the argument is otherwise identical. There are therefore $\frac{1}{2}n_g^2(n_g^2+1)$ of these terms as well. The same argument applies to $Q_{dd}$.

For $f=Q$ with the singlet color structure ($Q_{qq}^{(1)}$), the fields have both color and isospin. For overall symmetry, either all indices must be symmetrized over or exactly one set should be symmetrized and the other two antisymmetrized. This gives four possible terms: $Q_{j,I,\alpha}Q_{l,J,\beta}=(3,6,S)+(1,\bar{3},S)+(1,6,A)+(3,\bar{3},A)$, where, to control indices, the representation labels of the form $(SU(2),SU(3),SU(N_f))$ have been written instead of the tensors ($S$ and $A$ denote symmetric and anti-symmetric rank-$2$ $SU(N_f)$ tensors respectively). Exactly the same decomposition applies to the conjugate bilinear. Contracting the gauge indices of the decomposed bilinears, only the index contractions between pairs with the same symmetry parities are non-zero. This leaves four terms $(3\cdot 3,\bar{6}\cdot 6,S\otimes S)+(1\cdot 1,3\cdot\bar{3},S\otimes S)+(1\cdot 1,\bar{6}\cdot 6,A\otimes A)+(3\cdot 3,3\cdot\bar{3},A\otimes A)$, the first two and the last two add to give single tensors that are either symmetric in both $i\leftrightarrow k$ and $j\leftrightarrow l$ or antisymmetric in both. These therefore have $(\frac{1}{2}n_g(n_g+1))^2$ and $(\frac{1}{2}n_g(n_g-1))^2$ components respectively, giving a total of $\frac{1}{2}n_g^2(n_g^2+1)$.

This argument also gives the counting for the operator involving isospin triplet bilinears instead. This is because, by the isospin Fierz identity (a.k.a. Clifford algebra), this may be decomposed into a linear combination of the above operator and an identical version with the isospin pairings of the quarks switched. Subtracting the former component, which is accounted for above, an identical tensor decomposition may be performed on the remaining term, with isospin indices swapped $I\leftrightarrow J$ on the $Q$ pair. This simply introduces a relative negative sign in the two terms i.e. the decomposition is $(3\cdot 3,\bar{6}\cdot 6,S\otimes S)-(1\cdot 1,3\cdot\bar{3},S\otimes S)-(1\cdot 1,\bar{6}\cdot 6,A\otimes A)+(3\cdot 3,3\cdot\bar{3},A\otimes A)$. This operator has therefore been decomposed into two bisymmetric or biantisymmetric flavor tensors (independent to those from the isospin singlet operator above), so has the same number of independent components.

The remaining flavored operators (neglecting baryon and lepton number violation) may be enumerated similarly. Those of the form $(\bar{L}R)H^3$ clearly have $2n_g^2$ flavor components each, because the two fermions are distinct and the operator is not self-adjoint (hence the factor of $2$). An identical argument applies to the $(\bar{L}R)XH$ operators. Finally, in the $(\bar{L}R)(\bar{L}R)$ class, there are clearly $2n_g^4$ $Q_{lequ}$-type operators for each isospin configuration (by identical reasons as for the number of $Q_{ledq}$ operators), while for $Q_{quqd}$-type operators, the isospin contraction ensures that the $Q$ operators are never identical fermions, while the color contractions with the different right-handed quarks distinguishes them, so there are also $n_g^4$ of each of these.

In the operator basis chosen here, the addition of fermion flavor indices does not modify the dimension $4$ equations of motion beyond the inscription of flavor indices on the fields.

There are also no examples of operators that cannot exist for $n_g=1$ but can exist if the flavor provides an extra internal degree of freedom with which to distinguish what would otherwise be identical quanta (as would happen if e.g. right-handed neutrinos were included). This is simply because there are no such gauge-invariant combinations consistent with the Standard Model field content, which can be easily verified by counting operators with three fermions of the same type.

Having derived the number of independent flavored operators in the baryon and lepton conserving dimension $6$ Standard Model, these may be added together to give the total number. Following the enumeration tabulated, for $n_g=3$, there are $2499$.

\end{Lexercisenb}

\section{EFT for Nuclear and (some) Atomic Physics (van Kolck)}

These two lectures \cite{vanKolck:2019vge} introduced some
applications of effective field theory in the
context of nuclear and atomic physics.
Pionless EFT, a simple nuclear EFT
containing contact interactions, was presented,
and the prospects for including long-range forces were
discussed.

Equations from the lecture notes are referred to with a prefix L, e.g. eqn~(L1.1).

\begin{Lexercise}{5}{1}

For a spherical well potential,
\begin{equation*}
V (r) = - V_0 \Theta (R, r)
\end{equation*}
show that when the parameter $\alpha := \sqrt{m V_0} R$ is tuned close to the critical values $\alpha_c = (n + 1/2) \pi$ that the scattering length is given by,
$$ a_2 \sim \frac{R}{\alpha_c (\alpha - \alpha_c )} $$

\medskip

\noindent {\bf SOLUTION}

Eigenstates with energy $E = k^2/2m$ are described by,
\begin{equation}
\left[ -\frac{ \nabla^2}{2m} + V(r) \right] \psi = \frac{k^2}{2m} \psi
\end{equation}
and can be separated for central potentials into,
$$ \psi =\sum_{\ell = 0}^\infty R_{\ell} (r) P_{\ell} ( \cos \theta) $$
$$ \left[ \partial_r^2 + \frac{2}{r} \partial_r - \frac{\ell (\ell +1) }{ r^2 } - 2m V(r) + k^2 \right] R_{\ell} (r) = 0 $$
We define the $s$-wave scattering phase via the asymptotic behavior of the $\ell =0 $ mode,
\begin{equation}
R_0 ( r) \sim \frac{1}{kr} e^{i \delta_0 (k)} \sin \left( k r + \delta_0 (k) \right) \;\;\; \text{ as } \;\;\; r \to \infty
\label{eqn:swave}
\end{equation}
and so for a spherical wave potential we can solve,
\begin{equation}
\left[ \partial_r^2 + 2 m V_0 \Theta (R - r) + k^2 \right] ( r R_0 ) = 0
\end{equation}
using the boundary conditions $ r R_0 \to 0$ as $r \to 0$ and \eqref{eqn:swave},
\begin{equation}
r R_0 (r) = \begin{cases}
c \sin \left[ r \sqrt{ k^2 + 2 m V_0 } \right] \;\; & \;\; r < R , \\
\sin \left[ kr + \delta_0 \right] \;\; & \;\; r > R
\end{cases}
\end{equation}
where $c$ is a constant of integration. Demanding that the wave function and its first derivative are continuous across the $r=R$ boundary, we find,
\begin{align}
c &= \frac{ \sin \left[ k R + \delta_0 \right] }{\sin \left[ R \sqrt{ k^2 + 2 m V_0 } \right]} , \\
\sqrt{ k^2 + 2 m V_0 } \text{cot} \left[ R \sqrt{ k^2 + 2 mV_0 } \right] &= k \text{cot} \left[ k R + \delta_0 \right]
\end{align}
Taking the limit $k\to 0$, this gives,
\begin{equation}
k \text{cot} ( \delta_0 ) = \sqrt{2 m V_0} \text{cot} \left( R \sqrt{2 m V_0 } \right) + \mathcal{O} ( k^2 )
\end{equation}
Now using, $\text{cot} ( \alpha ) = - ( \alpha - (n+1/2) \pi ) + ...$ when $\alpha \approx (n+1/2) \pi $, we find,
\begin{equation}
- \frac{1}{a_2} = \lim_{k \to 0} k \text{cot} ( \delta_0 ) = - \frac{\alpha_c}{R} \left( \alpha - (n+1/2) \pi \right) + ...
\end{equation}
where $...$ are regular in the limit $\alpha \to \alpha_c$.

The divergence in the scattering length corresponds to a zero energy $s$-wave bound state accommodated by the potential when $\alpha \to \pi/2$.
At $\alpha = n \pi$, the scattering cross section vanishes identically (the Ramsauer-Townsend effect).

\end{Lexercise}

\begin{Lexercisenb}{5}{2}

Solve the three-dimensional Schrodinger equation with,
$$ V = \frac{4 \pi c_0}{2 m} \delta^3 (r) $$

\medskip

\noindent {\bf SOLUTION}

In momentum space, energy eigenstates obey,
\begin{align}
p^2 \psi (p) - 4 \pi c_0 \int \frac{d^3 k}{(2\pi)^3} \psi ( k ) = 2 m E \psi ( p )
\end{align}
Note that the integral interaction is divergent. If we rewrite,
\begin{equation}
\int d^3 k \left[ (k^2 - 2 m E) \delta^3 ( k - p ) - \frac{c_0}{2 \pi^2} \right] \psi ( k ) = 0
\end{equation}
then we see that the energy, $E$, is related to the momentum and the coupling $c_0$ by the condition,
\begin{equation}
1 = \frac{ c_0}{ 2 \pi^2} \int d^3 k \frac{1}{k^2 - 2 m E}
\end{equation}
Introducing a UV cutoff, $\Lambda$, this can be written as,
\begin{equation}
1 = - \frac{ c_0}{ 2 \pi^2} \; 4 \pi \left[ \Lambda - i \frac{\pi}{2} \sqrt{ 2 m E } + \mathcal{O} \left( \frac{\sqrt{ m E} }{ \Lambda } \right) \right]
\end{equation}
Suppose we measure a bound state at energy $E < 0$. Then, we renormalization the coupling $c_0$ so that,
\begin{equation}
\sqrt{- 2 m E } := \frac{1}{c_0^R} = \frac{1}{c_0 (\Lambda) } + \frac{2}{\pi} \Lambda
\end{equation}
That is, as the cutoff $\Lambda$ is taken to infinity, the coupling $c_0$ must run as above in order to maintain a bound state with energy $E$ in the spectrum of the theory. The theory is only predictive once this renormalization has been performed. To analyze other bound states, we must look for other solutions to,
\begin{equation}
1 = \frac{ c_0 (\Lambda ) }{ 2 \pi^2} \int d^3 k \frac{1}{k^2 - 2 m E' }
\end{equation}
\begin{align}
1 = c_0 ( \Lambda) \left[ \frac{2}{\pi} \Lambda + \sqrt{ -2 m E'} + ... \right]
\end{align}
which is satisfied iff $E' = E$, and so we conclude that there is only one bound state for the (renormalized) three-dimensional delta function.

The theory then predicts a wave function for this single bound state comprised of the spherical waves,
\begin{equation}
\left[ \partial_r^2 - ( c_0^R )^{-2} - \frac{\ell (\ell + 1)}{r} \right] ( r R_{\ell} ) = 0 \;\;\;\; r \neq 0
\end{equation}
and so we have an asymptotic wave function dominated by the $s$-wave,
$$ \psi \propto \frac{e ^{-r / c_0^R} }{r} \;\; \text{ for } \;\; r \to \infty $$

\end{Lexercisenb}

\section{EFT with Nambu-Goldstone Modes (Pich)}

The lectures \cite{Pich:2018ltt} discussed EFTs that are useful for describing the dynamics of massless modes that emerge after spontaneous symmetry breaking.
Chiral perturbation theory ($\chi$PT) and the electroweak sector of the Standard Model were emphasised.
For $\chi$PT, which is the focus of these exercises,
Ref.~\cite{Gasser:1983yg} paved the way forward to carry out an effective low-energy expansion.

\setcounter{exercise}{1}

\begin{exercise}
\item The quadratic mass term of the $\mathcal{O} (p^2)$ $\chi$PT Lagrangian generates a small mixing between the $\pi_3$ and $\eta_*$ fields, proportional to the quark mass difference $\Delta m = m_d - m_u$.

\begin{itemize}

\item[a)] Diagonalize the neutral meson mass matrix and find out the correct mass eigenstates and their masses.

\item[b)] When isospin is conserved, Bose symmetry forbids the decay $\eta \to \pi^0 \pi^+ \pi^-$ (why?).
Compute the decay amplitude to first-order in $\Delta m$.

\end{itemize}

\medskip

\noindent {\bf SOLUTION:}

\begin{itemize}

\item[a)]
Promoting the quark mass operator $ \bar q \mathcal{M} q / 2$, where $\mathcal{M} = \text{diag} (m_u, m_d, m_s)$, to a spurion field, the low energy chiral Lagrangian at order $p^2$ is,
\begin{equation}
\mathcal{L}_2 = \frac{f^2}{4} \langle \partial_\mu U \partial^\mu U^\dagger + B_0 \mathcal{M} U^\dagger + B_0 U \mathcal{M}^\dagger \rangle
\end{equation}
where $f$ and $B_0$ are EFT parameters, and the angled brackets denote a trace over the $SU(3)$ valued matrix,
\begin{equation}
U = e^{i \sqrt{2} \Phi / f} , \;\;\;\; \Phi = \left( \begin{array}{c c c}
\tfrac{1}{\sqrt{2}} \pi_3 + \frac{1}{\sqrt{6}} \eta_8 & \pi^+ & K^+ \\
\pi^- & - \tfrac{1}{\sqrt{2}} \pi_3 + \frac{1}{\sqrt{6}} \eta_8 & K^0 \\
K^- & \bar K^0 & - \tfrac{2}{\sqrt{6}} \eta_8
\end{array} \right) .
\label{eqn:Phicharge}
\end{equation}
The chiral Lagrangian contains the mass terms,
\begin{equation}
\mathcal{L}_2 \supset - \frac{1}{2} ( \pi_3 \;\; \eta_8 ) \left( \begin{array}{c c}
M_{\pi_3}^2 & - \Delta/2 \\
- \Delta/2 & M_{\eta_8}^2
\end{array} \right) \left( \begin{array}{c}
\pi_3 \\
\eta_8 \end{array} \right)
\end{equation}
$$ \Delta = \frac{2 B_0}{\sqrt{3}} ( m_d - m_u ) , \;\;\;\; M_{\pi_3}^2 = B_0 ( m_u + m_d ) , \;\;\;\; M_{\eta_8}^2 = \frac{B_0}{3} ( m_u + m_d + 4m_s ) $$
For convenience, we also define,
$$ M = M_{\eta_8}^2 - M_{\pi_3}^2 , \;\;\;\; a_{\pm} = M \pm \sqrt{ M^2 + \Delta^2} $$
The mass matrix is then diagonalized as,
\begin{equation}
\mathcal{L}_2 \supset - \frac{1}{2} ( \pi^0 \;\; \eta ) \left( \begin{array}{c c}
M_{\pi^0}^2 & 0 \\
0 & M_{\eta}^2
\end{array} \right) \left( \begin{array}{c}
\pi^0 \\
\eta \end{array} \right)
\end{equation}
$$ M_{\pi^0}^2 = M_{\pi_3}^2 + \tfrac{1}{2} a_- \approx M_{\pi_3}^2 - \frac{\Delta^2}{4 M} , \;\;\;\;
M_{\eta}^2 = M_{\pi_3}^2 + \tfrac{1}{2} a_+ \approx M_{\eta_8}^2 + \frac{\Delta^2}{4 M} $$
$$ \pi^0 = \frac{ a_+ \pi_3 + \Delta \eta_8 }{\sqrt{ a_+^2 + \Delta^2}} \approx \pi_3 + \frac{\Delta}{2 M} \eta_8 , \;\;\;\; \eta = \frac{ a_- \pi_3 + \Delta \eta_8 }{\sqrt{ a_-^2 + \Delta^2}} \approx \eta_8 - \frac{\Delta }{2 M } \pi_3 $$

\item[b)]
The final state quark content,
$$ \frac{1}{\sqrt{2}} ( u \bar u - d \bar d ) \;\; ( u \bar d ) \;\; ( \bar u d ) $$
is antisymmetric under exchanging $u \leftrightarrow d$, i.e. under isospin symmetry. This is because $\pi^0$ is an antisymmetric combination, and the $\pi^+ \pi^-$ pair must be symmetric because Bose symmetry mandates a symmetric wave function.
However, the original state $\eta = \tfrac{1}{\sqrt{6}} ( u\bar u + d \bar d - 2 s \bar s)$ is symmetric under $u,d$ isospin symmetry---and therefore the process $\eta \to \pi^0 \pi^+ \pi^-$ would violate isospin.

Explicitly, expanding $\mathcal{L}_2$ one finds the operators,
\begin{align}
\mathcal{L}_2 \supset& \frac{B_0}{\sqrt{3}} ( m_d - m_u ) \pi^0 \eta \left[ 1 - \frac{1}{3 f^2} \pi^+ \pi^- \right] \\
&+ \frac{1}{3 f^2} \left\{ ( \pi^0 \overset{\leftrightarrow}{\partial}_\mu \pi^+ ) ( \pi^- \overset{\leftrightarrow}{\partial}^\mu \pi^0 ) + \frac{1}{2} m_{\pi}^2 \pi^+ \pi^- ( \pi^{0 \, 2} + \eta^2 ) \right\} \nn
\end{align}
from which one can construct three independent Feynman diagrams,
\FloatBarrier
\begin{figure}[htbp!]
\qquad
\includegraphics[width=0.3 \textwidth]{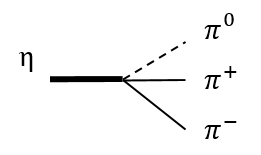}
\includegraphics[width=0.3 \textwidth]{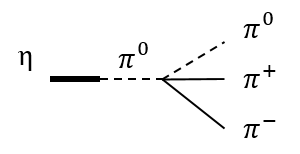}
\includegraphics[width=0.3 \textwidth]{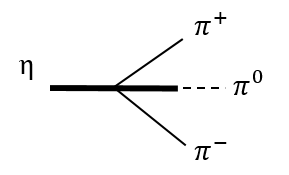}
\end{figure}
\FloatBarrier
which give a total decay amplitude,
\begin{equation}
\mathcal{M} ( \eta \to \pi^+ \pi^- \pi^0 ) = \frac{B_0}{3 \sqrt{3} f^2} (m_d - m_u ) \left\{ 1 + \frac{3 (p_2 + p_3 )^2 - 3 m_{\pi}^2 - m_{\eta}^2 }{ m_{\eta}^2 - m_{\pi}^2 } \right\} .
\end{equation}
which indeed vanishes in the isospin limit ($m_d = m_u$).

\end{itemize}

\end{exercise}
\begin{exercisenb}
~
\begin{itemize}

\item[a)] Compute the axial current at $\mathcal{O} (p^2)$ in $\chi$PT and check that $f_\pi = f$ at this order.

\item[b)] Expand the $\mathcal{O} (p^2)$ axial current to $\mathcal{O} (\Phi^3)$ and compute the 1-loop corrections to $f_\pi$.
Remember to include the pion wave-function renormalization.

\item[c)] Find the tree-level contribution of the $\mathcal{O} ( p^4 )$ $\chi$PT Lagrangian to the axial current.
Renormalize the UV loop divergences with the $\mathcal{O} ( p^4 )$ LECs.

\end{itemize}

\medskip

\noindent {\bf SOLUTION:}

Promoting an axial current $\bar q_L \slashed{\ell} q_L + \bar q_R \slashed{r} q_R$, where $\ell_\mu + r_\mu = 2 a_\mu$, to a spurion field,
\begin{align}
\ell_\mu &\to g_L \ell_\mu g_L^\dagger + i g_L \partial_\mu g_L^\dagger \\
r_\mu &\to g_R r_\mu g_R^\dagger + i g_R \partial_\mu g_R^\dagger
\end{align}
the low energy chiral Lagrangian at order $p^2$ is,
\begin{equation}
\mathcal{L}_2 = \frac{f^2}{4} \langle D_\mu U D^\mu U^\dagger \rangle , \;\;\;\; D_\mu U = \partial_\mu U - i r_\mu U + i U \ell_\mu
\end{equation}
The currents are then,
\begin{align}
J_L^\mu &= \frac{\partial}{\partial \ell_\mu } \mathcal{L}_2 = \tfrac{i}{2} f^2 D^\mu U^\dagger U = \frac{f}{\sqrt{2}} D^\mu \Phi + \frac{i}{2} [ D^\mu \Phi, \Phi ] - \frac{1}{3 \sqrt{2} f} [ [ D^\mu \Phi, \Phi ], \Phi ] + ... \\
J_R^\mu &= \frac{\partial}{\partial r_\mu } \mathcal{L}_2 = \tfrac{i}{2} f^2 D^\mu U U^\dagger = - \tfrac{f}{\sqrt{2}} D^\mu \Phi + \frac{i}{2} [ D^\mu \Phi, \Phi ] + \frac{1}{3 \sqrt{2} f} [ [ D^\mu \Phi, \Phi ], \Phi ] + ...
\end{align}
$$ \implies \;\; J_A^{\mu} = J_L^\mu + J_R^\mu = \sqrt{2} f \left( D^\mu \Phi - \frac{1}{3 f^2} [ [ D^\mu \Phi, \Phi ], \Phi ] + ... \right) $$

To leading order, the pion decay constant is then,
\begin{equation}
i \sqrt{2} f_\pi p^\mu = \langle 0 | u \gamma^\mu \gamma_5 \bar d | \pi^+ (p ) \rangle = \langle 0 | (J_A^\mu )_{12} | \pi^- (p) \rangle = i \sqrt{2} f p^\mu + ...
\end{equation}
where we have used the charge basis \eqref{eqn:Phicharge} to write the order $D^\mu \pi^+$ part\footnote{
Note that the terms in $\pi^+$ do not contribute to the one-loop amplitude, because terms like $\pi^0 D^\mu \pi^0 \pi^+$ contain loop integrands $k^\mu/(k^2-M_\pi^2)$ which vanish, and so we have omitted them from \eqref{eqn:JL}.
} of the current as,
\begin{equation}
(J_A^\mu )_{12} \supset - \sqrt{2} f \left[ 1 - \frac{1}{3 f^2} \left( 2 \pi^0 \pi^0 + 2 \pi^- \pi^+ + \bar K^0 K^0 + K^+ K^- \right) + ... \right] D^\mu \pi^+
\label{eqn:JL}
\end{equation}
(we're working in the isospin limit where $\pi_3 \approx \pi^0$).
The cubic vertices gives rise to the following one-loop contributions to the matrix element,
\FloatBarrier
\begin{figure}[htpb!]
\centering
\includegraphics[width=0.5\textwidth]{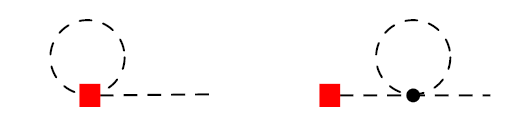}
\end{figure}
\FloatBarrier
\begin{align}
\frac{1}{f^2} \langle 0 | \pi^0 \pi^0 D^\mu \pi^+ | \pi^+ (p) \rangle &= i \frac{ \mu^{2\epsilon} }{f^2} \int \frac{d^d k}{(2\pi)^d} \frac{-i p^\mu}{k^2 - M_{\pi}^2} = \frac{2}{f^2} \langle 0 | \pi^+ \pi^- D^\mu \pi^+ | \pi^+ (p) \rangle \\
&= - i p^\mu \; \frac{M_{\pi}^2}{ (4 \pi f )^2} \left( \frac{4 \pi \mu^2}{M_\pi^2} \right)^{\epsilon} \Gamma ( -1 + \epsilon ) \\
&= - i p^\mu \frac{M_{\pi}^2}{ (4 \pi f )^2} \left[ \frac{1}{\epsilon} - \gamma_E + 1 + \text{log} \left( \frac{4 \pi \mu^2}{M_{\pi}^2} \right) + \mathcal{O} ( \epsilon) \right] \\
&= i p^\mu \frac{M_{\pi}^2}{ (4\pi f)^2} \text{log} \frac{M_{\pi}^2}{\mu^2} \;\;\;\; \text{ after subtraction},
\end{align}
where we've used (higher order) counterterms to subtract the $1/\epsilon - \gamma_E + 1 + \text{log} 4\pi $. Note that an overall factor of $1/2$ comes from the pion wave function normalization.
The kaon elements are identical, with $M_{\pi}$ replaced by $M_{K}$.
This gives a pion decay constant,
\begin{equation}
f_{\pi} = f \left( 1 - \frac{ M_{\pi}^2}{ (4\pi f)^2} \text{log} \frac{M_{\pi}^2}{\mu^2} - \frac{M_{K}^2}{ 2 (4\pi f)^2} \text{log} \frac{M_{K}^2}{\mu^2} \right) + \text{Tree level } \mathcal{L}_4
\end{equation}
where to reliably include the loop corrections (which are order $1/(4\pi f)^2$), we must also include the tree level corrections from the $\mathcal{O} ( p^4 )$ part of the Lagrangian.

The terms in $\mathcal{L}_4$ which contribute at tree level are,
\begin{align}
\mathcal{L}_4 \supset L_4 \langle D_\mu U^\dagger D^\mu U \rangle \langle U^\dagger \mathcal{M} + \mathcal{M}^\dagger U \rangle + L_5 \langle D_\mu U^\dagger D^\mu U \left( U^\dagger \mathcal{M} + \mathcal{M}^\dagger U \right) \rangle
\end{align}
where we've absorbed a factor of $B_0$ into the conventional definitions of $L_4$ and $L_5$. The contribution to the current is,
\begin{align}
J_L^\mu = \frac{\partial}{\partial \ell_\mu } \mathcal{L}_2 &= i 2 L_4 D^\mu U^\dagger U \langle U^\dagger \mathcal{M} + \mathcal{M}^\dagger U \rangle + i L_5 \{ D^\mu U^\dagger , U^\dagger \mathcal{M} + \mathcal{M}^\dagger U \} U \\
&= \frac{\sqrt{2}}{f} \left[ 4 L_4 \langle \mathcal{M} \rangle D^\mu \Phi + 2 L_5 \mathcal{M} D^\mu \Phi + 2 L_5 D^\mu \Phi \mathcal{M} \right] + ... \\
J_R^\mu = \frac{\partial}{\partial r_\mu } \mathcal{L}_2 &= - \frac{\sqrt{2}}{f} \left[ 4 L_4 \langle \mathcal{M} \rangle D^\mu \Phi + 2 L_5 \mathcal{M} D^\mu \Phi + 2 L_5 D^\mu \Phi \mathcal{M} \right] + ...
\end{align}
and so restoring the overall factor of $B_0$,
$$ ( J_A^{\mu} )_{12} = - \frac{8 \sqrt{2} B_0 }{f} \left[ 2 L_4 (m_u + m_d + m_s) + L_5 (m_u + m_d ) \right] D^\mu \pi^+ $$
and so by using the $L_4$ and $L_5$ coefficients to absorb the one-loop UV divergence and running with $\mu$,
we find a pion decay constant,
\begin{align}
f_{\pi} &= f \Big( 1 - \frac{M_{\pi}^2}{ (4\pi f)^2} \text{log} \frac{M_{\pi}^2}{\mu^2} - \frac{M_{K}^2}{ 2 (4\pi f)^2} \text{log} \frac{M_{K}^2}{\mu^2} \nonumber \\
&\qquad+ \frac{8 M_K^2 + 4 M_{\pi}^2 }{f^2} \bar L_4 (\mu ) + \frac{4 M_{\pi}^2 }{f^2} \bar L_5 ( \mu ) \Big)
\end{align}

\end{exercisenb}
\setcounter{exercise}{5}
\begin{exercisenb}

Assume the existence of a hypothetical light Higgs which couples to quarks with the Yukawa interaction
$$ \mathcal{L}_{h^0 \bar{q} q} = - \frac{h^0}{v} \sum_q k_q m_q \bar{q} q $$

\begin{itemize}

\item[a)] Determine at lowest-order in the $\chi$PT expansion the effective Lagrangian describing the
Higgs coupling to pseudoscalar mesons induced by the light-quark Yukawas.

\item[b)] Determine the effective $h^0 G_a^{\mu \nu} G_{\mu \nu}^a$ coupling induced by heavy quark loops.

\item[c)] The $G_a^{\mu \nu} G^a_{\mu \nu}$ operator can be related to the trace of the energy-momentum tensor, in the
3-flavor QCD theory:
\begin{equation*}
\Theta_\mu^\mu = \frac{ \beta_1 \alpha_s }{4 \pi} G_a^{\mu \nu} G^a_{\mu \nu} + \bar{q} \mathcal{M} q ,
\end{equation*}
where $\beta_1 = - \tfrac{9}{2}$ is the first coefficient of the $\beta$ function. Using this relation, determine the lowest-order $\chi$PT Lagrangian incorporating the Higgs coupling to pseudoscalar mesons induced by the heavy-quark Yukawas.

\item[d)] Compute the decay amplitudes $h^0 \to 2 \pi$ and $\eta \to h^0 \pi^0$.

\medskip

\noindent {\bf SOLUTION:}

\item[a)]
The light Higgs interaction term in the Lagrangian is, from the point of view of the
quark fields, the same as the mass term.
It is therefore convenient to (superficially) combine these terms,
\bea
{\cal L} &\supset&
- \sum_q m_q \, \bar q q
+{\cal L}_{h^0\bar q q}
\ = \
- \sum_q \wv m_q \, \bar q q \, , \nonumber \\
\wv m_q \ &\equiv&\ m_q \Big(\, 1 + \frac{h^0}v k_q \, \Big) \, .
\label{Pich 3 a - 1}
\eea
One can thus use the ordinary $\chi$PT expansion,
with minor modifications to the light quark mass matrix.
We shall consider $q=\{ u,d,s \}$, which was discussed thoroughly
in the lectures.
To be explicit, (angle brackets denote the trace)
\bea
{\cal L}^{\rm eff}_{h^0 \bar q q} &=&
\frac{f^2}4 \l<\,
D_\mu \bm U D^\mu \bm U^\dagger + \chi \bm U^\dagger + \bm U \chi^\dagger
\,\r> \, ,
\label{Pich 3 a - 2}
\eea
where $\bm U$ is a unitary ${\rm SU}(3)$ matrix, parametrized by the pseudoscalar
octet $\Phi\,$, and $f=f_\pi$ to this order (see problem 2).
The (light) quark mass-matrix (which enters in $\chi = 2 B_0 \wv {\cal M}$)
takes the form
$$
\wv {\cal M} = {\rm diag} ( \wv m_u , \wv m_d, \wv, m_s ) \, .
$$

\item[b)]
Coupling $h^0$ to the gluon-sector is provided via
vertex corrections (in full QCD) arising
from the heavy flavors.
The lowest order contribution is the triangle diagram shown in
Fig.~\ref{fig: h0gg}; the heavy quark of mass $m_Q$, couples to the gluons.
We consider here the frame in which the Higgs
has zero four-momentum (and are hence neglecting the Higgs mass).

\begin{figure}[!hbt]
\includegraphics[scale=1.3]{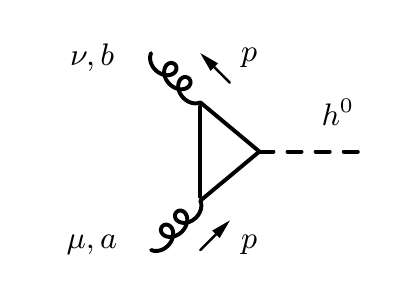}
\caption{\label{fig: h0gg}
Triangle diagram that couples the light Higgs to gluons.
Here we consider the special case where the Higgs field
has zero four-momentum.
There is also a diagram with the gluons crossed (not shown).
}
\end{figure}

Evaluating the diagram with standard techniques, we find
it to be
$$
- \frac{2\alpha_s}{3\pi} \frac{k_Q}{v} m_Q^2 \, \delta^{ab}
\Big(\, g_{\mu\nu} - \frac{p_\mu p_\nu}{p^2} \, \Big)
F \bigg( \frac{m_Q^2}{p^2} \bigg)\, .
$$
Here the function $F$ is what remains of the usual Feynman parametrization:
\bea
F(x) &\equiv&
\int_0^1 dy \frac{y}{y(1-y) - x}
\ \simeq \ - \, \frac{1}{2x} + \ldots \, ; \quad x\to \infty \, .
\eea
Taking the limit $m_Q^2 \gg p^2$ (i.e. $x\to \infty$)
is needed for the low-energy theory.
These heavy-quark loops are induced by the assumed Yukawa interaction
with $h^0$.
The Higgs-gluon interaction therefore follows from a term in the chiral Lagrangian,
\bea
{\cal L}_{h^0gg} &=&
\frac{\alpha_s}{4\pi} \frac{h^0}{v} \bar k \, G_a^{\mu\nu} G^a_{\mu\nu} \, ,
\label{Pich 3 a - 3}
\eea
where we have summed over the relevant heavy degrees of freedom and
defined the `average' Yukawa coupling $\bar k \equiv (k_c+k_b+k_t)/3$.
\item[c)]
The trace of the energy-momentum tensor that follows from the effective
chiral Lagrangian \eq{Pich 3 a - 2}, but now omitting the Higgs-quark coupling
($\wv {\cal M} \to {\cal M}$),
\bea
\Theta_\mu^{\ \mu} &=&
- \frac{f^2}2 \l< \,
D_\mu \bm U D^\mu \bm U^\dagger + B_0 ({\cal M} \bm U^\dagger + \bm U {\cal M})
\, \r> \, .
\label{Pich 3 a - 4}
\eea
By identifying this expression with the corresponding $\Theta_\mu^{\ \mu}$ from
full QCD, we may rewrite the $G_a^{\mu\nu}G^a_{\mu\nu}$ operator
in terms of light quark and pseudoscalar fields.
We also recall that the quark mass term $\bar q {\cal M} q$ is
associated with factors proportional to $B_0$ in
in the chiral Lagrangian.
This is the matching procedure that
reveals the low-energy representation of eqn~\eq{Pich 3 a - 3},
namely
\bea
{\cal L}^{\rm eff}_{h^0gg} &=&
-\, \frac{f^2}{2\beta_1} \frac{h^0}{v} \bar k
\l< \,
D_\mu \bm U D^\mu \bm U^\dagger + 3B_0 ({\cal M} \bm U^\dagger + \bm U {\cal M})
\, \r> \, ,
\label{Pich 3 a - 5}
\eea
where $\beta = -\tfrac92$ is the first coefficient of the $\beta$-function.
\item[d)]
Equations \eq{Pich 3 a - 2} and \eq{Pich 3 a - 4} together give the
interaction of the light Higgs with the Goldstone bosons.
Keeping then the terms proportional to $h^0$, we find that
\bea
{\cal L}^{\rm eff}_{h^0} &=&
-2 \bar k\frac{h^0}{\beta_1 v} \big[\
\partial_\mu \pi^+ \partial^\mu \pi^- +
\tfrac12 \partial_\mu \pi^0 \partial^\mu \pi^0 + \ldots\ \big]
\label{Pich 3 a - 6} \\
& & - B_0 \frac{h^0}{v} \Big\{\, (c_u m_u + c_d m_d)
\big[ \, \pi^+\pi^- + \tfrac12 \pi^0 \pi^0 \, \big] \nonu \\
& & \qquad \qquad \quad + \
\frac1{\sqrt{3}} (c_u m_u - c_d m_d) \pi^0 \eta \ +\ \ldots
\, \Big\} \, , \nonu
\eea
after expanding the matrix $\bm U = \exp (i\sqrt{2}\Phi/2)$
and defining $c_q \equiv k_q - 3\bar k /\beta_1\,$.
In \eq{Pich 3 a - 6} we have only kept the terms that are needed for
decay amplitudes $h^0 \to 2\pi$ and $\eta \to h^0 \pi^0$; the
Higgs also couples to strange mesons from the off-diagonal pieces in $\Phi$.

In the isospin limit, $m_u=m_d\equiv \hat m$ and thus
$$
B_0 = \frac{M_\pi^2}{m_u+m_d}
\ \to \ \frac{M_\pi^2}{2\hat m}\, ,
$$
where $M_\pi$ is the pion mass.
Then it is easy to read the decay amplitudes directly from the
appropriate terms in \eq{Pich 3 a - 6},
\bean
{\cal T}(h^0 \to 2 \pi ) &=&
\frac3{\beta_1v} \bar k \, (p_1\cdot p_2)
- B_0 \frac{3\hat m}{2v} (c_u + c_d) \\
&=&
-
\frac{M_\pi^2}{12v}
\big( \, 9(k_u+k_d) + 4 \bar k \, \big) \, ,
\eean
where $p_1$ and $p_2$ are the outgoing four-momenta of the pions.
We used the fact that $ p_1 \cdot p_2 \approx -M_\pi^2\,$, neglecting
the Higgs mass.
The second decay amplitude is
\bean
{\cal T}(\eta \to h^0 \pi^0) &=&
- B_0 \frac{\hat m}{\sqrt{3}\, v} (c_u - c_d)
\ = \
\frac{M_\pi^2}{2\sqrt{3} \, v} (k_u-k_d) \, .
\eean
This channel is particularly interesting, since it is only
possible if the up and down quark couple differently to $h^0$.
In the $O(p^2)$ chiral Lagrangian,
without the Yukawa coupling in \eq{Pich 3 a - 1}, the $\eta$-$\pi^0$ term
vanishes.

\end{itemize}

\end{exercisenb}

\setcounter{section}{4}

\section{Effective Field Theories and Inflation (Burgess)}

These three lectures \cite{Burgess:2017ytm}
introduced inflationary cosmology, focusing
on some uses of effective field theories
in its analysis.

Equations from the lecture notes are referred to with a prefix L, e.g. eqn~(L1.1).

\setcounter{exercise}{0}
\begin{exercise}

{\bf Slow growth of fluctuations during radiation domination}

The equation governing the growth of density fluctuations for non-relativistic matter in a spatially flat FRW geometry is
\begin{equation} \label{deltaeq}
\ddot \delta_{\bf k} + 2 H \, \dot \delta_{\bf k} + \left( \frac{c_s^2 {\bf k}^2}{a^2} - 4\pi G \rho_{m0} \right) \delta_{\bf k}= 0 \,,
\end{equation}
where $\delta = \delta \rho_m/\rho_{m0}$ is the fractional fluctuation in the matter density, ${\bf k}$ is its Fourier label while $a$ is the scale factor and $H = \dot a/a$ and so $H^2 = 8 \pi G \rho_0/3$.

For a matter-dominated universe, for which $\rho_0 \simeq \rho_{m0} \propto 1/a^3$ and $a \propto t^{2/3}$ show that as $c_s {\bf k} \to 0$ eq (\ref{deltaeq}) gives power-law solutions of the form $\delta_0 \propto t^n$ with $n = \frac23$ or $n = -1$. (The growing mode verifies the claim in class that $\delta_0 \propto a$ during matter domination.)

Consider now the transition between radiation and matter domination, for which $\rho_0 = \rho_{m0} + \rho_{r0}$ and so
\begin{equation}
H^2(a) = \frac{8 \pi G \rho_{0}}{3} = \frac{H_{\rm eq}^2}{2} \left[ \left( \frac{a_{\rm eq}}{a} \right)^3 + \left( \frac{a_{\rm eq}}{a} \right)^4 \right] \,,
\end{equation}
where radiation-matter equality occurs when $a = a_{\rm eq}$, at which point $H(a=a_{\rm eq}) = H_{\rm eq}$. The matter part of this expansion comes from
\begin{equation}
H_m^2 := \frac{8 \pi G \rho_{m0}}{3} = \frac{H_{\rm eq}^2}{2} \left( \frac{a_{\rm eq}}{a} \right)^3 \,.
\end{equation}
Verify that $\delta_0(x)$ satisfies
\begin{equation}
2x(1+x) \, \delta_0'' + (3x+2) \, \delta_0' - 3 \, \delta_0 = 0 \,,
\end{equation}
where the scale factor, $x = a/a_{\rm eq}$, is used as a proxy for time and primes denote differentiation with respect to $x$. Show that this is solved by $\delta_0 \propto \left( x + \frac23 \right)$, and thereby show how the growing mode during matter domination does not grow during radiation domination. ({\em Bonus:} show that the linearly independent solutions to this one only grow logarithmically with $x$ deep in the radiation-dominated era, for which $x \ll 1$.)

\medskip

\noindent {\bf SOLUTION}: Slow growth of fluctuations during radiation domination

\begin{equation}
\ddot \delta_{\k} + 2 H \dot \delta_{\k} + \left( \frac{c_s^2 \k^2}{a^2} - 4 \pi G \rho_{m0} \right) \delta_{\k} = 0
\end{equation}
In a matter dominated Universe, we can use the Friedmann equation, $3H^2 = 8 \pi G \rho_0$, to write the evolution for long wavelength modes as,
\begin{equation}
\ddot \delta_{0} + 2 H \dot \delta_{0} - \frac{3}{2} H^2 \delta_{0} = 0
\end{equation}
Changing the dependent variable to $a(t)$, we can use, $d a = a H dt$ and $H (t) = H_0 a^{-3/2}$ to write,
\begin{align}
0 &= H_0 a^{-1/2} ( H_0 a^{-1/2} \delta_{0}' )' + 2 H_0^2 a^{-2} \delta_{0}' - \frac{3}{2} H_0^2 a^{-3} \delta_{0} \\
&= H_0^2 a^{-3} \left( a^2 \partial_a^2 + \frac{3}{2} a \partial_a - \frac{3}{2} \right) \delta_0 \\
&= H^2 \sum_n \left( n + \frac{3}{2} \right) \left( n - 1 \right) \delta_0^{(n)} a^n
\end{align}
where we've written $\delta_0 (a)$ as $\sum_n \delta_0^{(n)} a^n$, and can conclude that the general solution is,
\begin{equation}
\delta_0 (t) = c_1 a + c_2 a^{-3/2} \propto c_1 t^{2/3} + c_2 t^{-1}
\end{equation}
where $c_1$, $c_2$ are constants of integration (c.f. method of Frobenius).

Now consider,
\begin{equation}
H^2 (a) = \frac{H_{\rm eq}^2}{2} \left[ x^{-3} + x^{-4} \right] , \;\;\;\; 4 \pi G \rho_{m0} = \frac{3 H_{\rm eq}^2 }{4} x^{-3}
\end{equation}
where $x = a/a_{\rm eq}$ is the dependent variable, and so we have,
\begin{align}
0 &= \frac{H_{\rm eq}^2}{2} \left[ \frac{ \sqrt{1 + x} }{x} \left( \frac{ \sqrt{1 + x} }{x} \delta_{0}' \right)' + 2 \frac{ 1 + x }{x^2} \delta_{0}' - \frac{3}{2 x^3} \delta_{0} \right] \\
&= \frac{H_{\rm eq}^2}{4 x^3} \left[ 2x (1 + x) \partial_x^2 + (3 x + 2) \partial_x - 3 \right] \delta_0 \\
&= 2 H_m^2 \sum_n \left( ( 2 n+ 3) (n-1) \delta_0^{(n)} + 2 n^2 \delta_0^{(n-1)} \right)
\end{align}
One solution to this is, $\delta \propto ( x + 2/3 )$. During radiation domination, we have $x < 1$, and so the constant $2/3$ piece of the $\delta_0$ fluctuations is important.

The other linearly independent solution is,
\begin{equation}
\delta_0 = 3 y - \left( 1 + \frac{3x}{2} \right) \text{log} \frac{1+y}{1-y} , \;\;\;\; y = \sqrt{1+x}
\end{equation}
which grows like $\text{log} ( x)$ for $x \ll 1 $.

\begin{figure}[b!]
\centering
\includegraphics[width=0.65\textwidth]{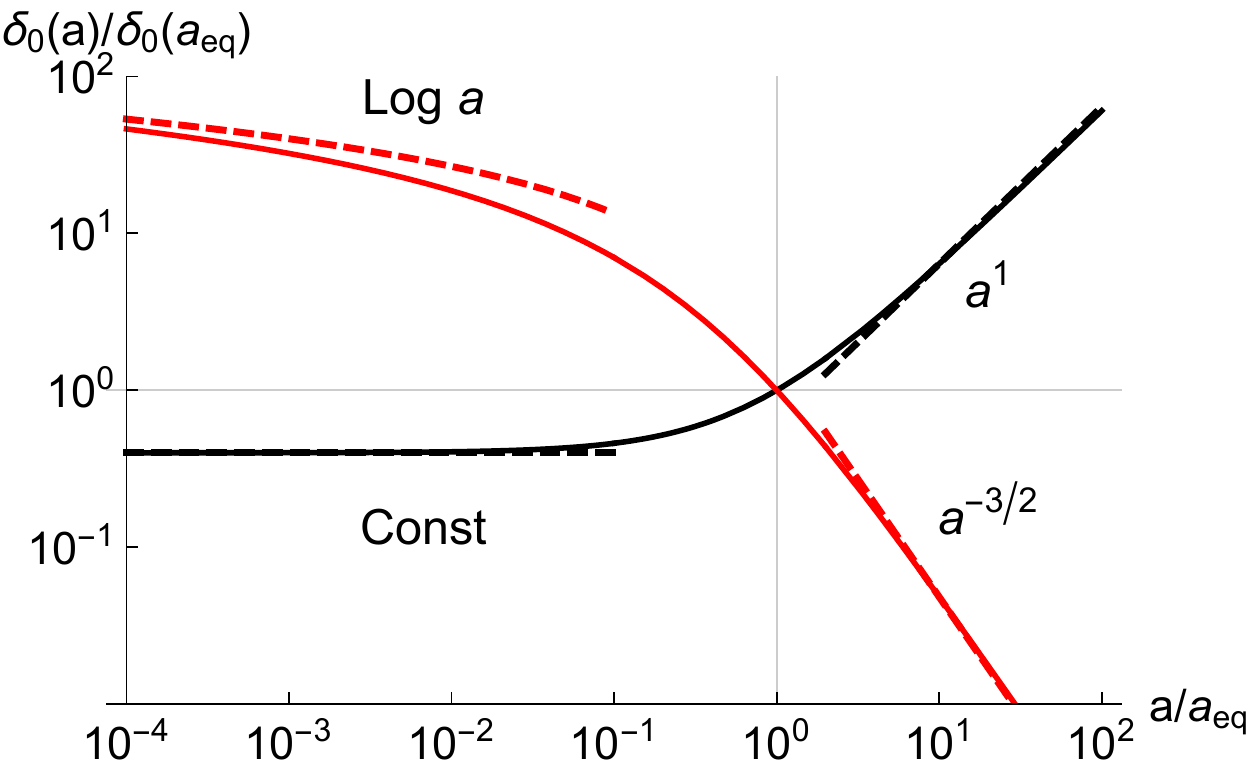}
\caption{The two independent solutions for growth of long wavelength fluctuations, $\delta_0 (t)$.}
\end{figure}

\end{exercise}

\begin{exercisenb}

{\bf Calculation of vacuum energy for a scalar field in a static spacetime}

There are a variety of ways commonly used to compute quantum corrections to the vacuum energy, and this tutorial is meant to show how they are related. For the purposes of the exercise a free real scalar field is used, with action $S = \int {\hbox{d}} t \; L = \int {\hbox{d}}^4x \; {\cal L}$ and Lagrangian $L = \int {\hbox{d}}^3x \; {\cal L}$. The Lagrangian density is
\begin{equation} \label{Ldens}
{\cal L} = - \sqrt{-g} \; \left[ \frac12 \, \partial_\mu \phi \, \partial^\mu \phi + \frac12 \, m^2 \, \phi^2 \right] \,,
\end{equation}
and the resulting field equation is the Klein-Gordon equation
\begin{equation}
(- \Box + m^2) \phi = ( - g^{\mu\nu} \nabla_\mu \nabla_\nu + m^2) \phi = 0 \,.
\end{equation}
But the relationship between the calculations described below is more general than just for this one example.

{\bf Canonical calculation} The simplest approach to calculating the vacuum energy is the same calculation that identifies all of the energy eigenstates and eigenvalues. This starts by assuming a static background spacetime with metric ${\hbox{d}} s^2 = - {\hbox{d}} t^2 + g_{ij} \, {\hbox{d}} x^i \, {\hbox{d}} x^j$, for which $g_{ij}$ is time-independent and a conserved energy can be formulated. Using the above action the field's canonical momentum is
\begin{equation}
\pi(x) = \frac{\delta S}{\delta \dot \phi(x)} = \sqrt{-g} \; \dot \phi(x) \,,
\end{equation}
(where an over-dot as in $\dot\phi$ denotes $\partial_t$) and so the Hamiltonian density is
\begin{equation}
{\cal H} = \pi \, \dot \phi - {\cal L} = \frac{\pi^2}{2 \sqrt{-g}} + \frac12 \sqrt{-g} \; \Bigl[ g^{ij} \, \nabla_i \phi \, \nabla_j \phi + m^2 \, \phi^2 \Bigr] \,.
\end{equation}

{\bf Background about quantization and mode functions}

Because this is quadratic in the fields it is essentially a fancy harmonic oscillator. To diagonalize it we expand the fields in terms of creation and annihilation operators
\begin{equation}
\phi(x) = \sum_n \Bigl[ a_n \, U_n(x) + a_n^\star \, U^*_n(x) \Bigr] \,,
\end{equation}
where we choose the mode functions, $U_n(x)$, to be simultaneous eigenstates of $- g^{ij} \nabla_i \nabla_j $ and $i \partial_t$. That is they satisfy the Klein-Gordan equation, $(- \Box + m^2) U_n = 0$, in a basis that also satisfies
\begin{equation} \label{modeeqs}
- g^{ij} \nabla_i \nabla_j U_n(x) = \omega_n^2 U_n(x) \qquad \hbox{and} \qquad i \dot U_n = \varepsilon_n \, U_n(x) \,,
\end{equation}
for eigenvalues $\omega_n^2$ and $\varepsilon_n$. The Klein-Gordon equation imposes a relation between these eigenvalues since $-\Box U_n = \ddot U_n - g^{ij} \nabla_i \nabla_j U_n = (- \varepsilon_n^2 - g^{ij} \nabla_i \nabla_j ) U_n$ and so
\begin{equation}
\Bigl( - g^{ij} \nabla_i \nabla_j + m^2 \Bigr) U_n = (\omega_n^2 + m^2) U_n = \varepsilon_n^2 \, U_n \,.
\end{equation}
This shows how $\varepsilon_n^2 = \omega_n^2 + m^2$ gets determined by the spectrum of $g^{ij} \nabla_i \nabla_j$ for the spacetime of interest.

So we may write
\begin{equation}
U_n({\bf x},t) = \frac{1}{\sqrt{2 \varepsilon_n}} \; u_n({\bf x}) \, e^{-i \varepsilon_n t} \,,
\end{equation}
where the prefactor is chosen for later convenience. Similarly
\begin{equation}
\pi(x) = \sqrt{-g} \;\dot \phi = -i \sqrt{-g} \;\sum_n \varepsilon_n \Bigl[ a_n \, U_n(x) - a_n^\star \, U^*_n(x) \Bigr] \,.
\end{equation}

The covariant normalization condition for the modes is defined using the Wronskian by
\bea \label{Wnorm}
W_\Sigma(U_n, U_m) &:=& -i\int_\Sigma {\hbox{d}}^3 {\bf x} \sqrt{-g} \; \Bigl[ \dot U_n^*(x) \, U_m(x) - U_n^* (x)\, \dot U_m(x) \Bigr] \nn\\
&=& \int_\Sigma {\hbox{d}}^3 {\bf x} \sqrt{-g} \; \Bigl[ \varepsilon_n U_n^* \, U_m + \varepsilon_m U_n^* \, U_m \Bigr] = \delta_{mn} \,,
\eea
where $\Sigma$ is a slice of fixed $t$. Similarly, because (\ref{modeeqs}) tell us $i \dot U_n^* = - \varepsilon_n U_n^*$, we see that $U_n^*$ and $U_n$ are eigenstates for different energy eigenvalues (notice for $m\ne 0$ there are no zero eigenvalues), and so are also orthogonal
\bea \label{Wnormstar}
W_\Sigma(U_n^*, U_m) &:=& -i\int_\Sigma {\hbox{d}}^3 {\bf x} \sqrt{-g} \; \Bigl[ \dot U_n(x) \, U_m(x) - U_n (x)\, \dot U_m(x) \Bigr] \nn\\
&=& (\varepsilon_m - \varepsilon_n) \int_\Sigma {\hbox{d}}^3 {\bf x} \sqrt{-g} \; U_n \, U_m = 0 \,.
\eea

It doesn't matter which $t$ we choose for $W$ when evaluating these orthogonality conditions provided the falloff of $U_n$ is sufficiently good at spatial infinity (if this exists), and this is the point of why $W$ is defined the way it is. To see why notice $(- \Box + m^2) U_n = 0$ implies the following chain of equalities
\bea
0 &=& -i\int_\Sigma^{\Sigma'} {\hbox{d}}^4x \; \sqrt{-g} \Bigl[ [(- \Box + m^2) U_n]^* U_m - U_n^* [(-\Box + m^2)U_m] \Bigr] \nn\\
&=& i\int_{\Sigma}^{\Sigma'} {\hbox{d}}^4 x \; \sqrt{-g} \;\nabla_\mu \Bigl[ (\nabla^\mu U_n)^* U_m - U_n^* (\nabla^\mu U_m) \Bigr]\\
&=& i\oint_\Sigma^{\Sigma'} {\hbox{d}}^3x \; \sqrt{-g} \; n_\mu \Bigl[ (\nabla^\mu U_n)^* U_m - U_n^* (\nabla^\mu U_m) \Bigr] \nn\\
&=& W_\Sigma(U_n, U_m) - W_{\Sigma'}(U_n, U_m) \,.\nn
\eea
Here the integration in the first line is over a slab of spacetime lying between two constant-$t$ slices, $\Sigma$ and $\Sigma'$. The second line then integrates both terms by parts and the third line uses Gauss' theorem to write the result in terms of a surface integral over the boundaries of the spacetime region of interest, with $n_\mu$ being the outward-pointing normal. If there are no spatial boundaries (or if the boundary conditions are chosen at spatial infinity appropriately) then the only boundaries contributing to the integrals are $\Sigma$ and $\Sigma'$. Then $n_\mu {\hbox{d}} x^\mu = \pm {\hbox{d}} t$ for the two surfaces, and the last line follows by recognizing that the surface integrals are precisely the Wronskians for each of the bounding constant-$t$ surfaces. Comparing first and last lines shows that $W_\Sigma(U_n, U_m)$ does not depend on $\Sigma$.

Given the above conventions and normalization condition, completeness of the modes implies
\begin{equation} \label{completeness}
\sum_n u_n({\bf x}) \, u_n^*({\bf y}) = \frac{\delta^3({\bf x} , {\bf y}) }{[-g({\bf x})]^{1/4} [-g({\bf y})]^{1/4}}\,,
\end{equation}
where the delta function transforms as a bi-density distribution that vanishes when ${\bf x} \ne {\bf y}$ and satisfies the defining condition
\begin{equation}
f({\bf x},t) = \int {\hbox{d}}^3 {\bf y} \; \delta^3({\bf x}, {\bf y}) \, f({\bf y},t)
\end{equation}
for all $f$ without any metrics. The completeness condition is related to the normalization condition because multiplying (\ref{completeness}) by $\sqrt{-g({\bf y})} \; u_m({\bf y})$ and integrating over ${\bf y}$ must give the tautology $u_m({\bf x}) = u_m({\bf x})$, which it does but only because the $u_m$'s are orthogonal and (\ref{Wnorm}) implies each mode satisfies the normalization condition
\begin{equation}
\int_\Sigma {\hbox{d}}^3 {\bf x} \sqrt{-g} \; u_n^*({\bf x}) u_n ({\bf x}) = 2 \varepsilon_n \int_\Sigma {\hbox{d}}^3 {\bf x} \sqrt{-g} \;U_n^*({\bf x}) U_n ({\bf x}) = 1 \,.
\end{equation}

Finally, the harmonic oscillator (or creation and annihilation) operator algebra is equivalent to the canonical quantization conditions because
\begin{equation}
[a_n, a_m] = 0 \qquad \hbox{and} \qquad [a_n, a_m^\star] = \delta_{nm} \,,
\end{equation}
imply
\bea
[ \phi(x), \phi(y)] &=& \sum_{nm} \Bigl\{ U_n(x) \, U^*_m(y) [a_n, a^\star_m] + U^*_n(x) U_m(y) [a_n^\star, a_m] \Bigr\} \nn\\
&=& \sum_{n} \frac{1}{2 \varepsilon_n} \Bigl\{ u_n({\bf x}) \, u^*_n({\bf y}) - u^*_n({\bf x}) u_n({\bf y}) \Bigr\} = 0 \,,
\eea
and
\bea
[ \pi({\bf x},t), \phi({\bf y},t)] &=& -i \sqrt{-g({\bf x})} \;\sum_{nm} \varepsilon_n \Bigl\{ U_n(x) \, U^*_m(y) [a_n, a^\star_m] - U^*_n(x) U_m(y) [a_n^\star, a_m] \Bigr\} \nn\\
&=& -\frac{i}{2} \sqrt{-g({\bf x})} \;\sum_{n} \Bigl\{ u_n({\bf x}) \, u^*_n({\bf y}) + u^*_n({\bf x}) u_n({\bf y}) \Bigr\} \\
&=& -i \frac{ [-g({\bf x})]^{1/4}}{[-g({\bf y})]^{1/4}} \; \delta^3({\bf x}, {\bf y}) = -i\, \delta^3({\bf x}, {\bf y}) \,.
\eea

{\bf Calculation of the energy eigenvalues and eigenstates}

\begin{enumerate}
\item The point of the above is that the energy is diagonal when expressed in terms of the eigenstates of $a^\star_n a_n$, as we see by evaluating the Hamiltonian in terms of $a_n$ and $a_n^\star$. To this end write
\bea
H = \int {\hbox{d}}^3 {\bf x} \; {\cal H}
&=& \int {\hbox{d}}^3{\bf x} \left\{ \frac{\pi^2}{2 \sqrt{-g}} + \frac12 \sqrt{-g} \; \Bigl[ g^{ij} \, \nabla_i \phi \, \nabla_j \phi + m^2 \, \phi^2 \Bigr] \right\} \nn\\
&=& \int {\hbox{d}}^3{\bf x} \left\{ \frac{\pi^2}{2 \sqrt{-g}} + \frac12 \sqrt{-g} \; \phi \Bigl[ - g^{ij} \, \nabla_i \nabla_j \phi + m^2 \, \phi \Bigr] \right\} \,,
\eea
and show that it can be written
\begin{equation}
H = \frac12 \sum_n \varepsilon_n \Bigl( a_n^\star a_n + a_n a_n^\star \Bigr) \,.
\end{equation}

\item The previous question shows that $H$ is diagonal in the basis for which the operators $a_n^\star a_n$ are diagonal for all $n$. Show this by using the commutation relation $[a_n, a_m^\star] = \delta_{nm}$ to rewrite $H$ as
\begin{equation}
H = E_0 + \sum_n \varepsilon_n \, a_n^\star a_n \,,
\end{equation}
with the constant $E_0$ being formally written as
\begin{equation}
E_0 = \frac12 \sum_n \varepsilon_n \,.
\end{equation}
This expression is `formal' because the sum typically diverges. It can be regularized in many ways (and you might reasonably wonder whether or not physical results depend on which way is used). One such is zeta-function regularization, which defines
\begin{equation}
\zeta(s) := \sum_n \varepsilon^{-s} \,,
\end{equation}
for complex $s$. This often converges where the real part of $s$ is sufficiently large and positive, and one tries to analytically extend this result down to the desired result $E_0 = \zeta(-1)$. Another way to proceed is instead to differentiate $E_0$ sufficiently many times with respect to $m^2$ that the sum converges, and then integrate the sum again to get $E_0$,

The energy eigenvalues for $H$ are clearly given by $H | \{ N_k \} \rangle = E | \{ N_k \} \rangle$ with
\begin{equation}
E = E_0 + \sum_n N_n \varepsilon_n \,,
\end{equation}
where the next exercise shows the allowed values for the $N_n$ are $N_n = 0,1,2,\cdots$. The state $|0\rangle$ denotes the ground state (or vacuum) for which $N_n = 0$ for all $n$ and has eigenvalue
\begin{equation}
H | 0 \rangle = E_0 | 0 \rangle \,.
\end{equation}

\item The basis diagonalizing $a_n^\star a_n$ for all $n$ is called the `occupation-number' basis and denoted $|\{N_k \}\rangle = |N_{n_1}, N_{n_2}, N_{n_3}, \cdots \rangle$ where the labels $N_n$ are the eigenvalues for $a_n^\star a_n$, for all possible values taken by $n$. That is, they satisfy
\begin{equation}
a_n^\star a_n | \{N_k \} \rangle = N_n | \{ N_k \} \rangle \,.
\end{equation}

Prove that the $N_n = 0, 1, 2, \cdots$ are non-negative integers as follows. First prove $[a^\star_n a_n , a_m] = -\delta_{nm} a_n$ and $[a^\star_n a_n , a_m^\star] = + \delta_{nm} a^\star_n$. Show that these relations imply that if $| \{ N_k \} \rangle$ is an eigenstate with eigenvector of $a_n^\star a_n$ with eigenvalue $N_n$ then $a_n | \{ N_k \} \rangle$ is also an eigenstate of $a_n^\star a_n$ but with eigenvalue $N_n - 1$ and $a_n^\star | \{ N_k \} \rangle$ is an eigenvector with eigenvalue $N_n + 1$.

Next prove $N_n \ge 0$ by evaluating $\langle \{ N_k \} | a^\star_n a_n | \{ N_k \} \rangle = N_n \langle \{ N_k \} | \{ N_k \} \rangle = N_n$ and recognizing that the left-hand side is non-negative because it is the norm of the vector $a_n | \{ N_k \} \rangle$. But this is inconsistent with the result that $a_n$ always lowers the eigenvalue by one unit unless there exists an eigenstate for which $a_n | \Psi \rangle = 0$. Repeating this argument for all labels $n$ shows there must be a state, $| 0 \rangle$, for which $a_n | 0 \rangle = 0$ for all $n$, and then all other eigenstates of $a_n^\star a_n$ are obtained by acting repeatedly on $|0 \rangle$ with $a_n^\star$. (For example consider the particular state $|2_{n_5}, 6_{n_{20}} \rangle$, for which the particle state labeled by $n_5$ has eigenvalue $N_{n_5} = 2$ for $a_{n_5}^\star a_{n_5}$ and the state labeled by $n_{20}$ has eigenvalue $N_{n_{20}} = 6$ for $a_{n_{20}}^\star a_{n_{20}}$. This is proportional to $(a_{n_5}^\star)^2 \; (a_{n_{20}}^\star)^6 | 0 \rangle$, and so on for any other choices for these eigenvalues.)

{\bf Path integral method of evaluating the vacuum energy}

An alternate way to proceed instead uses the path integral formulation for the effective action
\begin{equation}
e^{i \Gamma[g]} = \int {\cal D} \phi \, e^{i S[\phi, g]} \,,
\end{equation}
where the action $S[\phi,g]$ is given as the integral over (\ref{Ldens}), regarded as a function of the fields $\phi$ and $g_{\mu\nu}$. In this expression $\Gamma[g]$ is a contribution to the action for the metric, $g_{\mu\nu}$, obtained after integrating out the field $\phi$. It is to be added to other terms (like the Einstein-Hilbert term), but our interest is in anything of the form $\Gamma = - \int {\hbox{d}}^4 x \; \sqrt{-g} \; \rho_v$, because this gravitates like a cosmological constant (or vacuum energy). For time-translational invariant systems the integral over t diverges proportional to $\int_{-T}^T {\hbox{d}} t = T$ as $T \to \infty$ and it is the energy $E_0 = - \Gamma/T$ that should remain finite in this limit.

Because the functional integral is Gaussian it can be evaluated in terms of a functional determinant of the quadratic operator appearing in the action: $\Delta = (- \Box + m^2) \delta^4(x-y)$.
\begin{equation}
e^{i \Gamma} = \left[ {\det} \Bigl( - \Box + m^2 - i \epsilon \Bigr) \right]^{-1/2}\,,
\end{equation}
and so
\begin{equation}
\Gamma = \frac{i}{2} \ln \det \Bigl( - \Box + m^2 - i \epsilon\Bigr) = \frac{i}{2} \hbox{Tr} \, \ln \Bigl( - \Box + m^2 -i\epsilon \Bigr) \,.
\end{equation}
Here $\epsilon$ is a positive quantity that is taken to zero at the end, and imposes (as usual for a Feynman propagator) the right boundary conditions to describe matrix elements in the vacuum. We suppress the $i \epsilon$ in what follows, but recall it when needed by regarding $m^2$ as having a small negative imaginary part.

To evaluate this again choose eigenfunctions that diagonalize $- g^{ij} \nabla_i \nabla_j$ and $i \partial_t$. That is choose a basis of functions, $V_n(x)$, for which
\begin{equation}
- g^{ij} \nabla_i \nabla_j V_n = \omega_n^2 V_n \;\; \hbox{ and } \;\; i \partial_t V_n = \varepsilon V_n \,,
\end{equation}
and so
\begin{equation}
(- \Box+m^2) V_n = (- \varepsilon^2 + \omega_n^2 + m^2) V_n
\end{equation}
is diagonalized with eigenvalues $\lambda_n = - \varepsilon^2 + \omega_n^2 + m^2= - \varepsilon^2 + \varepsilon^2_n$. Notice that unlike the previous section we do not also have $(- \Box + m^2) U_n = 0$ and so we {\em cannot} identify $\varepsilon^2$ with $\varepsilon_n^2 := \omega_n^2 + m^2$. Instead $\varepsilon$ is the Fourier transform variable for time, arising generically for time-translationally invariant systems.

In terms of this our operator in this basis is
\begin{equation}
\langle n \varepsilon | \Delta | r \varepsilon' \rangle = (- \varepsilon^2 + \omega^2_n + m^2) \;2 \pi \delta(\varepsilon - \varepsilon') \delta_{nr} \,,
\end{equation}
and so the trace may be given by taking diagonal elements and summing over their eigenvalues, with
\begin{equation}
\Gamma (m^2) = \frac{i}2 \hbox{Tr} \, \ln \Bigl( - \Box + m^2 \Bigr) = \frac{i}2 \sum_n \int_{-\infty}^\infty \frac{ {\hbox{d}} \varepsilon}{2\pi} \;\ln (- \varepsilon^2 + \omega_n^2 + m^2 ) \; 2\pi \delta(0) \,.
\end{equation}
The factor of $\delta(0)$ arises due to time translation invariance, as may be seen by writing
\begin{equation}
2\pi \delta(E) = \lim_{T \to \infty} \int_{-T}^T {\hbox{d}} t\; e^{-i \varepsilon t} \quad \hbox{and so} \quad
2\pi \delta (0) = \lim_{T \to \infty} T \,,
\end{equation}
and so the well-behaved quantity is the energy
\begin{equation}
E_0 = - \lim_{T \to \infty} \frac{ \Gamma}{T} = - \frac{i}2 \sum_n \int_{-\infty}^\infty \frac{ {\hbox{d}} \varepsilon}{2\pi} \;\ln (- \varepsilon^2 + \omega_n^2 + m^2 ) \,.
\end{equation}

Again the remaining sums and integrals diverge. The integration over $\varepsilon$ passes through singularities at $\pm \varepsilon_n$, which we should navigate by Wick rotating. That is, keeping in mind (as usual) that $m^2 \to m^2 - i \epsilon$ is required for the Feynman propagator, we can rotate our contour of integration counter-clockwise by 90 degrees in the complex $\varepsilon$ plane by writing $\varepsilon \to i \, \varepsilon_{\scriptscriptstyle E}$ with $\varepsilon_{\scriptscriptstyle E}$ also running from $-\infty$ to $\infty$. The integral converges if we first differentiate with respect to $m^2$, so show that
\bea
\frac{\partial E_0}{\partial m^2} &=& \frac{1}2 \sum_n \int_{-\infty}^\infty \frac{{\hbox{d}} \varepsilon_{\scriptscriptstyle E}}{2\pi} \left( \frac{1}{ \varepsilon_{\scriptscriptstyle E}^2 + \omega_n^2 + m^2} \right) \nn\\
&=& \frac{1}{4\pi} \sum_n \left[ \frac{1}{\varepsilon_n} \, \arctan\left( \frac{\varepsilon}{\varepsilon_n} \right) \right]^{\infty}_{-\infty} = \frac{1}{4} \sum_n \frac{1}{\varepsilon_n} \,,
\eea
where, as above, $\varepsilon_n = \sqrt{\omega_n^2 + m^2}$. Integrating again with respect to $m^2$ then gives
\begin{equation}
E_0(m^2) = \frac{1}{2} \sum_n \varepsilon_n \,,
\end{equation}
up to an arbitrary $m^2$-independent constant. This is the same sum as was obtained in the canonical calculation earlier.

{\bf Flat space evaluation}

As a particularly simple case consider the case of a flat geometry, for which $- g^{ij} \nabla_i \nabla_j = -\nabla^2$ can be diagonalized in Fourier space, with eigenfunctions $\exp(i {\bf p} \cdot {\bf x})$ and eigenvalues ${\bf p}^2$.

In terms of this the required operator in this basis is
\begin{equation}
\langle p | \Delta | q \rangle = ( p_\mu p^\mu + m^2) \; (2 \pi)^4 \delta^4(p - q) \,,
\end{equation}
and so the trace may be given by taking diagonal elements and summing over their eigenvalues, with
\begin{equation}
\Gamma (m^2) = \frac{i}2 \hbox{Tr} \, \ln \Bigl( - \Box + m^2 \Bigr) = \frac{i}2 \int_{-\infty}^\infty \frac{ {\hbox{d}}^4 p}{(2\pi)^4} \;\ln (p_\mu p^\mu + m^2 ) \; (2\pi)^4 \delta^4(0) \,.
\end{equation}
The additional factor of $\delta^3(0)$ arises due to spatial translation invariance, as may be seen by writing (as we did before for time)
\begin{equation}
(2\pi)^3 \delta^3({\bf p}) = \lim_{L \to \infty} \int_{-L}^L {\hbox{d}}^3{\bf x} \; e^{i {\bf p} \cdot {\bf x} } \quad \hbox{and so} \quad
(2\pi)^3 \delta^3 (0) = \lim_{L \to \infty} L^3 \,,
\end{equation}
and so is proportional to the volume of space (as well as the previous proportionality to $T$). The well-behaved quantity for infinite translationally invariant systems is therefore the energy {\em density},
\begin{equation}
\rho_v = \lim_{L \to \infty} \frac{ E_0}{L^3} = - \lim_{L,T \to \infty} \frac{ \Gamma}{TL^3} = - \frac{i}2 \int_{-\infty}^\infty \frac{ {\hbox{d}}^4 p}{(2\pi)^4} \;\ln (p_\mu p^\mu + m^2 ) \,.
\end{equation}

To avoid the singularities at $p^0 = \pm \sqrt{{\bf p}^2 + m^2}$, we again Wick rotate. In the resulting euclidean integral the angular integrals can be done once and for all, giving a factor of the volume of the unit 3-sphere: $2\pi^2$. The remaining integral converges if we first differentiate with respect to $m^2$ thrice, so show that
\begin{equation}
\left({\frac{\partial}{\partial m^2}} \right)^3 \rho_v = \frac{2\pi^2}2 \int_{0}^\infty \frac{p_{\scriptscriptstyle E}^3 {\hbox{d}} p_{\scriptscriptstyle E}}{(2\pi)^4} \; \frac{2}{( p_{\scriptscriptstyle E}^2 + m^2)^3} = \frac{1}{32\pi^2 m^2} \,,
\end{equation}
and so integrating three times with respect to $m^2$ then gives
\begin{equation}
\rho_v = \frac{m^4}{64\pi^2} \ln \left( \frac{m^2}{\mu^2} \right) + A m^4 + B m^2 + C \,,
\end{equation}
where $\mu$, $A$, $B$ and $C$ are arbitrary $m^2$-independent constants. Although the values of $A$, $B$ and $C$ can depend on how the integrals were regulated, the logarithmic term cannot.
\end{enumerate}

\medskip

\noindent{\bf SOLUTION}: Calculation of vacuum energy for a scalar field in a static spacetime.

\begin{itemize}

\item[1.] The Hamiltonian can be written as,
\begin{align}
H &= \int d^3 x \left\{ \frac{\pi^2}{2 \sqrt{-g}} + \frac{1}{2} \sqrt{-g} \phi \left[ - g^{ij} \nabla_i \nabla_j + m^2 \right] \phi \right\}
\end{align}
where the canonical field and it's conjugate momenta may be expanded in terms of annihilation and creation operators,
\begin{align}
\phi (x) &= \sum_n \left[ a_n U_n (x) + a_n^* U_n^* (x) \right] \\
\pi (x) = -i \sqrt{-g} \sum_n \epsilon_n \left[ a_n U_n (x) - a_n^* U_n^* (x) \right]
\end{align}
where the mode functions $U_n$ are orthogonal and normalized with respect to the Klein-Gordon norm.
This gives,
\begin{equation}
H = \frac{1}{2} \sum_n \epsilon_n \left( a_n^* a_n + a_n a_n^* \right)
\end{equation}
which shows that number eigenstates (of $a_n^* a_n$) are also energy eigenstates.

\item[2.] Using $\left[ a_n, a_n^* \right] = \delta_{nm}$, the Hamiltonian can be written,
\begin{equation}
H = E_0 + \sum_n \epsilon_n a_n^* a_n ,
\end{equation}
where $E_0 = \tfrac{1}{2} \sum_n \epsilon_n$ is the zero point energy.

\item[3.] Using the canonical commutation relations, one has that,
\begin{align}
[ a_n^* a_n, a_m ] &= - \delta_{nm} a_n \\
[ a_n^* a_n, a_m^* ] &= + \delta_{nm} a_n^*
\end{align}
Then for a state with definite occupation numbers, $a_n^* a_n | \{ N_k \} \rangle = N_n | \{ N_n \} \rangle$, we have that,
\begin{align}
a_m^* a_m \left( a_n | \{ N_k \} \rangle \right) = a_n a_m a_m^* | \{ N_n \} \rangle - \delta_{nm} a_n | \{ N_k \} \rangle = (N_m - \delta_{nm} ) a_n | \{ N_k \} \rangle
\end{align}
and so $a_n$ lowers the occupation number $N_n$ by one, and similarly,
\begin{align}
a_m^* a_m \left( a_n^* | \{ N_k \} \rangle \right) = a_n^* a_m a_m^* | \{ N_n \} \rangle + \delta_{nm} a_n^* | \{ N_k \} \rangle = (N_m + \delta_{nm} ) a_n^* | \{ N_k \} \rangle
\end{align}
and so $a_n^*$ increases $N_n$ by one.

In order for the Hilbert space to have positive-definite norm, there must be a null state, $a_n | 0 \rangle = 0$, which prevents any $N_n$ from becoming negative (such states would have negative norm). From this single state, repeated action of $a_n^*$ generates the entire Fock space of possible states.

\end{itemize}

\end{exercisenb}

\begin{exercisenb}

{\bf Quantum fluctuations of a scalar field in a class of inflationary spacetimes}

For a change of pace we work in the Schr\"odinger picture, rather than the Heisenberg picture, and so compute the vacuum wavefunctional, $\Psi[\varphi,\,t]$, for a scalar field.

{\bf Action and Hamiltonian}

Our starting point is the Lagrangian density for a spectator scalar
\begin{equation}
L = \int {\hbox{d}}^3 x\,a(t)^3\,\left[\frac12\,\dot{\phi}^2
-\frac{1}{2\,a^2(t)} \left(\nabla \phi \right)^2
-\frac{m^2(t)}{2}\,\phi^2 \right] \,,
\end{equation}
in an FRW spacetime with metric
\begin{equation}
{\hbox{d}} s^2\,=\,- {\hbox{d}} t^2 + a^2(t) {\hbox{d}} \vec{x}^2
\end{equation}
and Hubble parameter $H(t)\,=\,\dot{a}/a$. Here $m(t)$ denotes the (possibly slowly time-dependent) mass.

Find the canonical momentum, $\pi_k$, for each Fourier mode, $\varphi_k$, of the scalar field. Given the quantization condition $\pi_k = - i \delta/\delta \varphi_k$, show that the Hamiltonian density in Schr\"odinger representation can be expressed in Fourier space as
\begin{equation}
{\cal H} = {\cal H}_0 + \sum_k {\cal H}_k \,,
\end{equation}
with ${\cal H}_k$ for $k \ne 0$ given by
\begin{equation}
{\cal H}_k\,=\,-\frac{1}{a^3}\,\frac{\delta^2}{\delta {\varphi}_k\, \delta {\varphi}_{-k}}+
{a^3}\,\left[
\frac{c_s^2\,k^2}{a^2}+ m^2
\right] \,\varphi_k \varphi_{-k}
\end{equation}
where $\varphi^*_k \, = \, \varphi_{-k}$.

{\bf Ground state wave functional}

Use this Hamiltonian to evolve the state wave-functional, $\Psi = \prod_k \Psi_k$, according to the Schr\"odinger equation,
\begin{equation} \label{scheq}
i\,\frac{\partial \Psi_k}{\partial t}\,=\,{\cal H}_k\,\Psi_k \,,
\end{equation}
and for free fields seek solutions subject to a Gaussian ansatz,
\begin{equation}
\Psi[\varphi] = \prod_k \Psi_k[\varphi] \,=\, \prod_{k}
{\cal N}_k(t)\, \exp \Bigl\{ -a^3(t)\,\Bigl[
\,\alpha_k(t)\,\varphi_k \,\varphi_{-k}
\Bigr] \Bigr\}
\label{defPsi}
\end{equation}
and show that the variance of $\varphi_k$, $\langle |\varphi_k|^2 \rangle$, is given by
$[a^3(\alpha_k + \alpha_k^*)]^{-1}$. Determine the evolution equations for the functions ${\cal N}_k(t)$, $\alpha_k(t)$ by substituting into (\ref{scheq}). Show that they imply $\alpha_k$ must satisfy
\begin{equation}
0= \dot{\alpha}_k+i \,\alpha_k^2+3\,H\,\alpha_k-i\left( \frac{k^2}{a^2}+m^2\right)
\hskip1cm {\rm for}\,\,k\ge0
\label{evea}
\end{equation}
where all quantities (including the Hubble parameter) can be time dependent, and the dot denotes derivative
with respect to time. The additional equation for ${\cal N}_k$ ensures it evolves in a way that is consistent with normalization, but is not needed in what follows.

The solution for $\alpha_k$ can be made very explicit if we assume power-law expansion, $a = a_0 (t/t_0)^p$ (so that $H = p/t$ and $\epsilon = -\dot H/H^2 = 1/p$) and a time-independent ratio $m/H$. (Show that de Sitter space can be obtained as the special case where $p \to \infty$ and so $\epsilon \to 0$.)

Equations of the form of (\ref{evea}) are integrated by changing variables from $\alpha_k$ to $u_k$ where
\begin{equation}\label{reak1}
\alpha_k = -i \left( \frac{\dot u_k}{u_k} \right) = \-i\,a\,H\, \left[ \frac{\partial_a\, u_k(a)}{u_k(a)} \right] \,.
\end{equation}
Show that (\ref{evea}) is then satisfied if $u_k$ solves the Klein-Gordon equation,
\begin{equation} \label{KGmodes}
\ddot u_k + 3H \, \dot u_k + \left( \frac{ k^2}{a^2} + m^2 \right) u_k = 0 \,.
\end{equation}
For constant $\epsilon$ and $m^2/H^2$ show that this is solved by
\begin{equation}
u_k(a) = \tilde {\cal C}_k \, y^q \,\sigma_k (y),
\end{equation}
where $\tilde {\cal C}_k$ is $a$-independent, provided $q$ and $y$ are chosen as
\begin{equation} \label{qdef}
q = \frac{3-\epsilon}{2\, (1-\epsilon)} \,,
\end{equation}
and
\begin{equation}
y(a,k) := \frac{1}{(1-\epsilon)}\left( \frac{k}{a H} \right) = \frac{1}{(1-\epsilon)}\left( \frac{k}{a_0 H_0} \right) \left( \frac{a_0}{a}
\right)^{1-\epsilon} \,.
\end{equation}
The point of these changes of variables is that they turn eqn~(\ref{KGmodes}) into the Bessel equation for $\sigma_k$:
\begin{equation}
y^2\,\sigma_k''+y\,\sigma_k'+\left(y^2-\nu^2 \right)\,\sigma_k = 0 \,,
\end{equation}
where primes here denote derivatives with respect to $y$. Show that the order $\nu$ is given by
\begin{equation} \label{eqfnu}
\nu^2=\frac{1}{(1-\epsilon)^2}\left[ \frac{(3-\epsilon)^2}{4}-\frac{m^2}{H^2}\right] \,.
\end{equation}

The solutions for $\sigma_k$ are (naturally) Bessel functions, and demanding agreement with the adiabatic vacuum before horizon exit tells us
\begin{equation}\label{bdsol}
u_k \propto \exp \left[ \mp i \int {\hbox{d}}t \left( \frac{k}{a} \right) \right] \propto e^{ \pm iy } \qquad \hbox{for $k/a \gg H$} \,,
\end{equation}
of which we choose the lower sign since this turns out below to ensure the real part of $\alpha_k$ is positive (as required to ensure $\Psi_k$ can be normalized). Show that this fixes the mode functions to be
\begin{equation} \label{mods1}
u_k (a) = \tilde {\cal C}_k \, y^q(a,k) \,H^{(2)}_{\nu}\left[y(a,k)\right] = \frac{{\cal C}_k}{\sqrt{a^3 H}} \; H^{(2)}_{\nu}\left[y(a,k)\right]
\end{equation}
where ${\cal C}_k \propto k^q \tilde {\cal C}_k$ relabels the integration constants and $H^{(2)}_\nu$ is the Hankel function of the second kind. The second equality in (\ref{mods1}) follows from eqn~(\ref{qdef}), which implies $a^3Hy^{2q}$ is time-independent. Notice this reduces to the solution for a massive field in de Sitter space in the limit $\epsilon \to 0$.

Although ${\cal C}_k$ drops out of (\ref{reak1}) and (so does not contribute directly to $\alpha_k$), some later formulae are simpler if we choose ${\cal C}_k$ so that the Wronskian,
\begin{equation} \label{Wronskiandef}
{\cal W}(u,v) := a^3 (u^* \dot v - v^* \dot u) \,,
\end{equation}
satisfies ${\cal W}(u,u) = i$. Prove that in this case is the expression for the real and imaginary parts of $\alpha_k$ become
\bea
\alpha_k+\alpha_k^*&=& -i \left( \frac{u_k^* \dot u_k - u_k \dot u_k^*}{|u_k|^2} \right) = \frac{1}{a^3\,|u_k|^2} \label{omefir}
\\
\hbox{and} \quad \alpha_k-\alpha_k^*&=&-i\,a\,H\,\left[\frac{\partial_a\,\left(|u_k|^2\right)}{|u_k|^2} \right] \,. \label{omesec}
\eea
What does the first of these imply for the variance of $\varphi_k$ in terms of $u_k$?

Because ${\cal W}$ is independent of time (when evaluated with solutions to (\ref{KGmodes}) it is convenient to compute the implications for ${\cal C}_k$ in the remote past, where $ k \gg aH$, in which case the Hankel function has the asymptotic form
\begin{equation}
H^{(2)}_\nu(y) \to \sqrt{\frac{2}{\pi y}} \; e^{-iy +\frac{i\pi}{2} \left(\nu+\frac12 \right)} \quad \hbox{for $y \to \infty$} \,.
\end{equation}
Use this to show
\begin{equation}
\left| {\cal C}_k \right|^2 = \frac{\pi}{4(1-\epsilon)} \,,
\end{equation}
for all $k$ and $\nu$.

Consequently the quantity relevant to fluctuations in the lectures is
\begin{equation} \label{smp20}
| u_k|^2 = \frac{\pi}{4(1-\epsilon) a^3 H } \, | H_\nu^{(2)}(y) |^2 \,.
\end{equation}
Use the asymptotic expression
\begin{equation}
H^{(2)}_{\nu} \left(y\right) \to \frac{i \Gamma(\nu)}{\pi}\,\left( \frac{y}{2} \right)^{-\nu} \quad \hbox{for $y \to 0$} \,,
\end{equation}
to derive the small-$k$ limit
\begin{equation} \label{smp2}
| u_k|^2 \to \frac{2^{2\nu-2} |\Gamma(\nu)|^2 (1-\epsilon)^{2\nu-1}}{\pi a^3 H } \, \left( \frac{aH}{k}
\right)^{2\nu} \,.
\end{equation}

Evaluate this for the case $\nu = \frac32$ (which for the de Sitter case $\epsilon = 0$ is a massless scalar field) and show that it agrees with the result obtained using the mode function directly, which in the case $\nu = \frac32$ is very simple:
\begin{equation}
u_k =- (1 -\epsilon) \, \frac{H}{\sqrt{2 k^3}} \; (y - i) e^{-iy} \qquad \hbox{for $\nu = \frac32$}
\end{equation}
up to an irrelevant phase. Prove that this does solve the Klein Gordon equation in the case $\nu = \frac32$.

The power spectrum $\Delta^2(k)$ is proportional to $k^3 |u_k|^2$ evaluated for $k \ll aH$. For de Sitter space $H$ is constant, and in this case what is the predicted $k$-dependence for $k^3 |u_k|^2$ when $\nu = \frac32$? When $\epsilon \ne 0$ $H$ is time dependent and we are supposed to evaluate $H$ at the moment where $aH = k$. If this were the whole story (and it is not quite), and if $\Delta^2(k) \propto A (k/k_0)^{n_s-1}$, what is the prediction for $n_s$ as a function of $\epsilon$?

\medskip

\noindent{\bf SOLUTION}

{\bf Action and Hamiltonian:}

\begin{equation}
\pi = \frac{\delta S}{\delta \dot \phi} = a^3 \dot \varphi
\end{equation}
Na\"{i}vely, one would define,
$$ \pi_k = - i \omega_k a^3 \varphi_k $$
which is an explicitly time-dependent momentum. Unlike on flat space, where translation invariance guarantees constant momenta, on FLRW we can have time-dependence. The on-shell condition is,
$$ \omega_k^2 = c_s^2 k^2 + m^2 $$
where $c_s$ is the sound speed for the scalar fluctuations on this background.

The Hamiltonian is given by,
\begin{align}
H := \pi \varphi - \mathcal{L} &= a^3 \left[ \frac{1}{2a^3} \pi^2 + \phi \frac{ -\nabla^2/a^2 + m^2 }{2} \phi \right] \\
&= a^3 \left[ \frac{1}{a^3} \pi_k \pi_{-k} + \left( \frac{c_s^2 k^2}{a^2} + m^2 \right) \varphi_k \varphi_{-k} \right]
\end{align}

{\bf Ground state wave functional:}

Using the Gaussian ansatz,
\begin{equation}
\Psi [ \varphi ] = \prod_k \mathcal{N}_k (t) \text{exp} \left\{ - a^3 (t) \left[ \alpha_k (t) \varphi_k \varphi_{-k} \right] \right\}
\end{equation}
we can solve the Schr\"{o}dinger equation,
\begin{equation}
i \frac{\partial \Psi_k }{ \partial t } = \mathcal{H}_k \Psi_k
\end{equation}
to find
\begin{equation}
i \frac{ \dot{\mathcal{N}}_k }{\mathcal{N}_k } - i a^3 \left( 3 H \alpha_k + \dot \alpha_k \right) \varphi_k \varphi_{-k} = \alpha_k + a^3 \left( - \alpha_k^2 + \frac{c_s^2 k^2 }{a^2} + m^2 \right) \varphi_k \varphi_{-k}
\end{equation}
As this holds for all $\varphi_k$, we must have that,
\begin{equation}
0 = \dot{\alpha_k} + i \alpha_k^2 + 3 H \alpha_k - i \left( \frac{k^2}{a^2} + m^2 \right) \;\;\;\; \text{for } k \geq 0
\end{equation}
along with the condition that $\dot{\mathcal{N}}_k = i \alpha_k \mathcal{N}_k$.

{\bf Power-law solutions:}
If $a = a_0 (t/t_0)^p$, then $H = p/t$. We can take $p \to \infty$ in order to set $\dot H/ H^2 \to 0$, and simultaneously take the limit $t = t_0 + \delta t \to t_0$ so that,
\begin{equation}
a (t_0) = \lim_{ \substack{ \delta t \to 0 \\ H \to \infty } } \text{exp} \left( H ( t_0 + \delta t ) \text{log} ( 1 + \frac{\delta t}{t_0} ) \right) = \begin{cases}
\;\; a_0 \;\;\;\; &\text{ if } H \delta t \to 0 , \\
a_0 \; e^{ H_0 t_0} \;\;\;\; &\text{ if } H \delta t \to H_0 t_0 , \\
\;\; \; 0 \;\;\;\; &\text{ if } H \delta t \to \infty
\end{cases}
\end{equation}
where $H_0$ is then the constant curvature of the dS space.

Using $\dot \alpha_k = - i \ddot u_k / u_k - i \alpha_k^2$, we get the Klein-Gordon equation for $u_k$. Then changing variables with $\dot y = - (1-\epsilon) H y$, we find,
\begin{align}
\tilde{\mathcal{C}}_k (1-\epsilon)^2 H^2 y^q &\Big\{ y^2 \sigma_k'' + y \sigma_k' \left( 1 + 2q + \frac{\epsilon-3}{1-\epsilon} \right) \nonumber \\
&\qquad + \sigma_k \left( y^2 + \frac{m^2}{H^2 ( 1- \epsilon)^2} + q^2 + \frac{q (\epsilon - 3 )}{1- \epsilon} \right) \Big\} = 0
\end{align}
and so choosing $q = (3-\epsilon)/2(1-\epsilon)$ we arrive at Bessel's equation for $\sigma_k$, with order
\begin{equation}
\nu^2 = \frac{1}{ (1-\epsilon)^2 } \left[ \frac{ (3 - \epsilon)^2 }{4} - \frac{m^2}{H^2} \right]
\end{equation}

The Bessel functions of first and second kind have the following asymptotic behavior,
\begin{align}
\lim_{z \to \infty } J_n ( z) &\sim \sqrt{ \frac{2}{\pi z} } \, \text{cos} \left( z - \frac{ m \pi }{2} - \frac{\pi}{4} \right) \\
\lim_{z \to \infty } Y_n ( z) &\sim \sqrt{ \frac{2}{\pi z} } \, \text{sin} \left( z - \frac{ m \pi }{2} - \frac{\pi}{4} \right)
\end{align}
and therefore the desired asymptotic behavior for $\sigma$ is given by a Hankel functions of the second kind,
$$ H^{(2)}_n (z) = J_n (z) - i Y_n (z) $$
This gives,
$$ u_k = \mathcal{C}_k / \sqrt{a^3 H} \, H_{\nu}^{(2)} ( \frac{1}{1-\epsilon} \frac{k}{a H} ) $$

The Wronskian,
\begin{equation}
\mathcal{W} (u , u ) = a^3 \left[ u^* \dot u - u \dot u^* \right] = i
\end{equation}

The corresponding $\alpha_k$ are,
$$ \alpha_k = - i \frac{ \dot u_k }{ u_k } \;\; \implies \;\; \alpha_k + \alpha_k^* = - i \frac{ \dot u_k u_k^* - \dot u_k^* u_k }{ |u_k |^2 } = \frac{1}{a^3 | u_k |^2} $$
$$ \alpha_k = - i a H \frac{ \partial_a u_k }{ u_k } \;\; \implies \;\; \alpha_k - \alpha_k^* = -i a H \frac{ \partial_a u_k u_k^* + \partial_a u_k^* u_k }{ |u_k |^2 } $$

The variance of $\varphi_k$ is,
$$ \langle | \varphi_k |^2 \rangle = \frac{1}{ a^3 ( \alpha_k + \alpha_k^* ) } = | u_k |^2 $$

To find $\mathcal{C}_k$, we use the asymptotics of the Hankel function to write,
\begin{equation}
u_k \sim \frac{\mathcal{C}_k}{ \sqrt{ a^3 H} } \sqrt{ \frac{2}{\pi y} } e^{-i y + \tfrac{i \pi}{2} ( \nu + \tfrac{1}{2} ) }
\end{equation}
\begin{equation}
\mathcal{W} (u, u) = a^3 \left( \frac{ | \mathcal{C}_k |^2 }{ a^3 H } \frac{2}{\pi y} \; 2 i (1- \epsilon) H y \right) = i
\end{equation}
$$ \implies \;\;\;\; | \mathcal{C}_k |^2 = \frac{\pi}{4 (1-\epsilon )} $$
The variance is then given by,
\begin{align}
\lim_{y \to 0} | u_k |^2 &\to \frac{\pi}{4 ( 1 - \epsilon) a^3 H} \left| \frac{ i \Gamma (\nu) }{ \pi } \left( \frac{y}{2} \right)^{-\nu} \right|^2 \\
&= \frac{ 2^{2\nu-2} | \Gamma (\nu ) |^2 ( 1 - \epsilon )^{2\nu-1} }{ \pi a^3 H } \left( \frac{a H}{k} \right)^{2 \nu}
\end{align}

{\bf Massless modes:}
In the de Sitter limit, $\epsilon = 0$, we find that,
$$ \nu^2 = \frac{9}{4} - \frac{ m^2}{H^2} $$
and so massless fields on de Sitter correspond to $\nu = 3/2$.

Specializing to $\nu=3/2$, and using $\Gamma (3/2 ) = \sqrt{\pi}/2$,
\begin{equation}
| u_k |^2 \to \frac{ ( 1- \epsilon)^2 H^2 }{ 2 k^3}
\end{equation}
Note that this agrees with the exact mode function,
$$ u_k = - ( 1- \epsilon) \frac{H}{ \sqrt{2 k^3} } ( y - i ) e^{-i y} \;\; \implies \;\; | u_k |^2 = | 1-\epsilon|^2 \frac{H}{2 k^3} | 1 + i y |^2 $$
This corresponds to a $\sigma_k$,
\begin{equation}
\sigma_k = - \sqrt{ \frac{2}{\pi} } y^{-3/2} ( y - i ) e^{-iy}
\end{equation}
$$ \implies \;\; y^2 \sigma''_k + y \sigma_k' = - (y^2 - \frac{3}{2} ) \sigma_k $$
and therefore $\sigma_k$ satisfies the Bessel equation with $\nu = 3/2$.

Considering,
$$ k^3 | u_k |^2 \Big|_{\nu = 3/2 } \to \frac{1}{2} ( 1 - \epsilon )^2 H^2 \;\;\;\; \text{for} \;\; k \ll a H$$
we conclude that $k^3 | u_k |^2 $ is $k$-independent on de Sitter,
$$ k^3 | u_k |^2 \Big|_{\nu = 3/2 } \to \frac{1}{2} H_0^2 $$
On FLRW, if we treat $\epsilon$ as a constant then we can solve,
$$ a (t) = k_0 \left( \epsilon( t - t_0 ) \right)^{1/\epsilon} $$
$$ a H = k \;\; \implies \;\; H = ( k / k_0 )^{- \frac{ \epsilon }{ 1-\epsilon} } $$
and so we find that,
\begin{equation}
n_s - 1 = \frac{- 2 \epsilon}{1 - \epsilon} + \mathcal{O} ( \dot \epsilon )
\end{equation}
This is the effect that a spectator field has on the spectral tilt---note that there must also be a inflation field to drive the expansion.

\end{exercisenb}

\setcounter{section}{5}

\section{EFT of Large-Scale Structure (Baldauf)}

The series of four lectures \cite{Baldauf:lectures} provide an introduction to the topic of
Large-Scale Structure (LSS) with an emphasis on the EFT approach. The lecture starts with
a general introduction to LSS phenomenology and relevant statistical tools.
Then, the standard cosmological perturbation theory, which allows to compute relevant observables,
is developed. Building on shortcomings of standard perturbation theory, the EFT approach
is introduced with emphasis on different problems it successfully addresses.
Eventually, relation to observations of biased tracers is discussed together with
the need to account for redshift-space distortion.

This section contains five introductory problems useful for a familiarization with the different
aspects of LSS theory. The first two problems deal with standard phenomenology in cosmology as well as
some statistical aspects of Gaussian fields. Problems three and four cover fluid mechanic aspects
in an expanding Universe and problem five is an application of perturbation theory.
Particular aspects of EFT are briefly touched via questions about counterterms and
UV-sensitivity in problem one and five.
Equations from the lecture notes are referred to with a prefix L, e.g. eqn~(L1.1).

\setcounter{exercise}{0}
\begin{exercise}
{\bf Clustering of Fixed Height Subsamples} Consider a Gaussian random
field $\delta$ described by a power
spectrum $P$. The corresponding real space correlation
function is $\xi$ and the variance $\sigma^2 = \xi(0)$. Consider
the PDF of fluctuations
\begin{equation}
\mathbb{P}_{\tx{2pt}}(\delta_1,\delta_2| r) = \frac{1}{(2\pi)^2 |C(r)|}
\exp\left[-\frac{1}{2} Y^T C^{-1}(r) Y\right]
\end{equation}
where
\begin{subequations}
\begin{empheq}[left=\empheqlbrace]{align}
Y &= \begin{pmatrix} \delta_1 \\ \delta_2 \end{pmatrix}
= \sigma \begin{pmatrix} \nu_1 \\ \nu_2 \end{pmatrix} \\
C_{ij}(r) &=
\langle\delta_i \delta_j \rangle =\sigma^2 \delta^{(K)} + (1-\delta^{(K)})\xi(r)
\end{empheq}
\end{subequations}
are respectively the state vector and the covariance matrix.
The correlation matrix of the field can be recovered as
\begin{equation}
\xi(r) = \langle\delta(\tb{x}) \delta(\tb{x}+ \tb{r}) \rangle =
\sigma^4 \int \mathrm{d}\nu_1 \int \mathrm{d}\nu_2 \ \nu_1\nu_2 \
\mathbb{P}_{\tx{2pt}}(\sigma\nu_1, \sigma\nu_2 | r)
\end{equation}
Consider the subset of fluctuations of fixed amplitude
$\nu_c$ and calculate their correlation function
\begin{equation}\label{Def_xi_c}
\xi_c(r)= \frac{\mathbb{P}_{\tx{2pt}}(\sigma\nu_1, \sigma\nu_2 | r)}
{\mathbb{P}_{\tx{1pt}}(\sigma\nu_c) \mathbb{P}_{\tx{1pt}}(\sigma\nu_c)}
-1
\end{equation}
For large separations this allows an expansion in the small
quantity $\xi/\sigma^2$. Write down this expansion to
second order. The prefactors of this expansion are
called bias parameters. Fourier transform the expression
to k-space and consider the low-k limit.
Can you identify contributions that would require a counterterm?

\emph{Remark:} The above model can be used to model regions
that will eventually form dark matter halos, i.e.,
formation sites of galaxies. In this context the field is smoothed
on a scale $R$ to consider fluctuations of a
given mass $M \propto \overline{\rho} R^3$
\begin{equation}
\delta_R(\tb{x}) =
\int
d^3x W_R(|\tb{x}-\tb{x}'|)\delta(\tb{x}') ~.
\end{equation}
where $W_R(r)$ is a Gaussian or top hat filter.

\medskip

\noindent {\bf SOLUTION:}

\noindent {\bf Computation of $\bm{\xi_c(r)}$}

As $\delta$ is a Gaussian random field of variance $\xi(\tb{r})$,
$\delta_1 = \delta(\tb{x})$ and $\delta_2 = \delta(\tb{x}+\tb{r})$ are,
by definition, Gaussian random variables of variance $\xi(0) = \sigma^2$.
Thus,
\begin{equation}
\mathbb{P}_{\tx{1pt}}(\sigma \nu_1) = \frac{1}{\sigma\sqrt{2\pi}}\exp\left[-\frac{\nu_1^2}{2}\right] \ .
\end{equation}

Moreover,
\begin{equation}
C(r) = \left(
\begin{matrix}
\sigma^2 & \xi(r)\\
\xi(r) & \sigma^2
\end{matrix}
\right)
\end{equation}
so that,
\begin{subequations}\label{Cprop}
\begin{empheq}[left=\empheqlbrace]{align}
\abs{C(r)} &= \sigma^4 - \xi(r)^2 \\
C^{-1}(r) &= \frac{1}{\abs{C(r)}}\left(
\begin{matrix}
\sigma^2 & -\xi(r)\\
-\xi(r) & \sigma^2
\end{matrix}
\right)
\ .
\end{empheq}
\end{subequations}
Using \eqref{Cprop} to compute $\mathbb{P}_{\tx{2pt}}$ and plugging it into \eqref{Def_xi_c}
(together with the explicit form of $\mathbb{P}_{\tx{1pt}}$,) one gets
\begin{equation}
\xi_c(r) = \sqrt{\frac{1}{1-\left(\frac{\xi(r)}{\sigma^2}\right)^2}}
\exp\left[ \nu_c^2 \frac{\frac{\xi(r)}{\sigma^2}}{1+\frac{\xi(r)}{\sigma^2}}\right] - 1 \ .
\end{equation}

In the following, the explicit $r$ dependence in $\xi$ is dropped off.

\noindent{\bf Expansion in $\bm{\frac{\xi}{\sigma^2}}$}

Let us assume $\xi \ll \sigma^2$. Expanding in this case,
\begin{align}
\xi_c &= \left(1-\left(\xx\right)^2\right)^{-\frac{1}{2}} \exp\left[ \nu_c^2 \frac{\xx}{1+\xx}\right] - 1\\
&= \left\{ 1 + \frac{1}{2}\left(\xx\right)^2 + O\left(\xx\right)^3\right\} \nonumber \\
&\phantom{= } \times \exp \left[ \nu_c^2
\left\{ \xx - \left(\xx\right)^2 + O\left(\xx\right)^3\right\}
\right] - 1 \\
&= \left\{ 1 + \frac{1}{2}\left(\xx\right)^2 + O\left(\xx\right)^3\right\} \nonumber \\
&\phantom{= } \times \left\{ 1 + \nu_c^2 \xx + \left(\nu_c^2 - \frac{1}{2}\nu_c^4\right) \left(\xx\right)^2
+ O\left(\xx\right)^3\right\} - 1
\end{align}
and finally,
\begin{equation}\label{biasExp}
\xi_c = \nu_c^2 \xx + \left(\frac{1}{2} + \nu_c^2 - \frac{1}{2}\nu_c^4\right)\left(\xx\right)^2
+ O\left(\xx\right)^3 \ .
\end{equation}

\noindent{\bf Low-$k$ limit}

As $P(\tb{k}) = \tx{FT}[\xi](\tb{k})$, the Fourier transformed of \eqref{biasExp} reads
\begin{equation}\label{biasFT}
P_c(\tb{k}) = \nu_c^2 \frac{P(\tb{k})}{\sigma^2}
+ \left(\frac{1}{2} + \nu_c^2 - \frac{1}{2}\nu_c^4\right)\frac{\left(P * P\right)(\tb{k})}{\sigma^4} \ .
\end{equation}
where $P_c$ is the Fourier transformed of $\xi_c$.
Taking the low-$k$ limit, we have
\begin{equation}
P(\tb{k}) \propto k^{n_S}
\end{equation}
and
\begin{align}
(P*P)(\tb{k}) &= \int \frac{\mathrm{d}^3 \tb{q}}{(2\pi)^3} P(\tb{q})P(\tb{k}-\tb{q}) \\
&\simeq \int \frac{\mathrm{d}^3 \tb{q}}{(2\pi)^3} P(\tb{q})P(-\tb{q}) \label{PsquareInt}\\
&\propto k^0 \ .
\end{align}
Therefore, considering $n_S >0$, the first term in \eqref{biasFT} goes to $0$ in the low-$k$ limit, while the second term goes to a constant, provided that \eqref{PsquareInt} is converging to a non-zero constant.

The necessity (or not) of a counterterm is analyzed by looking at the UV sensitivity of each term.
Consider a certain regularization/smoothing of the $\delta$ field. The power spectrum $P = P_R$ picks up
a dependence on the smoothing scale $R$ which, when integrated in a UV divergent integral,
creates a dependence of $P_c$ on $R$.
Thus the $P*P$ term will require a counterterm to cancel this dependence (while the $P$ term does not)\footnote{Even if
the integral would be convergent, the fact that it runs over non-perturbative wavenumber requires
potential counterterms to capture effectively the physics out of the domain of validity, say for $k > k_{\tx{NL}}$.}.
More explicitly, let us consider a family of regulator $W_R(\tb{x})$ such that $\lim_{R\to0} W_R(\tb{x}) = \delta^{(D)}(\tb{x})$ or equivalently
$\lim_{R\to0} W_R(\tb{q}) = 1$.
By definition the regularized fluctuation field is,
\begin{equation}
\delta_R(\tb{x}) = \int \mathrm{d}^3\tb{y} \ W_R(\tb{x}-\tb{y})\delta(\tb{y}) \ .
\end{equation}
Thus
\begin{align}
\xi_R(r) &= \mean{\delta_R(\tb{x})\delta_R(\tb{x}+\tb{r})} \\
&= \int \mathrm{d}^3\tb{y}\mathrm{d}^3\tb{y'} \ W_R(\tb{x}-\tb{y})W_R(\tb{x}+\tb{r}-\tb{y'})
\mean{\delta(\tb{y})\delta(\tb{y'})}\nonumber \\
&= \int \mathrm{d}^3\tb{y}\mathrm{d}^3\tb{r'} \ W_R(\tb{x}-\tb{y})W_R(\tb{x}-\tb{y}+\tb{r}-\tb{r'})
\mean{\delta(\tb{y})\delta(\tb{y}+\tb{r'})}\nonumber \\
&= \int \mathrm{d}^3\tb{y}\mathrm{d}^3\tb{r'} \ W_R(\tb{x}-\tb{y})W_R(\tb{x}-\tb{y}+\tb{r}-\tb{r'}) \xi(\tb{r'})\nonumber \\
&= \int \mathrm{d}^3\tb{y} \ W_R(\tb{x}-\tb{y})\left(W_R * \xi\right)(\tb{x}-\tb{y}+\tb{r})\nonumber \\
&= \int \mathrm{d}^3\tb{y} \ W_R(\tb{y})\left(W_R * \xi\right)(\tb{y}+\tb{r}) \nonumber \\
&= \left( \tilde{W}_R * W_R * \xi \right)(\tb{r}) \label{regCorr}
\end{align}
where for the last line we defined
\begin{equation}
\tilde{W}_R(\tb{x}) \equiv W_R(-\tb{x}) \ .
\end{equation}
Assuming $W_R(\tb{x})$ to be real and even, the Fourier transformed of \eqref{regCorr} reads
\begin{equation}\label{regPower}
P_R(\tb{q}) = W_R^2(\tb{q}) P(\tb{q}) \ .
\end{equation}
Now getting back to \eqref{biasFT}, the first term proportional to $P_R$ converges to $P$
so that the $R$ dependence vanishes (regardless of any assumption on $P$). However in the second term,
one cannot simply take the $R\to0$ limit inside the integral
\begin{equation}
P_R * P_R (\tb{k}) = \int \frac{\mathrm{d}^3 \tb{q}}{(2\pi)^3} W_R^4(\tb{q}) P(\tb{q})P(\tb{k}-\tb{q})
\end{equation}
as it diverges when assuming boldly the low-$k$ form of $P$ to hold for arbitrarily high momentum $q$\footnote{The
conclusion is the same even when taking into account the transfer function $T(\tb{q})$.}.
Therefore, either we have to take into account the full $k$ dependence
of $P$ (but its UV part is not known) or we can employ an EFT approach.
In the latter case one assumes the decoupling between the UV physics and checks it \textit{a posteriori}
(unless it can be proven from a known underlying theory.) Such decoupling has for consequence that
observables are independent of the UV part of $P$. As such, one can use an arbitrary regularized $P_R$
to the price of adding counterterms in the theory. The $R$ dependence of these counterterms are such
that they cancel the arbitrariness of the regularization and allow to recover the independence of observables
regarding the UV physics. In particular, they will cancel the divergence in $P_R*P_R$ when taking
$R \to 0$. Eventually, the remaining $R$-independent terms can be matched to experimental data
at a given $k$ momentum. In this case the counterterm should have a $k^0$ dependence.
Such counterterm corresponds to a stochasticity correction $\epsilon_{2,R}$ in the power spectrum
of fixed amplitude fluctuations such that
\begin{equation}
\lim_{R\to0} \epsilon_{2,R} + \left(\frac{1}{2} + \nu_c^2 - \frac{1}{2}\nu_c^4\right)\frac{\left(P_R * P_R\right)(\tb{k})}{\sigma^4}
\end{equation}
is finite and denoted as
\begin{equation}
\epsilon_{2} + \left(\frac{1}{2} + \nu_c^2 - \frac{1}{2}\nu_c^4\right)\frac{\left(P * P\right)(\tb{k})}{\sigma^4}\ .
\end{equation}
Consequently, the model is modified by including the stochasticicy counterterm and
the power spectrum of fixed amplitude fluctuations now reads as
\begin{equation}
P_{c}(\tb{k}) = \epsilon_{2} + \nu_c^2 \frac{P(\tb{k})}{\sigma^2}
+ \left(\frac{1}{2} + \nu_c^2 - \frac{1}{2}\nu_c^4\right)\frac{\left(P * P\right)(\tb{k})}{\sigma^4}\ .
\end{equation}

\end{exercise}
\begin{exercisenb}

{\bf Equality Scale} Integrating the Bose-Einstein distribution,
we get that the radiation energy density
is related to the temperature $T_{CMB}$ of the CMB photons by
$\rho_{rad} = \frac{\pi^2}{15} T_{CMB}^4$. Use the measured
values of the CMB temperature of $T_{CMB} = 2.725 $K and matter density
$\Omega_{m,0} = 0.28$ to calculate the scale
factor of matter-radiation equality $a_{eq}$.
Calculate the size of the horizon at $a_{eq}$ and the wavenumber
$k_{eq}$
of fluctuations entering at matter-radiation equality.
This is the characteristic scale, at which the transfer
function transitions from the large scale $k_0$ behavior
to the small scale $\ln(k)/k^2$ behavior

\medskip

\noindent {\bf SOLUTION:}

\noindent{\bf Scale factor $a_{\tx{eq}}$ at matter-radiation equality}

At matter-radiation equality one has,
\begin{align}
\Omega_r(a_{\tx{eq}}) &= \Omega_m(a_{\tx{eq}})\\
\Omega_{r,0} a_{\tx{eq}}^{-4} &= \Omega_{m,0} a_{\tx{eq}}^{-3}\\
a_{\tx{eq}} &= \frac{\Omega_{r,0}}{\Omega_{m,0}} \\
a_{\tx{eq}} &= \frac{\rho_{r,0}}{\rho_c\Omega_{m,0}} \\
\end{align}

Assuming only photons contribute to radiation energy-density, one gets
$\rho_{r,0} = \frac{\pi^2}{15}T_{\tx{CMB}}^4$ and
\begin{align}
a_{\tx{eq}} &= \frac{\pi^2}{15}\frac{T_{\tx{CMB}}^4}{\rho_c\Omega_{m,0}} \\
a_{\tx{eq}} &= 8.8 \times 10^{-5} h^{-2} \ .
\end{align}

\noindent{\bf Contribution of Neutrinos}

To be more precise in the computation of $a_\tx{eq}$, we should take neutrinos into account.
Before $a_\tx{eq}$, the temperature is sufficiently large so that we can neglect
the contribution from their masses. As for photons, neutrino's energy density is obtained
by integrating its distribution function (Fermi-Dirac this time)
\begin{equation}
\rho_\nu = 2 \int \frac{\mathrm{d}^3\tb{p}}{(2\pi)^3} \frac{\sqrt{\tb{p}^2 + m_\nu^2}}{e^{\frac{p}{T_\nu}} + 1} \ .
\end{equation}
To see why one can disregard its mass, one can make the change of variable $x=\frac{p}{T_\nu}$
\begin{equation}
\rho_\nu = \frac{1}{\pi^2} T_\nu^4 \int_0^{+\infty} \mathrm{d}x \ x^2 \frac{\sqrt{x^2 + \left(\frac{m_\nu}{T_\nu}\right)^2}}{e^x + 1}
\end{equation}
so that for $T_\nu \gg m_\nu$
\begin{equation}
\rho_\nu = \frac{1}{\pi^2} T_\nu^4 \int_0^{+\infty} \mathrm{d}x \ \frac{x^3}{e^x + 1} \ .
\end{equation}
After integration, taking into account 3 families of neutrinos leads to
\begin{equation}
\rho_\nu = 3 \ \frac{7}{8} \ \frac{\pi^2}{15} \ T_\nu^4 \ .
\end{equation}
To get the energy density contribution of neutrinos, we still need to know their temperature $T_\nu$.
It can be related to the temperature of photons $T_\gamma$ from entropy conservation before and after electrons and
positrons annihilate. This is a standard result in cosmology which gives $T_\nu = \left(\frac{4}{11}\right)^{\frac{1}{3}}T_\gamma$.
Therefore,
\begin{align}
\rho_\nu &= 3 \ \frac{7}{8} \left(\frac{4}{11}\right)^{\frac{4}{3}} \ \rho_\gamma \\
\rho_r &= \left(1 + 3 \ \frac{7}{8} \left(\frac{4}{11}\right)^{\frac{4}{3}}\right) \rho_r^{\slashed{\nu}} \label{corrNeutrino}
\end{align}
and the correction to $a_\tx{eq}$ reads as
\begin{align}
a_\tx{eq} &= \left(1 + 3 \ \frac{7}{8} \left(\frac{4}{11}\right)^{\frac{4}{3}}\right) a_\tx{eq}^{\slashed{\nu}} \\
a_\tx{eq} &= 1.5 \times 10^{-4} h^{-2} \ .
\end{align}

\noindent{\bf Horizon $\chi(a_{\tx{eq}})$}

The size of the horizon at $a_{\tx{eq}}$ is
\begin{align}
\chi(a_{\tx{eq}}) &= c \int_{0}^{t_\tx{eq}} \frac{\mathrm{d}t}{a(t)} \\
&= c \int_{0}^{a_\tx{eq}} \frac{\mathrm{d}a}{a^2H(a)} \ .
\end{align}
In radiative dominated era,
\begin{equation}
H(a) \simeq H_0 \sqrt{\Omega_{r,0} a^{-4}} \label{radApprox}
\end{equation}
so that,
\begin{align}
\chi(a_{\tx{eq}}) &= \frac{c \ a_{\tx{eq}}}{H_0 \sqrt{\Omega_{r,0}}} \\
&= \frac{c}{H_0 \Omega_{m,0}} \sqrt{\Omega_{r,0}} \\
&= \sqrt{1 + 3 \ \frac{7}{8} \left(\frac{4}{11}\right)^{\frac{4}{3}}}
\frac{c}{H_0 \Omega_{m,0}} \frac{\pi \ T_{\tx{CMB}}^2}{\sqrt{15 \ \rho_c}} \\
\chi(a_{\tx{eq}}) &= 5.1 \times 10^{2} h^{-2} \tx{Mpc}
\end{align}
and the corresponding wavenumber $k_\tx{eq}$ gives
\begin{align}
k_\tx{eq} &= \frac{2\pi}{\chi(a_{\tx{eq}})} \\
k_\tx{eq} &= 1.2 \times 10^{-2} h^2 \tx{Mpc}^{-1} \ .
\end{align}

\noindent{\bf Influence of Matter around $a_{\tx{eq}}$}

The approximation made in \eqref{radApprox} might be too crude, especially near
matter-radiation equality. To be more precise, let us do the calculation taking into account
matter energy density. Thus \eqref{radApprox} becomes
\begin{equation}
H(a) \simeq H_0 \sqrt{\Omega_{r,0} a^{-4} + \Omega_{m,0} a^{-3} } \label{mattApprox}
\end{equation}
so that
\begin{align}
\chi(a_{\tx{eq}}) &= c \int_{0}^{a_\tx{eq}} \frac{\mathrm{d}a}{a^2H(a)} \\
&= \frac{c}{H_0\sqrt{\Omega_{r,0}}} \int_{0}^{a_\tx{eq}} \frac{\mathrm{d}a}{\sqrt{1 + \frac{\Omega_{m,0}}{\Omega_{r,0}} a }} \\
&= \frac{c}{H_0\sqrt{\Omega_{r,0}}} \int_{0}^{a_\tx{eq}} \frac{\mathrm{d}a}{\sqrt{1 + \frac{a}{a_\tx{eq}} }} \\
&= \frac{c \ a_\tx{eq}}{H_0\sqrt{\Omega_{r,0}}} \int_{0}^{1} \frac{\mathrm{d}x}{\sqrt{1 + x }} \\
&= \frac{c \ a_\tx{eq}}{H_0\sqrt{\Omega_{r,0}}} \left(2\sqrt{2} - 2 \right) \\
\chi(a_{\tx{eq}}) &= 4.2 \times 10^{2} h^{-2} \tx{Mpc}
\end{align}
and the wavenumber at equivalence becomes
\begin{equation}
k_\tx{eq} = 1.4 \times 10^{-2} h^2 \tx{Mpc}^{-1} \ .
\end{equation}

\end{exercisenb}
\begin{exercisenb}

{\bf Fluid Equations}
Using the definitions of density, mean streaming
velocity and velocity dispersion in
terms of the distribution function $f (\tb{x}, \tb{p}, \tau)$
and the conservation of phase space density $df/ d\tau = 0$, derive
the continuity and Euler equations for collisionless dark matter.
You will need to use the energy momentum
conservation of the homogeneous background Universe $
\overline{\rho}' + 3{\cal H} \overline{\rho} = 0$.

\medskip

\noindent {\bf SOLUTION:}

\noindent{\bf Vlasov's equation}

Using Liouville's theorem for the dark matter distribution function $f(\tb{x},\tb{p},\tau)$,
\begin{equation}
\frac{\mathrm{d}f}{\mathrm{d}\tau} = \frac{\partial f}{\partial \tau}
+ \frac{\mathrm{d}x_i}{\mathrm{d}\tau} \frac{\partial f}{\partial x_i}
+ \frac{\mathrm{d}p_i}{\mathrm{d}\tau} \frac{\partial f}{\partial p_i} = 0 \ ,
\end{equation}
where summation on repeated indices are implied. Besides, the equations of motion for collisionless dark matter reads
\begin{subequations}
\begin{empheq}[left=\empheqlbrace]{align}
x'_i &= \frac{p_i}{a m} \\
p'_i &= - a m \nabla_i \phi
\end{empheq}
\end{subequations}
with $\phi$ the peculiar potential. Hence the collisionless Boltzmann's equation (or Vlasov's equation)
\begin{equation}\label{BoltzLess}
\underbrace{
\left[
\frac{\partial}{\partial \tau}
+ \frac{p_i}{a m} \frac{\partial}{\partial x_i}
- a m \nabla_i \phi \frac{\partial}{\partial p_i}
\right]
}_{\mathcal{L}}
f = 0
\end{equation}
where we have defined the Liouville operator $\mathcal{L}$ for notation convenience.

\noindent{\bf Continuity equation}

Now taking the $0^\tx{th}$ order comoving velocity moment of \eqref{BoltzLess}
\begin{equation}
\int\mathrm{d}^3\tb{p} \ \mathcal{L}f = 0
\end{equation}
one gets,
\begin{equation}
\frac{\partial}{\partial \tau}\left( \int\mathrm{d}^3\tb{p} f\right)
+ \int\mathrm{d}^3\tb{p} \left( \frac{p_i}{a m} \frac{\partial f}{\partial x_i}\right)
- \int\mathrm{d}^3\tb{p} \left( a m \nabla_i \phi \frac{\partial f}{\partial p_i}\right)
= 0 \ .
\end{equation}
Using vanishing boundary conditions on $f$ one can discards the $3^\tx{rd}$ term and one obtains,
using definitions of $\rho$ and $v_i$,
\begin{equation}
\frac{\partial}{\partial \tau}\left( \frac{a^3}{m} \rho \right)
+ \frac{\partial}{\partial x_i} \left( v_i \frac{a^3}{m} \rho \right) = 0 \ .
\end{equation}
Developing $\rho = \bar{\rho}(1+\delta)$ and using the energy-momentum conservation of the
homogeneous background universe, i.e.\ $\bar{\rho}' + 3 \mathcal{H}\bar{\rho} = 0$,
\begin{align}
\frac{a^3}{m}\bar{\rho}'(1+\delta) + \frac{a^3}{m}\bar{\rho}\delta' + 3 \frac{a^2 a'}{m}\bar{\rho}(1+\delta)
+ \bar{\rho} \frac{a^3}{m} \frac{\partial}{\partial x_i} \left( v_i (1+\delta) \right) &= 0 \\
\frac{a^3}{m}\bar{\rho} \left\{ -3 \mathcal{H} + 3 \mathcal{H}
+ \frac{\partial}{\partial x_i} \left( v_i (1+\delta) \right) + \delta'
+ 3 \mathcal{H}\delta - 3 \mathcal{H}\delta \right\} &= 0
\end{align}
and finally we recover the continuity equation
\begin{equation}\label{continuity}
\delta' + \frac{\partial}{\partial x_i} \left( \left(1+\delta\right) v_i \right) = 0 \ .
\end{equation}

\noindent{\bf Euler equation}

Similarly, for the Euler equation, one needs to take the $1^\tx{st}$ comoving velocity moments
\begin{equation}
\int\mathrm{d}^3\tb{p} \ \frac{p_i}{am} \mathcal{L}f = 0
\end{equation}
which reads
\begin{equation}\label{1stmom}
\left( \int\mathrm{d}^3\tb{p} \ \frac{p_i}{am} \frac{\partial f}{\partial \tau} \right)
+ \int\mathrm{d}^3\tb{p} \left( \ \frac{p_i}{am} \frac{p_j}{a m} \frac{\partial f}{\partial x_j}\right)
- \int\mathrm{d}^3\tb{p} \left( \frac{p_i}{am} a m \nabla_j \phi \frac{\partial f}{\partial p_j}\right)
= 0 \ .
\end{equation}
Integrating by part, the $1^\tx{st}$ term of \eqref{1stmom} reads as
\begin{equation}
\frac{\partial}{\partial \tau}\left( \frac{a^3}{m} \rho v_i \right) + \frac{a^3\rho}{m}\mathcal{H}v_i
\end{equation}
while the $2^\tx{nd}$ term reads as
\begin{equation}
\frac{a^3}{m}\frac{\partial}{\partial x_j}\left(\sigma_{ij} \rho \right)
+ \frac{a^3}{m}\frac{\partial}{\partial x_j}\left(v_i v_j \rho \right)
\end{equation}
and the last term as
\begin{equation}
-\nabla_j \phi \int \mathrm{d}^3\tb{p} \
\left[ \frac{\partial}{\partial p_j}\left(p_i f\right)
- \frac{\partial p_i}{\partial p_j} f \right] = \nabla_j(\phi) \frac{a^3}{m}\rho \ ,
\end{equation}
again assuming vanishing boundary conditions.
Dividing \eqref{1stmom} by $\frac{a^3}{m}\rho$ one gets
\begin{equation}
\frac{1}{a^3\rho} \frac{\partial}{\partial \tau}\left(a^3 \rho v_i \right)
+ \frac{1}{\rho} \frac{\partial}{\partial x_j}\left(v_i v_j \rho \right) + \mathcal{H} v_i
= -\nabla_i(\phi) - \frac{1}{\rho}\frac{\partial}{\partial x_j}\left(\sigma_{ij} \rho \right) \ .
\end{equation}
Developing the first two terms and canceling the background contributions using again $\bar{\rho}' + 3 \mathcal{H}\bar{\rho} = 0$,
one gets
\begin{multline}
\frac{1}{a^3\rho} \frac{\partial}{\partial \tau}\left(a^3 \rho v_i \right)
+ \frac{1}{\rho} \frac{\partial}{\partial x_j}\left(v_i v_j \rho \right) \\
= \frac{\partial v_i}{\partial \tau} + \frac{v_i}{1+\delta}\left[ \delta' + \frac{\partial\delta}{\partial x_j} v_ j \right]
+ v_i \left(\frac{\partial}{\partial x_j} v_j \right) + \left(v_j \frac{\partial}{\partial x_j}\right) v_i \ .
\end{multline}
Eventually, using the continuity relation \eqref{continuity} which reads
\begin{equation}
\delta' + \frac{\partial\delta}{\partial x_j} v_ j = - \left(\frac{\partial}{\partial x_j} v_j \right)\times(1+\delta)
\end{equation}
one gets the Euler equation
\begin{equation}
\frac{\partial v_i}{\partial \tau} + \left(v_j \frac{\partial}{\partial x_j}\right) v_i + \mathcal{H} v_i
= -\nabla_i(\phi) - \frac{1}{\rho}\frac{\partial}{\partial x_j}\left(\sigma_{ij} \rho \right) \ .
\end{equation}

\end{exercisenb}

\begin{exercisenb}

\noindent{\bf Recursion Relations}
Starting from the $k$-space version of the continuity and
Euler equations in a
matter-only EdS Universe, and the ansatz
$$
\delta (\bm{k},\tau) =
\sum_{i=1}^\infty
a^i (\tau)\delta^{(i)}(\bm{k})
~~~~~~~~~~~,~~~~~~~~~~~
\theta (\bm{k},\tau) =
-{\cal H}(\tau) \sum_{i=1}^\infty
a^i (\tau) \theta^{(i)}(\bm{k})
$$
derive the recursion relations for the gravitational coupling
kernels $F_n$ and $G_n$ relating
the $n$-th order fields
to the linear density fields
\bea
\delta^{(n)} (\bm{k})& =&
\prod_{m=1}^n\left\{ \int
\frac{d^3 q_m}{(2\pi)^3}
\delta^{(1)}(\bm{q}_m) \right\} F_n(\bm{q}_1,...,\bm{q}_n)
\delta^{(D)} (\bm{k} - \bm{q}|_1^n)
\nonumber \\
\theta^{(n)} (\bm{k})& =&
\prod_{m=1}^n\left\{ \int
\frac{d^3 q_m}{(2\pi)^3}
\delta^{(1)}(\bm{q}_m) \right\} G_n(\bm{q}_1,...,\bm{q}_n)
\delta^{(D)} (\bm{k} - \bm{q}|_1^n)
\nonumber
\eea

\medskip

\noindent {\bf SOLUTION:}

Let us consider a matter-only EdS Universe, neglecting vorticity, the decaying mode and taking the ansatz
\begin{subequations}\label{ansatz}
\begin{empheq}[left=\empheqlbrace]{align}
\delta(\tb{k},\tau) &= \sum_{n=1}^{\infty} a^n(\tau)\delta^{(n)}(\tb{k}) \\
\theta(\tb{k},\tau) &= - \mathcal{H}(\tau) \sum_{n=1}^{\infty} a^n(\tau)\theta^{(n)}(\tb{k})
\end{empheq}
\end{subequations}
as solution of Euler and continuity equations
\begin{subequations}
\begin{align}
\delta'(\tb{k}) + \theta(\tb{k}) &=
-\int \frac{\mathrm{d}^3\tb{q}}{2\pi} \frac{\mathrm{d}^3\tb{q'}}{2\pi}
\delta^{(D)}(\tb{k} - \tb{q} - \tb{q'}) \\
&\phantom{= -\int \frac{\mathrm{d}^3\tb{q}}{2\pi}} \times \alpha(\tb{q},\tb{q'})\theta(\tb{q})\delta(\tb{q'}) \nonumber\\
\theta'(\tb{k}) + \mathcal{H} \theta(\tb{k}) + \frac{3}{2} \Omega_m \mathcal{H}\delta(\tb{k}) &=
-\int \frac{\mathrm{d}^3\tb{q}}{2\pi} \frac{\mathrm{d}^3\tb{q'}}{2\pi}
\delta^{(D)}(\tb{k} - \tb{q} - \tb{q'}) \nonumber\\
&\phantom{= -\int \frac{\mathrm{d}^3\tb{q}}{2\pi}} \times \beta(\tb{q},\tb{q'})\theta(\tb{q})\theta(\tb{q'}) \ .
\end{align}
\end{subequations}

In this case $\Omega_m = 1$ and Friedman's equation gives us
\begin{equation}
\mathcal{H}'(\tau) = - \frac{1}{2} \mathcal{H}^2(\tau) \ .
\end{equation}
Therefore,
\begin{subequations}\label{series}
\begin{equation}
\delta'(\tb{k},\tau) = \sum_{n=1}^{\infty} a^n(\tau)\left[ n \mathcal{H}(\tau) \delta^{(n)}(\tb{k})\right]
\end{equation}
and
\begin{equation}
\theta'(\tb{k},\tau) = \sum_{n=1}^{\infty} a^n(\tau)\left[ \left(\frac{1}{2} - n\right) \mathcal{H}^2(\tau) \theta^{(n)}(\tb{k})\right] \ .
\end{equation}
Since we are only interested in the finite perturbation expansion\footnote{
Note that it is not necessary the case but one could extend the derivations for a given
resummation scheme. For example, using the usual limit of partial sums or
a Borel resummation would lead to similar results as they are both compatible with
product, sum and derivative of series.
The only care to be taken in these cases would be to consider the domain of convergence.},
$\theta \times \delta$ at order $n$ is just the Cauchy product i.e.\
\begin{align}
\theta(\tb{q},\tau)\delta(\tb{q'},\tau) &= \sum_{n=1}^{\infty} a^n(\tau)
\left[ -\mathcal{H}(\tau) \sum_{p=1}^{n-1} \theta^{(p)}(\tb{q})\delta^{(n-p)}(\tb{q'})\right] \\
\theta(\tb{q},\tau)\theta(\tb{q'},\tau) &= \sum_{n=1}^{\infty} a^n(\tau)
\left[ \mathcal{H}^2(\tau) \sum_{p=1}^{n-1} \theta^{(p)}(\tb{q})\theta^{(n-p)}(\tb{q'})\right] \ .
\end{align}
\end{subequations}

Inserting \eqref{series} in the continuity equation one gets that for all $n \geq 1$,
\begin{multline}
n \delta^{(n)}(\tb{k}) - \theta^{(n)}(\tb{k})
= \int \frac{\mathrm{d}^3\tb{q}}{2\pi} \frac{\mathrm{d}^3\tb{q'}}{2\pi}
\delta^{(D)}(\tb{k} - \tb{q} - \tb{q'}) \alpha(\tb{q},\tb{q'}) \\
\times \left[ \sum_{p=1}^{n-1} \theta^{(p)}(\tb{q})\delta^{(n-p)}(\tb{q'}) \right] \ .
\end{multline}

Now inserting the explicit dependence of $\delta^{(n)}$ and $\theta^{(n)}$ in terms of $\delta^{(1)}$ and the kernels
$F_n$ and $G_n$, one gets the equality for the integrand of $\tb{q}_1,\ldots,\tb{q}_n$
\begin{multline}
\left[n F_n(\tb{q}_1, \ldots, \tb{q}_n) - G_n(\tb{q}_1, \ldots, \tb{q}_n)\right] \delta^{(D)}(\tb{k} - \tb{q}|_1^n) \\
= \int \frac{\mathrm{d}^3\tb{q}}{2\pi} \frac{\mathrm{d}^3\tb{q'}}{2\pi}
\delta^{(D)}(\tb{k} - \tb{q} - \tb{q'}) \alpha(\tb{q},\tb{q'})
\times \left[ \sum_{p=1}^{n-1} G_p(\tb{q}_1, \ldots, \tb{q}_p) \delta^{(D)}(\tb{q} - \tb{q}|_1^p) \right.\\
\left. F_{n-p}(\tb{q}_{p+1}, \ldots, \tb{q}_n) \delta^{(D)}(\tb{q} - \tb{q}|_{p+1}^n) \right] \ .
\end{multline}
Killing the integral on $\tb{q}$ and $\tb{q'}$, one gets the first recursion relation on the kernels
\begin{subequations}\label{recursions}
\begin{multline}\label{recursion1}
n F_n(\tb{q}_1, \ldots, \tb{q}_n) - G_n(\tb{q}_1, \ldots, \tb{q}_n) =
\sum_{p=1}^{n-1} \alpha\left(\tb{q}|_1^p,\tb{q}|_{p+1}^n\right)
G_p(\tb{q}_1, \ldots, \tb{q}_p) \\
\times F_{n-p}(\tb{q}_{p+1}, \ldots, \tb{q}_n) \ .
\end{multline}
Similarly, by inserting \eqref{series} into the Euler equation, one gets the second recursion relation
on the kernels which reads,
\begin{multline}\label{recursion2}
\left(2 n + 1\right) G_n(\tb{q}_1, \ldots, \tb{q}_n) - 3 F_n(\tb{q}_1, \ldots, \tb{q}_n) =
\sum_{p=1}^{n-1} 2 \beta\left(\tb{q}|_1^p,\tb{q}|_{p+1}^n\right)
G_p(\tb{q}_1, \ldots, \tb{q}_p) \\ \times G_{n-p}(\tb{q}_{p+1}, \ldots, \tb{q}_n) \ .
\end{multline}
\end{subequations}

Finally, by combining \eqref{recursion1} and \eqref{recursion2} one obtains
the recursion relations for the gravitational kernels
\begin{align}
F_n(\tb{q}_1, \ldots, \tb{q}_n) &= \sum_{p=1}^{n-1} \frac{G_p(\tb{q}_1, \ldots, \tb{q}_p)}{(2n+3)(n-1)}
\left\{ (2n+1)\alpha\left(\tb{q}|_1^p,\tb{q}|_{p+1}^n\right)F_{n-p}(\tb{q}_{p+1}, \ldots, \tb{q}_n) \right.\nonumber \\
&\phantom{= \sum_{p=1}^{n-1} \frac{G_p(\tb{q}_1, \ldots, \tb{q}_p)}{(2n+3)(n-1)} \{}
\left. + 2 \beta\left(\tb{q}|_1^p,\tb{q}|_{p+1}^n\right) G_{n-p}(\tb{q}_{p+1}, \ldots, \tb{q}_n)\right\} \\
G_n(\tb{q}_1, \ldots, \tb{q}_n) &= \sum_{p=1}^{n-1} \frac{G_p(\tb{q}_1, \ldots, \tb{q}_p)}{(2n+3)(n-1)}
\left\{ 3 \alpha\left(\tb{q}|_1^p,\tb{q}|_{p+1}^n\right)F_{n-p}(\tb{q}_{p+1}, \ldots, \tb{q}_n) \right.\nonumber \\
&\phantom{= \sum_{p=1}^{n-1} \frac{G_p(\tb{q}_1, \ldots, \tb{q}_p)}{(2n+3)(n-1)} \{}
\left. + 2 n \beta\left(\tb{q}|_1^p,\tb{q}|_{p+1}^n\right) G_{n-p}(\tb{q}_{p+1}, \ldots, \tb{q}_n)\right\}
\end{align}

\end{exercisenb}
\begin{exercisenb}

{\bf Two-loop power spectrum}
Write down the the diagrams and integrals contributing to the two-loop
matter power spectrum in terms of the gravitational coupling kernels $F_n$.
Try to identify the diagrams with
the strongest UV-sensitivity.

\medskip

\noindent {\bf SOLUTION:}

\noindent{\bf Diagrams and integral at two-loop order}

In perturbation theory, connected diagrams contributing to the power spectrum $P(\tb{k}) \equiv \mean{\delta(\tb{k}) \delta(-\tb{k})}$ contains only two vertices $F_{n_1}$ and $F_{n_2}$
arising from the expansion of $\delta$ functions using the ansatz \eqref{ansatz}. Using the topological identity
for connected planar graphs $L = I - V + 1$ where $L$ is the number of loops, $I$ the number of internal lines and $V$ the number of vertices, leads to
\begin{equation}\label{topo2Loop}
n_1 + n_2 = 6
\end{equation}
for two-loop diagrams. Two-loop contributions to $P(\tb{k})$ corresponds to any contribution verifying \eqref{topo2Loop}.

Therefore\footnote{The time-dependence is omitted for clarity. Assuming the time-evolution factorizes at linear order leads to a simple power of linear growth factor},
\begin{align}
P_\tx{2-loop}(\tb{k}) &= 2\mean{\delta^{(1)}(\tb{k}) \delta^{(5)}(-\tb{k})}
+ 2\mean{\delta^{(2)}(\tb{k}) \delta^{(4)}(-\tb{k})}
+ \mean{\delta^{(3)}(\tb{k}) \delta^{(3)}(-\tb{k})} \ , \\
&\equiv P_{15}(\tb{k}) + P_{24}(\tb{k}) + P_{33}(\tb{k}) \ ,
\end{align}
where $P_{n_1 n_2}(\tb{k})$ corresponds to the sum of all diagrams with vertices $F_{n_1}$ and $F_{n_2}$.
Assuming Gaussianity of $\delta^{(1)}(\tb{k})$ one can apply the Wick theorem leading to four distinct diagrams depicted in Fig.~\ref{2loopDiags}.
Defining $F^{(s)}_n$ as the symmetrized kernel $F_n$, associated integrals read as
\begin{subequations}\label{2loopPowerSpectra}
\begin{align}
P_{15}(\tb{k}) &= 30 \int \frac{\mathrm{d}^3 \tb{q}_1}{(2\pi)^3} \frac{\mathrm{d}^3 \tb{q}_2}{(2\pi)^3} \
F^{(s)}_{5}\left(\tb{q}_1,-\tb{q}_1,\tb{q}_2,-\tb{q}_2,\tb{k}\right) \nonumber\\
&\phantom{= 30 \int \frac{\mathrm{d}^3 \tb{q}_1}{(2\pi)^3} \frac{\mathrm{d}^3 \tb{q}_2}{(2\pi)^3}} \
\times P_{11}(\tb{q}_1) P_{11}(\tb{q}_2) P_{11}(\tb{k}) \ , \\
P_{24}(\tb{k}) &= 24 \int \frac{\mathrm{d}^3 \tb{q}_1}{(2\pi)^3} \frac{\mathrm{d}^3 \tb{q}_2}{(2\pi)^3} \
F^{(s)}_{4}\left(\tb{q}_1,\tb{k}-\tb{q}_1,\tb{q}_2,-\tb{q}_2\right) F^{(s)}_{2}\left(-\tb{q}_1,-(\tb{k}-\tb{q}_1)\right) \nonumber \\
&\phantom{= 30 \int \frac{\mathrm{d}^3 \tb{q}_1}{(2\pi)^3} \frac{\mathrm{d}^3 \tb{q}_2}{(2\pi)^3}} \
\times P_{11}(\tb{q}_1) P_{11}(\tb{k} - \tb{q}_1) P_{11}(\tb{q}_2) \ , \\
P^{\tx{R}}_{33}(\tb{k}) &= 9 \int \frac{\mathrm{d}^3 \tb{q}_1}{(2\pi)^3} \frac{\mathrm{d}^3 \tb{q}_2}{(2\pi)^3} \
F^{(s)}_{3}\left(\tb{q}_1,-\tb{q}_1,\tb{k}\right) F^{(s)}_{3}\left(\tb{q}_2,-\tb{q}_2,-\tb{k}\right) \nonumber \\
&\phantom{= 30 \int \frac{\mathrm{d}^3 \tb{q}_1}{(2\pi)^3} \frac{\mathrm{d}^3 \tb{q}_2}{(2\pi)^3}} \
\times P_{11}(\tb{q}_1) P_{11}(\tb{q}_2) P_{11}(\tb{k}) \ , \\
P^{\tx{I}}_{33}(\tb{k}) &= 6 \int \frac{\mathrm{d}^3 \tb{q}_1}{(2\pi)^3} \frac{\mathrm{d}^3 \tb{q}_2}{(2\pi)^3} \
F^{(s)}_{3}\left(\tb{q}_1,\tb{q}_2,\tb{k} - \tb{q}_1 - \tb{q}_2\right) F^{(s)}_{3}\left(-\tb{q}_1,-\tb{q}_2,-(\tb{k}-\tb{q}_1-\tb{q}_2)\right) \nonumber \\
&\phantom{= 30 \int \frac{\mathrm{d}^3 \tb{q}_1}{(2\pi)^3} \frac{\mathrm{d}^3 \tb{q}_2}{(2\pi)^3}} \
\times P_{11}(\tb{q}_1) P_{11}(\tb{q}_2) P_{11}(\tb{k} - \tb{q}_1 - \tb{q}_2) \ ,
\end{align}
\end{subequations}
where $P^{\tx{R}}_{33}(\tb{k})$ and $P^{\tx{I}}_{33}(\tb{k})$ correspond respectively to the $1$PR and $1$PI diagram with two $F_3$ vertices.

\begin{figure}
\begin{center}
\includegraphics[width=10cm]{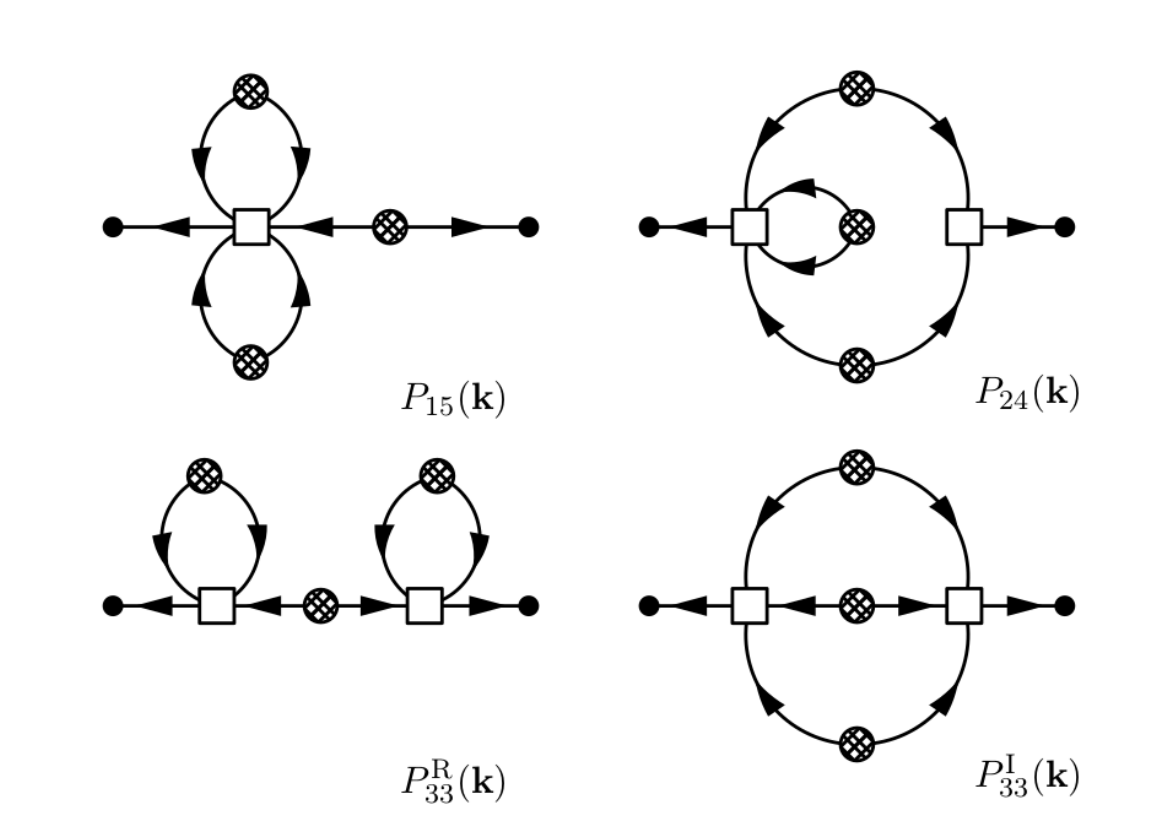}
\end{center}
\caption{Two-loop diagrams contributing to the power spectrum.}
\label{2loopDiags}
\end{figure}

\noindent{\bf Ultraviolet sensitivity}

To analyze systematically the ultraviolet sensitivity of integrals in \eqref{2loopPowerSpectra} one can in principle simply apply Weinberg's asymptotic theorem \cite{Weinberg:1959nj} to the integrands.
However it quickly becomes cumbersome to analyze all possible subspaces of integration and the associated asymptotic coefficients of the gravitational coupling kernels $F^{(s)}_n$ in these subspaces.
Instead one can derive an upper bound on the superficial degree of divergence $D_{\tx{UV}}$ for any diagram.

The linear power spectrum $P_{11}$ scales as a power $n$ of the momentum in the UV limit. Combined with a factor $3$ for each loop integration leads to
\begin{equation}
D_{\tx{UV}} \leq 3L - n I + \sum_i D( F^{(s)}_{n_i} )
\end{equation}
where the sum is on the set of vertices (inside a loop) and $D( F^{(s)}_{n_i} )$ is an upper bound on the asymptotic coefficients of $F^{(s)}_{n_i}$.
Since for $p \gg k$ \cite{Baldauf:lectures}
\begin{equation}\label{RefineUVdiv}
F^{(s)}_{n_i}(\tb{q}_1, \ldots, \tb{q}_{n_i - 2}, \tb{p}, -\tb{p}) \propto \frac{k^2}{p^2} \ ,
\end{equation}
it is clear that $D( F^{(s)}_{n_i} ) \geq -2$. However, the asymptotic behavior of the kernel could be different when taking
more than two momenta in the UV limit. To easily obtain an upper bound on $D( F^{(s)}_{n_i} )$ one typically uses a decoupling argument from the UV physics
so that $D( F^{(s)}_{n_i} ) < 0$ (otherwise the mass fluctuation derived in perturbation theory would strongly depend on the UV initial conditions.) See for example \cite{Goroff:1986ep} where it is claimed that $D( F^{(s)}_{n_i} ) < 0$ has been checked by explicit computation for $n_i \leq 6$ (which is enough in our case.)
Then, because $F^{(s)}_{n_i}$ is a rational function in the momenta $D( F^{(s)}_{n_i} )$ can only be an integer so that $D( F^{(s)}_{n_i} ) \leq -1$.
Thus,
\begin{equation}\label{UVdiv}
D_{\tx{UV}} \leq 3L - n I - V \ ,
\end{equation}
where $V$ is the number of vertices (inside a loop) of the diagram.
Eventually, using \eqref{RefineUVdiv}, \eqref{UVdiv} can be refined to
\begin{equation}\label{UVdivT}
D_{\tx{UV}} \leq 3L - n I - V - T \ ,
\end{equation}
where $T$ denotes the number of tadpoles of the diagram.

Applying \eqref{UVdivT} to two-loop diagrams leads to
\begin{subequations}\label{UVbound}
\begin{align}
D_{\tx{UV}}(P_{15}) &\leq 3 + 2 n \ , \\
D_{\tx{UV}}(P_{24}) &\leq 3 + 3 n \ , \\
D_{\tx{UV}}(P^{\tx{R}}_{33}) &\leq 2 + 2 n \ , \\
D_{\tx{UV}}(P^{\tx{I}}_{33}) &\leq 4 + 3 n \ .
\end{align}
\end{subequations}
As seen in \eqref{UVbound} the UV-sensitivity of a diagram depends on $n$.
From now on we assume the inequalities \eqref{UVbound} to be saturated.
Regardless of $n$
\begin{subequations}
\begin{align}
D_{\tx{UV}}(P^{\tx{R}}_{33}) &\leq D_{\tx{UV}}(P_{15}) \ , \\
D_{\tx{UV}}(P_{24}) &\leq D_{\tx{UV}}(P^{\tx{I}}_{33}) \ ,
\end{align}
\end{subequations}
so that the most UV-sensitive two-loop diagram is either $P_{15}$ or $P^{\tx{I}}_{33}$.
One conclude that the most UV-sensitive diagram depends on the model.
\emph{If $n > -1$ then the most UV-sensitive diagram is $P^{\tx{I}}_{33}$ else
it is $P_{15}$.}
Note that to be complete, one should also check the superficial degree of divergence of any sub-diagram.

\end{exercisenb}

\newpage

\section*{Acknowlegements}

We would like to thank all the lecturers of the school, namely
T.~Baldauf,
T.~Becher,
C.~Burgess,
S.~Caron-Huot,
J.~Hisano,
U.~van Kolck,
T.~Mannel,
A.~Manohar,
M.~Neubert,
A.~Pich,
L.~Silvestrini,
R.~Sommer and
P.~Vanhove,
as well as the organizers, Sacha Davidson, Paolo Gambino, Mikko Laine and Matthias Neubert for making this extremely interesting, fun and adventurous school possible. We all gained a lot of knowledge and understanding in these intense four weeks. Especially, we want to thank Sacha Davidson, who always looked after us, helped whenever it was needed and motivated us to finish this write up. Without her it would not exist in this form.

\bibliographystyle{OUPnum}
\bibliography{leshouches}

\end{document}